\documentclass[final,1p,times,authoryear]{elsarticle}

\usepackage{graphicx}
\usepackage[table]{xcolor}
\usepackage{multirow}

\usepackage{amssymb}
\usepackage{amsmath}
\usepackage{mathrsfs}  % for mathscr font
\usepackage{cancel}
\usepackage{breqn}
\usepackage{siunitx}
\usepackage[normalem]{ulem}
\usepackage{caption}
\usepackage{subcaption}
\usepackage{lineno,hyperref}
\modulolinenumbers[5]

\renewcommand{\l}{\ell}

\newcommand{\R}{\mathbb{R}}

\newcommand{\vc}[1]{\mathbf{#1}}

\newcommand{\matrixII}[4]{\left(\begin{array}{cc}#1&#2\\#3&#4\end{array}\right)}

\newcommand{\sgn}{\text{sgn}}
\newcommand{\const}{\text{const}}

\newsavebox{\measurebox}

\begin{document}

\title{A continuum geometric approach for inverse design of origami structures}

\author[inst1]{Alon Sardas}
\author[inst1]{Michael Moshe}
\author[inst2]{Cy Maor}

\affiliation[inst1]{organization={Racah Institute of Physics, The Hebrew University of Jerusalem},%Department and Organization
            addressline={Address One}, 
            city={Jerusalem},
            postcode={9190401}, 
            country={Israel}}
\affiliation[inst2]{organization={Einstein Institute of Mathematics, The Hebrew University of Jerusalem},%Department and Organization
            addressline={Address One}, 
            city={Jerusalem},
            postcode={9190401}, 
            country={Israel}}

\begin{abstract}
Miura-Ori, a celebrated origami pattern that facilitates functionality in matter, has found multiple applications in the field of mechanical metamaterials. 
Modifications of Miura-Ori pattern can produce curved configurations during folding, thereby enhancing its potential functionalities. Thus, a key challenge in designing generalized Miura-Ori structures is to tailor their folding patterns to achieve desired geometries. In this work, we address this inverse-design problem by developing a new continuum framework for the differential geometry of generalized Miura-Ori. 
By assuming that the perturbation to the classical Miura-Ori is slowly varying in space, we derive analytical relations between geometrical properties and the perturbation field.
These relationships are shown to be invertible, allowing us to design complex curved geometries.
Our framework enables porting knowledge, methods and tools from continuum theories of matter and differential geometry to the field of origami metamaterials.
\end{abstract}

\maketitle

\section{Introduction} 
\begin{figure}
	\centering
	\includegraphics[width = 0.7\linewidth]
	{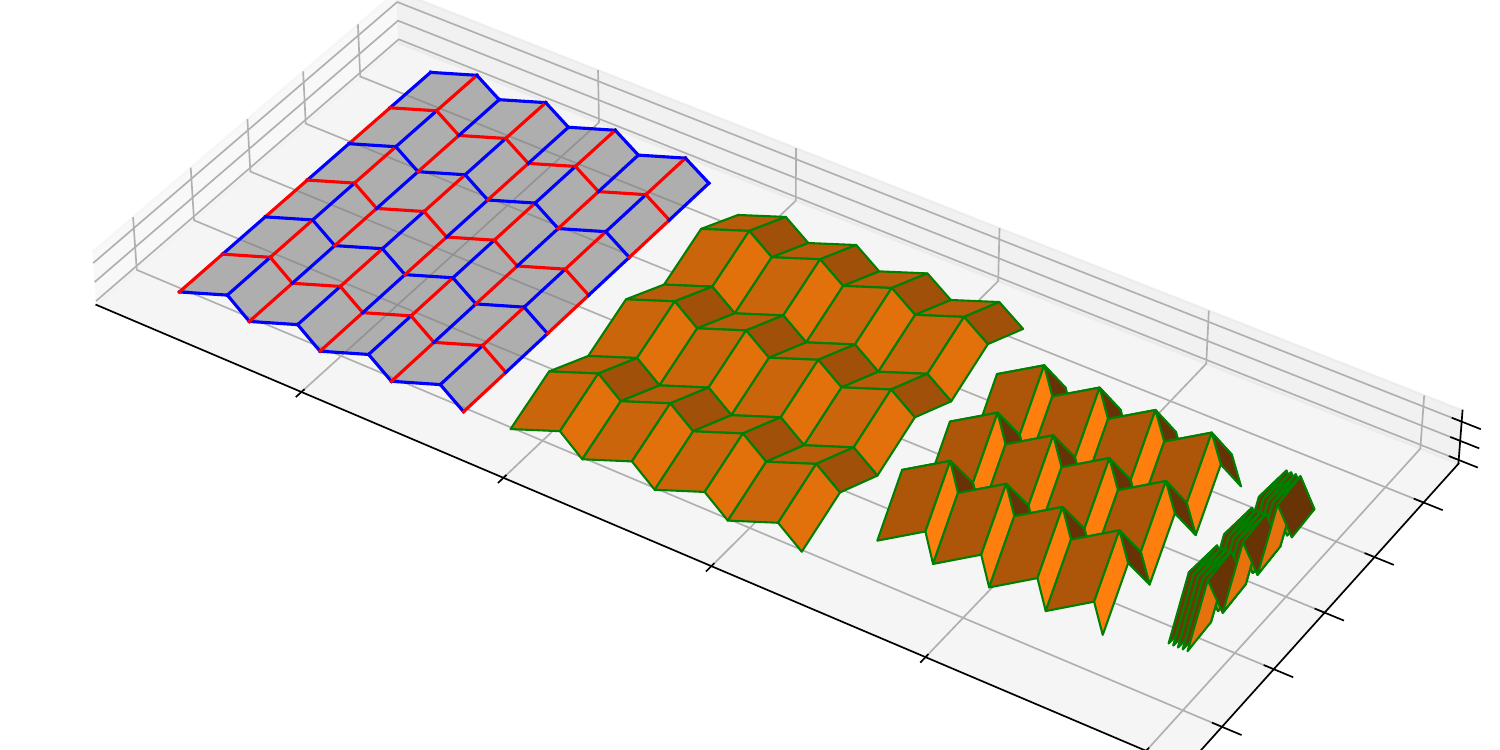}
	
    \caption{
        A Miura-Ori pattern in its flat configuration (left), its kinematics when it is partially-folded (2 patterns in the middle) and its fully-folded configuration, where it is flat and compact (right).
        In the flat configuration, the mountain and valley assignments are shown in red and blue lines, respectively.
    }
    \label{fig:miura-ori}
\end{figure}

When a sheet of paper is tightly confined to a small volume, such as when it is crumpled, it develops a disordered pattern of creases and folds. However, the emergent pattern is far from optimal for the task of volume reduction.
Indeed, ordered origami patterns may be much more efficient and can fold to a completely flat configuration.
A celebrated type of such an origami pattern is the Miura-Ori pattern (fig.~\ref{fig:miura-ori}).
The advantages of Miura-Ori extend beyond its efficiency: it induces a one-parameter family of configurations that spans from the initial flat configuration to the final flat and fully-folded one. 
As such, the folding process is reversible and highly controlled, making it of potential use for designing material's functionality. An additional advantage of Miura-Ori, as opposed to a general origami pattern, is that it significantly simplifies the mathematical description of the folding pattern and its analysis.
Indeed, Miura-Ori (MO) has found multiple potential applications from the macro-scale of satellite's solar panels to nano-origami of graphene \cite{miura1985method, ho2020complex}.

A central property of the MO pattern is that during the folding process the pattern remains flat on average (Fig.~\ref{fig:miura-ori}).
The modification of MO beyond its classical pattern may allow for curved configuration along the folding process, and thus widen the possible functionalities. In a series of recent works it was shown that generalized Miura-Ori patterns (GMO), in which the creases are non-uniform, induce curved configurations during the folding process \cite{tachi2009generalization, lang2018rigidly, feng2020designs}. As a result, GMO provides a shaping mechanism that, upon activation, curves an initially flat sheet. 
As such, it defines two interesting geometric problems: First, given a foldable creases pattern, what is the induced curved geometry? Second, it defines the inverse problem of designing a pattern that induces a desired curved geometry.

These problems are not unique to GMO, and are relevant to any shaping mechanism. Examples for such shaping mechanisms include static and responsive kirigami \cite{sussman2015algorithmic, neville2016shape,cheng2020kirigami, peng2024programming}, temperature-responsive NIPA gels \cite{klein2007shaping, kim2012designing, wu2013three}, and liquid crystal elastomers \cite{aharoni2014geometry,mostajeran2016encoding,aharoni2018universal}. 

In all these cases, the inverse design problem reduces to two main tasks: First, relating the emergent geometry with the microscopic shape mechanism, and second, optimizing the microscopic degrees of freedom to target the desired geometry. 
An example of an inverse design of GMO, which is at the focus of the current work, is presented in \cite{dudte2016programing}. In that work it was shown that upon treating the lengths and angles of the GMO pattern as variables, and defining a loss function that measures a distance from a desired configuration in $\R^3$, a numerical optimization process converges and yields a GMO that approximates well the desired geometry. 
In \cite{dang2022inverse}, a more geometric approach was adopted, where instead of targeting a configuration in space, one could target geometric properties, e.g., curvature and distance within the desired surface, and define a geometric loss function correspondingly.  

From a practical perspective, these approaches are successful: Given a desired geometry, expressed either as a configuration or in terms of its geometric properties, the optimization schemes are expected to converge. 
Still, certain questions remain out of the scope in these approaches due to their numerical basis.
For example, under which conditions a solution of a  GMO exists given a desired geometry.
While it is possible to test the methods for given geometries and find if a solution exists, a systematic analysis cannot rely on these numerical schemes --- calling for an analytic approach to this problem, similar to those developed for the inverse design of liquid crystal elastomers \cite{aharoni2014geometry}.

Previous attempts to develop analytical methods to design GMO were limited to systems with high level of symmetry, e.g., cylindrical or axisymmetric origami structures \cite{zhou2015design, wang2016folding, song2017design, dang2022deployment}, thus emphasizing the need in a more general approach to deal with other geometries.
In this work, we develop a new framework to attack this inverse design problem. First, we treat the perturbations as slowly varying in space and therefore the entire origami pattern is described by smooth fields of the lengths and sector angles, and second, we consider the local geometrical properties of the folded origami pattern, and these properties are used for the design algorithm.

\begin{figure}
	\centering
	\includegraphics[width = 1.0\linewidth]
	{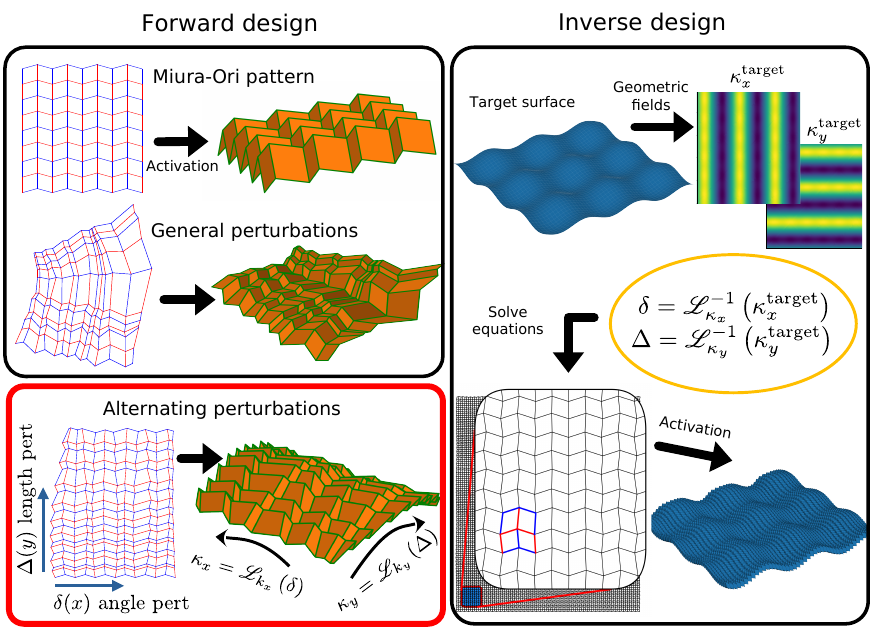}
	
	\caption{
        Illustration of the forward and inverse design processes using perturbed Miura-Ori patterns.
        The classical Miura-Ori is shown in its flat configuration and after activating it, i.e., folding it to a specific folded configuration.
        The Miura-Ori may be generalized to rigidly and flat-foldable quadrilateral mesh origami. Example of such pattern is shown under general perturbation, and upon activation, the pattern may be used to design complex surfaces.
        We suggest the alternating perturbation described by the perturbation fields $\delta(x)$ of the angles and $\Delta(y)$ of the lengths. These are applied to the L-shaped boundary of the pattern, and the rest of the pattern is determined uniquely due to the constraints of the pattern.
        An example of this pattern is shown inside the red frame.
        After folding, the pattern may approximate non-trivial surfaces, whose geometrical properties can be calculated as function of the perturbations and their derivatives.
        In particular, the principal curvatures $\kappa_x$ and $\kappa_y$ are obtained by applying the differential operators $\mathscr{L}_{k_x}$ and $\mathscr{L}_{k_y}$ to the perturbations $\delta(x),\Delta(y)$, respectively.
        In the inverse design process (right part of the figure), the target surface is given and it dictates the principal curvatures $\kappa_{x}^{\text{target}}$ and $\kappa_{y}^{\text{target}}$.
        The equations for the perturbations, induced by the differential operators, are solved to find the required perturbations $\delta$ and $\Delta$ for the desired curvatures(the algorithm is presented in detail in sec.~\ref{sec:inverse-design-results}).
        These perturbations are used to build the origami pattern, that upon activation should approximate the target surface.
    }
    \label{fig:forward-and-inverse-design}
\end{figure}

As a manifestation of our approach, we focus on a specific angle-length-based perturbation of MO where the perturbed MO is fully determined by the angles perturbation along one boundary, and the lengths perturbation along the perpendicular one (see fig.~\ref{fig:forward-and-inverse-design} left panel). 
Upon assuming that the perturbations to the MO are slowly varying in space, we develop a continuum formulation of GMO in which we relate the perturbation fields with the emergent geometry. We show that these analytical relations can be inverted, thus providing an analytical solution for the inverse problem, as illustrated in right panel of fig.~\ref{fig:forward-and-inverse-design} (See also fig.~\ref{fig:more-design-examples} for various examples and their physical realizations).
While here we mainly studied a specific perturbation class, other kinds of perturbations can be employed and our continuum design method could also yield other patterns, for example ones similar to the cylindrical and axisymmetric patterns mentioned above.

There are two major advantages of this framework: First, it provides a clear and intuitive relation between the microscopic perturbation and the emergent macroscopic geometry.
Second, in contrary to numerical optimization methods, it can be scaled without increasing complexity. This property allows to detect the onset of horizons in the designed GMO, beyond which, a solution does not exist.
Accordingly, we study the geometries that can be designed and targeted within this framework. 

The paper is structured as follows:
First, in sec.~\ref{sec:RFFQM-review} we briefly review the concept of rigidly and flat-foldable quadrilateral mesh (RFFQM) origami that generalizes the Miura-Ori pattern. We summarize the kinematic compatibility conditions satisfied by classical and generalized Miura-Ori fold patterns, and we explain the corresponding marching algorithm that specifies the complete GMO fold pattern from its boundary perturbations of angles and lengths relative to the classical Miura-Ori, on two boundaries (illustrated in fig.~\ref{fig:marching-algorithm}).
We then provide a minimal necessary review on differential geometry of surfaces (sec.~\ref{sec:differential-geometry-review}), followed by our proposed framework for the discrete differential geometry of generalized Miura-Ori (\ref{sec:differential-geometry-of-miura-ori}).
{In sec.~\ref{sec:slowly-varying-pert} we discuss a subfamily of GMO, which we call alternating perturbations, which will be used as a proof of concept for our analytical design algorithm. This subfamily of patterns is studied here under the assumption that the perturbation fields vary slowly, to gain analytical description of the pattern.}
We then introduce the forward and inverse problems (sec.~\ref{sec:forward-design-geometric-fields} and sec.~\ref{sec:inverse-design-results}), and focus on the inverse problem of elliptic, hyperbolic and mixed geometries. 
In particular, in sec.~\ref{sec:inverse-design-results} we introduce a weakly-nonlinear framework for solving the inverse problem by targeting the desired principal curvatures of GMO, and implement our method to design a variety of crease patterns that induce various geometries, from a simple hemispherical one to a more complex eggs tray-like surface. We conclude by discussing the relevance of our approach to a wider family of origami and other shaping mechanisms (sec.~\ref{sec:discussion}).

\section{Theoretical Framework}

\subsection{Origami} \label{sec:RFFQM-review}
\subsubsection{Miura-Ori}
We study a generalization of a Miura-Ori pattern. In Miura-Ori, all the panels are the same parallelogram repeated along the pattern \cite{miura1985method}, as illustrated in fig.~\ref{fig:miura-ori}.
This pattern can fold smoothly, without stretching or bending of the panels to a family of partially-folded states, and then collapse to a fully-folded state, in which the angles between adjacent panels collapses to zero, with all the panels lying on the same plane, as depicted in the rightmost configuration in fig.~\ref{fig:miura-ori}.
The folding process is characterized by a single degree of freedom that we choose to be the folding angle of the most left and bottom vertical crease line. We call it the \textbf{activation angle} of the pattern and denote it by $\omega$, where $\omega=0$ is the flat state of the paper (in which the angle between adjacent panels is $\pi$), $0<\omega<\pi$ is partially-folded and $\omega=\pi$ is the fully-folded state.\footnote{In general, the activation angle may also be negative, describing folding angle to the opposite direction. This also means that the mountain-valley assignment is flipped and it is equivalent to applying a positive activation angle and looking at the folded configuration from the bottom.}
The Miura-Ori pattern is therefore \textbf{rigidly-foldable}, since it folds smoothly by setting only one folding angle, without bending or stretching the panels, and \textbf{flat-foldable}, since it folds to a fully-folded state, in which adjacent panels collapse on each other.
These properties make the Miura-Ori pattern appealing to engineering because it can be deployed by the use of a single mechanical motor that sets the activation angle, and its fully-folded state is usually more compact compared to the flat sheet, which may facilitate the transportation of large origami designs.

In this work we study slight perturbations of the Miura-Ori pattern where the panels are allowed to be general quadrangles. We focus on a smaller class for which the rigidly-foldable and flat-foldable features are preserved, and therefore, the generalized patterns are called rigidly and flat-foldable quadrilateral mesh (RFFQM) origami.
These features enforce constraints on the pattern \cite{tachi2009generalization, lang2018rigidly, feng2020designs}. We provide here briefly these constraints and the process used to build a valid RFFQM pattern.
This is essentially a summary of the kinematic analysis and marching algorithm developed in \cite{feng2020designs}, which provides a convenient discrete parameterization of RFFQM that we will use to build our continuum approach to inverse-design.

\subsubsection{RFFQM compatibility conditions}
\begin{figure}
	\centering

     \subfloat[\label{fig:isolated-4-vertex}]{
        \centering
        \includegraphics[width=0.45\linewidth]
        {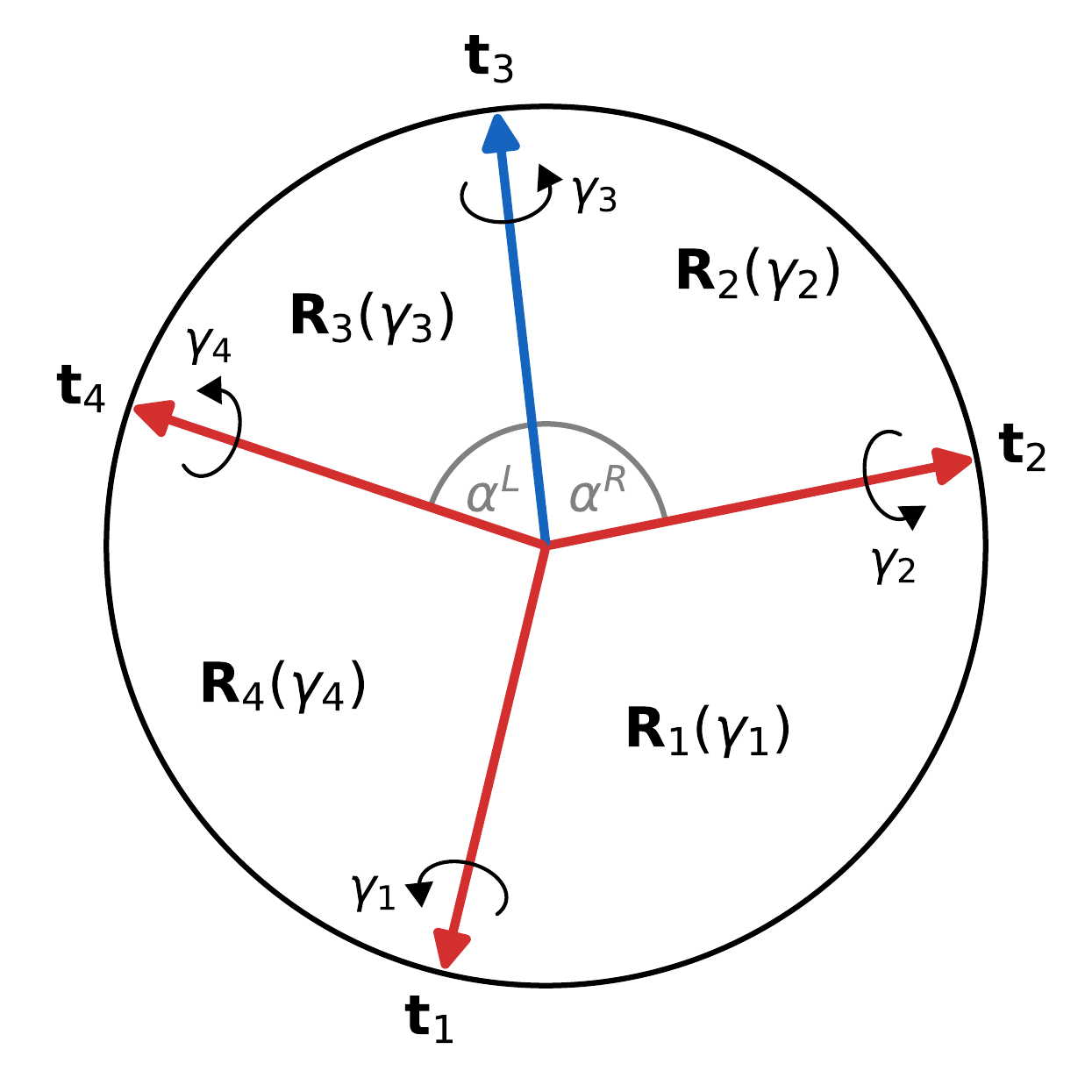}
     }
     \subfloat[\label{fig:panel-compatibility}]{
         \includegraphics[width = 0.5\linewidth]
    	{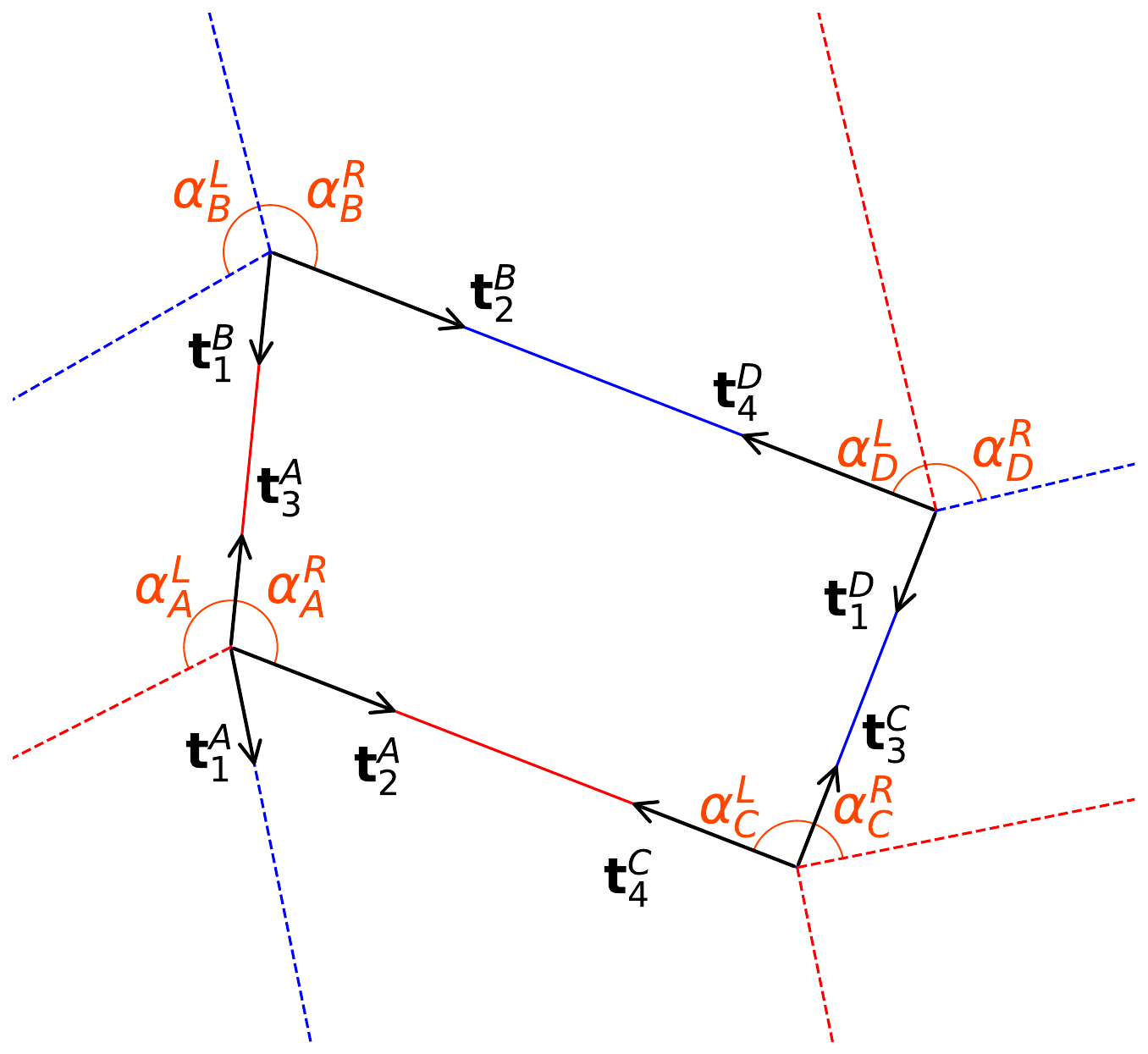}
     }
	
 \caption{
     (a) An isolated flat 4-vertex. The local geometry at the vertex is determined by 2 sector angles $\alpha^R, \alpha^L$. After folding, the rotation matrices $\vc R_i$ that rotate around the fold $\vc t_i$ by angle $\gamma_i$ are applied to the panels. The MV-assignment around a vertex depends on whether $\alpha^L+\alpha^R>\pi$ or $<\pi$,
     yet $\vc t_1$ and $\vc t_3$ are always of opposite assignment.
     (b) An isolated flat panel. The kinematics of such panel are expressed by rotating the 4-adjacent panels around the crease lines.
     }
\end{figure}

\paragraph{A single 4-vertex}
We begin by looking at a single 4-vertex from the pattern, as shown in fig.~\ref{fig:isolated-4-vertex}.
A necessary and sufficient condition for a single vertex to be flat-foldable is that the alternating sum of the sector angles around the vertex must vanish.
This condition is known as Kawasaki's condition \cite{kawasaki1991relation}.
In our quadrilateral mesh framework, Kawasaki's condition together with the demand that the paper starts in flat configuration reduce to the requirement that for any vertex, the sum of opposing angles must be $\pi$. 
That is, for any vertex on the crease, the 4 sector angles $\theta_i$ around it will satisfy:
\begin{equation} \label{eq:kawasaki-4-vertex}
	\theta_1 + \theta_3 = \theta_2+\theta_4=\pi.
\end{equation}
The important implication is that any 4-vertex is characterized by only 2 adjacent angles, denoted in fig.~\ref{fig:isolated-4-vertex} by $\alpha^L$ and $\alpha^R$, and the other 2 are obtained by eq.~\eqref{eq:kawasaki-4-vertex}.

\paragraph{4-vertex kinematics}
Next, we consider the kinematics, or the folding process, of a single 4-vertex.
We follow the notation from fig.~\ref{fig:isolated-4-vertex}, where the 2 angles that characterize the vertex are $\alpha^L,\alpha^R$; the 4 crease lines that come out of the vertex are $\vc t_{i}$ and the right-hand folding angle around each fold are denoted by $\gamma_i\in[-\pi,\pi]$.
The mountain-valley (MV) assignment, which is marked by red crease and blue crease, respectively, corresponds to the sign of the folding angle around that crease, where valley is for $\gamma_i>0$ and mountain for $\gamma_i<0$.

Following the derivations done by \cite{feng2020designs}, under our choice of MV assignment and assuming that $\alpha^L+\alpha^R \neq \pi$, the vertex will fold smoothly without tearing the paper if and only if the folding angles depend on $\gamma_1$ by the law:\footnotemark
\begin{equation} \label{eq:angles-around-vertex}
    \gamma_{2}=\bar{\gamma}_{2}\left(\gamma_{1};\alpha^L,\alpha^R\right),
    \quad
    \gamma_{3}=-\gamma_{1},
    \quad
    \gamma_{4}=\gamma_{2},
\end{equation}
where $\bar{\gamma}_{2}\left(\gamma_{1};\alpha^L,\alpha^R\right)$ is a function expressed by:
\begin{equation} \label{eq:gamma2-expr}
	\bar{\gamma}_{2}\left(\gamma_{1};\alpha^L,\alpha^R\right):=\sgn\left(\pi-\alpha^L-\alpha^R\right)\sgn\left(\gamma_{1}\right)\arccos\left(\frac{\left(-1-\cos\alpha^L \cos\alpha^R\right)\cos\gamma_{1}+\sin\alpha^L\sin\alpha^R}{-1-\cos\alpha^L\cos\alpha^R+\sin\alpha^L\sin\alpha^R\cos\gamma_{1}}\right).
\end{equation}
\footnotetext{
In the special case where $\alpha^L+\alpha^R =\pi$ the vertex may be folded in half while $\gamma_1=\gamma_3=0$ and then folded again by setting some angle $\gamma_1=-\gamma_3$. This is no longer rigid motion since a single folding angle is not enough to characterize the kinematics entirely. In this context we assume that $\alpha^L$ and $\alpha^R$ are small perturbations to the Miura-Ori angles, so this special case is not relevant. For the general discussion we refer the readers to \cite{feng2020designs}.}
This shows that for a single 4-vertex, the kinematics are determined only by a single folding angle $\gamma_1$.
In addition, we can see that if $\gamma_1 =\pm \pi$, then all the other folding angles are also fully-folded with $\gamma_{i}\in \left\{ -\pi,+\pi\right\}$, so it is flat-foldable.

\paragraph{Panel kinematics}
We zoom out to look at a single quadrilateral panel and its kinematics, as shown in fig.~\ref{fig:panel-compatibility}.
Demanding that the sum of angles in the quadrangle is $2\pi$ gives the condition:
\begin{equation} \label{eq:angles-in-quadrangles}
    \alpha_{D}^{R}=\alpha_{A}^{R}-\alpha_{B}^{L}+\alpha_{C}^{L},
\end{equation}
and demanding that the panel will fold smoothly gives the condition:
\begin{equation} \label{eq:alpha-D-L-by-mu-abc}
\alpha_{D}^{L}=
\arccos\left(-\frac{2\mu_{ABC}+\left(\mu_{ABC}^{2}+1\right)\cos\left(\alpha_{D}^{R}\right)}{2\mu_{ABC}\cos\left(\alpha_{D}^{R}\right)+\left(\mu_{ABC}^{2}+1\right)}\right)
,
\end{equation}
for $\mu_{ABC}$ that is
\begin{equation} \label{eq:mu-abc}
    \mu_{ABC}=
    -\frac{\cos\left(\frac{1}{2}\left(\alpha_{C}^{L}+\alpha_{C}^{R}\right)\right)}{\cos\left(\frac{1}{2}\left(\alpha_{C}^{L}-\alpha_{C}^{R}\right)\right)}\frac{\cos\left(\frac{1}{2}\left(\alpha_{B}^{L}+\alpha_{B}^{R}\right)\right)}{\cos\left(\frac{1}{2}\left(\alpha_{B}^{L}-\alpha_{B}^{R}\right)\right)}\frac{\cos\left(\frac{1}{2}\left(\alpha_{A}^{L}-\alpha_{A}^{R}\right)\right)}{\cos\left(\frac{1}{2}\left(\alpha_{A}^{L}+\alpha_{A}^{R}\right)\right)}
    .
\end{equation}
This expression for $\alpha_{D}^{L}$ is valid only if $\left|\mu_{ABC}\right|\neq1$.
Otherwise, we obtain $\alpha_{D}^{L}\in\left\{ -\pi,0,\pi\right\} $, so there is no valid value for $\alpha_{D}^{L}$ that will yield a quadrangular panel.

After determining all the angles inside our quadrangular panel we can calculate the lengths of the crease lines.
For each quadrangle, the right and upper edges $\l_{CD},\l_{BD}$ are calculated by the angles of the quadrangle and the other 2 edges, according to the relation:
\begin{equation} \label{eq:marching-lengths}
    \begin{bmatrix}\l_{CD}\\
    \l_{BD}
    \end{bmatrix}=
    \frac{1}{\sin\left(-\alpha_{A}^{R}+\alpha_{B}^{L}-\alpha_{C}^{L}\right)}\begin{bmatrix}-\sin\alpha_{B}^{L} & \sin\left(-\alpha_{A}^{R}+\alpha_{B}^{L}\right)\\
\sin\left(\alpha_{A}^{R}+\alpha_{C}^{L}\right) & -\sin\alpha_{C}^{L}
\end{bmatrix}\begin{bmatrix}\l_{AB}\\
\l_{AC}
\end{bmatrix}
    ,
\end{equation}
with the notation of the edges appearing in fig.~\ref{fig:panel-compatibility}.

\begin{figure}
	\centering
	
	\includegraphics[width = 0.55\linewidth]
	{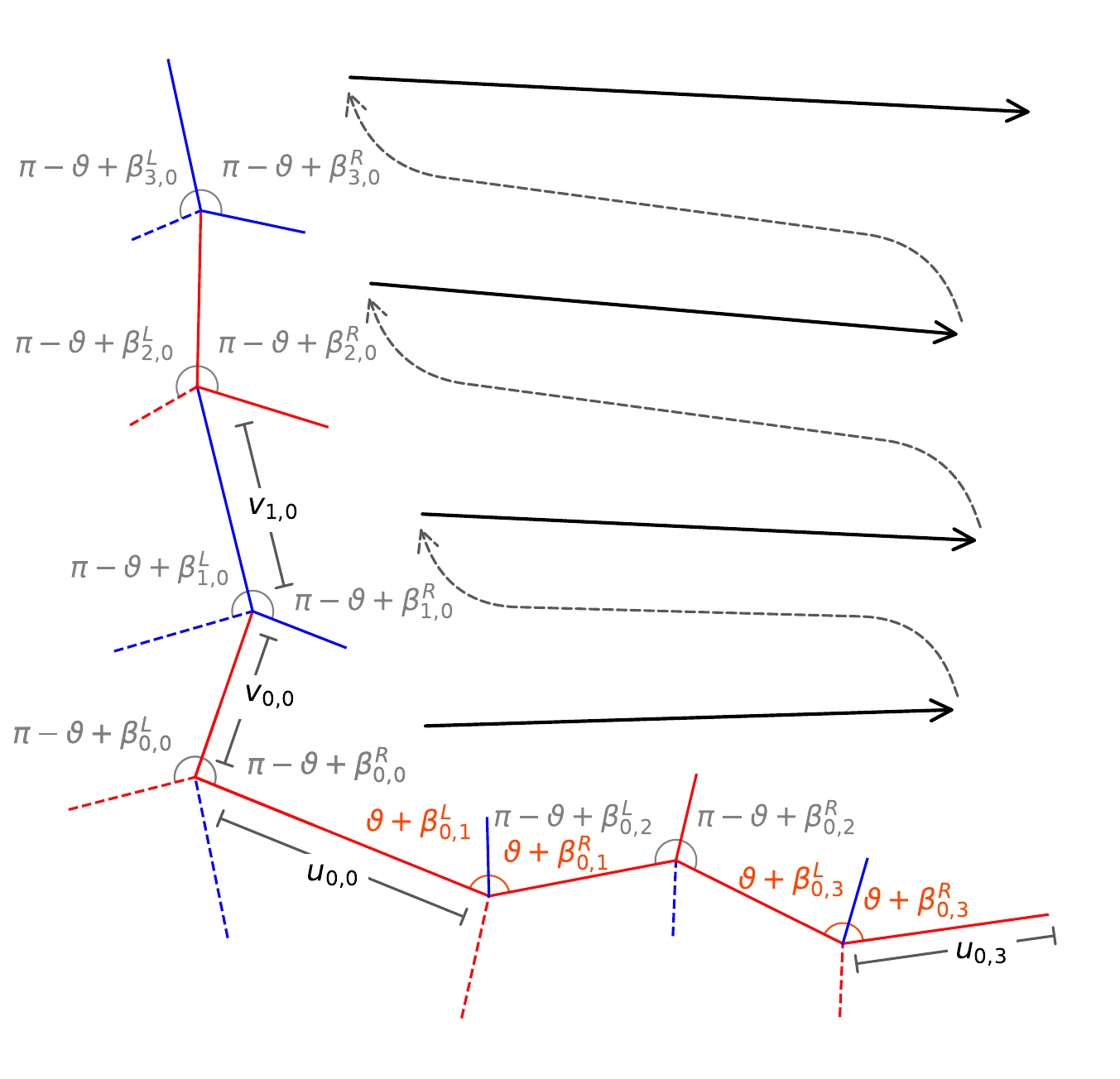}
 
	\caption{
		An illustration of the marching algorithm. We assume that the L-shaped boundary data is given. This includes the angles of the vertices, described by the deviations $\beta^L_{m,n},\beta^R_{m,n}$ from unperturbed Miura-Ori pattern with parallelogram angle $\vartheta$,  and the assignment of the creases, with red for mountain and blue for valley. The lengths of the crease lines at the boundary are given and denoted by $u_{0,n}, v_{m,0}$.
        The algorithm advances iteratively row by row as the arrows indicate.
  	}
	\label{fig:marching-algorithm}
\end{figure}

\paragraph{Marching algorithm} \label{sec:marching-algorithm}
It follows from \eqref{eq:angles-in-quadrangles}--\eqref{eq:marching-lengths} that the entire RFFQM pattern can be built by percolating the input data given on the L-shaped boundary of the crease pattern \cite{feng2020designs}, as illustrated in fig.~\ref{fig:marching-algorithm}.
At each step of the algorithm, the data of the left and bottom vertices are known, and we need to determine the data on the top-right vertex. The panel compatibility conditions are used to extract the angles of the 4th vertex. Then we can continue the iterations by moving to the next panel to the right at the same row, or to the next row.
Once all the sector angles are determined, the crease lengths of the entire pattern are calculated, where at each step the left and bottom edges of a quadrangle are known, and the lengths of the upper and right edges are calculated by \eqref{eq:marching-lengths}.

In this work we focus on small perturbations of a classical Miura-Ori pattern, with unperturbed parallelograms whose minor angle is $\vartheta\in (0,\pi/2)$.
Thus, for each vertex at row $m$ and column $n$, for $m,n\in\{0,1,2,...\}$, we write its upper angles denoted by $\alpha^L_{m,n},\alpha^R_{m,n}$ using the unperturbed Miura-Ori angles with the perturbations described by $\beta^L_{m,n}$ and $\beta^R_{m,n}$.
The angles along the pattern in this notation are:
\begin{equation} \label{eq:crease-angles-by-pert}
    \alpha^L_{m,n}=\begin{cases}
    \vartheta+\beta^L_{m,n}\\
    \pi-\vartheta+\beta^L_{m,n}
    \end{cases}
    \quad
    \alpha^R_{m,n}=\begin{cases}
    \vartheta+\beta^R_{m,n} &       n\text{ is odd}\\
    \pi-\vartheta+\beta^R_{m,n} &   n\text{ is even}
    \end{cases}
\end{equation}
The task becomes finding the perturbation angles $\beta^L_{m,n},\beta^R_{m,n}$ based on the given values along the L-shaped boundary, where $m=0$ or $n=0$.

The marching algorithm may fail if one of the quadrangles' angles has invalid value that is not in range $(0,\pi)$ or if one of the lengths is found to be non-positive.
Luckily, if the given input data is small enough perturbations to a Miura-Ori, with $\beta^L,\beta^R\ll \vartheta$, then the marching algorithm is guaranteed to succeed \cite{dang2022inverse}.

A crease pattern generated by the marching algorithm is guaranteed to be a valid RFFQM pattern. Thus, its entire kinematics is determined by a single folding angle. As before, the folding angle around the vertical crease line at the left-bottom edge of the pattern is our activation angle that controls the kinematics.

The angle compatibility conditions can be linearized, as shown in \ref{app:linearized-marching-algorithm}. 
This can be used as a simple but approximated substitute for the angles marching algorithm in the case of small perturbation to Miura-Ori pattern. 
We do not detail these linearized equations here since in this work we focus on a specific type of perturbations for which the full compatibility conditions can be explicitly solved without linearization.

\subsection{Geometry of surfaces} \label{sec:differential-geometry-review}

We give here a short overview on the geometry of 2D surfaces embedded in 3D Euclidean domain. This topic appears in standard text books such as \cite{do2016differential}.

\paragraph{Parametrized surfaces}
A surface $S\subset \mathbb{R}^3$ is called parametrized surface when it is equipped with a smooth and invertible map $\mathbb{X}:U\to S$ from 2D coordinate domain $U\subset \mathbb{R}^2$ to the surface.
The partial derivatives $\mathbb{X}_x,\mathbb{X}_y$, which are 2 vectors in 3D space, are further assumed to be independent, and they span the tangent plane of the surface at each point.

\paragraph{First fundamental form}
Our surface is embedded in the Euclidean space that dictates how to measure the magnitude of tangent vectors and angles between them.
A line element $ds$ on the surface is expressed in coordinates by:
\begin{equation}
ds^2 = a_{xx}\,dx^2+2a_{xy}\,dx\,dy+a_{yy}\,dy^2,
\end{equation}
with $dx,dy$ distances in coordinate space. $a_{xx}$ and $a_{yy}$ correspond to distance in directions of $dx$ and $dy$, respectively, and $a_{xy}$ is (roughly) a measure of the angle between $dx$ and $dy$.
This measure of distance is known as the \emph{metric} or \emph{first fundamental form}.
It is common to express it in a $2\times 2$ matrix form:
\[
a_{\alpha \beta}=
\begin{bmatrix}a_{xx} & a_{xy}\\
    a_{xy} & a_{yy}
\end{bmatrix}.
\]
We note that different surfaces may have the same metric, as is the case for a flat plane and a cylinder for example.
To make these surfaces distinguishable, we need to consider the behavior of the normal vector to the surface.

\paragraph{Second fundamental form}
The \emph{Gauss map} relates for each point on the surface the unit point vector $\vec N$ that is perpendicular to the tangent plane.
There is a freedom in the choice of orientation, but it does not affect the rest of the discussion.
The \emph{second fundamental form} is a measure of how the normal vector changes along the surface.
It can be calculated by:
\begin{equation} \label{eq:SFF-coordinates}
    b_{\alpha\beta}=\mathbb{X}_{\alpha\beta}\cdot\vec{N}=-\mathbb{X}_{\alpha}\cdot\vec{N}_{\beta}.
\end{equation}

The first and second fundamental forms together determine uniquely the entire shape of the parametrized surface \cite{do2016differential}. This theorem is called Bonnet theorem or the fundamental theorem for the local theory of surfaces.

A similar measurement for the change of the normal is the \emph{shape operator}, which is related to the fundamental forms according to:
\begin{equation} \label{eq:shape-op-by-forms}
    \begin{bmatrix}C_{xx} & C_{xy}\\
C_{yx} & C_{yy}
\end{bmatrix}=-\begin{bmatrix}a_{xx} & a_{xy}\\
a_{xy} & a_{yy}
\end{bmatrix}^{-1}\begin{bmatrix}b_{xx} & b_{xy}\\
b_{yx} & b_{yy}
\end{bmatrix}.
\end{equation}
The shape operator is symmetric, and therefore it can be orthogonally diagonalized.
The orthogonal eigenvectors are the \emph{principal directions} and the eigenvalues denoted by $\kappa_1,\kappa_2$ are called the \emph{principal curvatures}.

Looking at a plane that contains the normal to the surface at a point and one of the principal directions, its intersection with the surface forms a curve in the plane.
The radius of curvature of this curve is equal to the inverse of the principal curvature associated with that direction.

\paragraph{Gaussian curvature}
The \emph{Gaussian curvature} is the product of the principal curvatures $K_G=\kappa_1 \kappa_2$. 
It can be extracted from the fundamental forms according to:
\begin{equation} \label{eq:gaussian-by-fundamental-forms}
    K_G=\frac{\det(b_{\alpha \beta})}{\det(a_{\alpha \beta})}.
\end{equation}
A sphere for example has constant positive Gaussian curvature, a plane and a cylinder have zero Gaussian curvature and a hyperboloid has constant negative Gaussian curvature.
The Gaussian curvature is an intrinsic geometrical property in the sense that isometries --- transformations that change the embedding of the surface while preserving the metric --- do not change the Gaussian curvature. 
This is a result of Gauss's Theorema Egregium, according to which the Gaussian curvature may be obtained only by the metric and its derivatives, by the formula:
\begin{equation} \label{eq:K-by-metric-gauss}
    K_G = -\frac{1}{a_{xx}} \left( \frac{\partial}{\partial x}\Gamma_{xy}^y - \frac{\partial}{\partial y}\Gamma_{xx}^y + \Gamma_{xy}^x\Gamma_{xx}^y - \Gamma_{xx}^x\Gamma_{xy}^y + \Gamma_{xy}^y\Gamma_{xy}^y - \Gamma_{xx}^y\Gamma_{yy}^y\right),
\end{equation}
where ${\Gamma^{\alpha}}_{\beta\gamma}=\frac{1}{2} a^{\alpha\delta}\left( \partial_\gamma a_{\delta\beta}+ \partial_\beta a_{\delta\gamma}-\partial_\delta a_{\beta\gamma}\right)$ are the Christoffel symbols, $a^{\alpha\delta}$ with upper indices is the inverse of the metric.

\section{Differential geometry of Miura-Ori: local calculations} \label{sec:differential-geometry-of-miura-ori}
\begin{figure}
	\centering
	
	\begin{subfigure}[b]{0.42\linewidth}
		\centering
		\includegraphics[width=\linewidth]
		{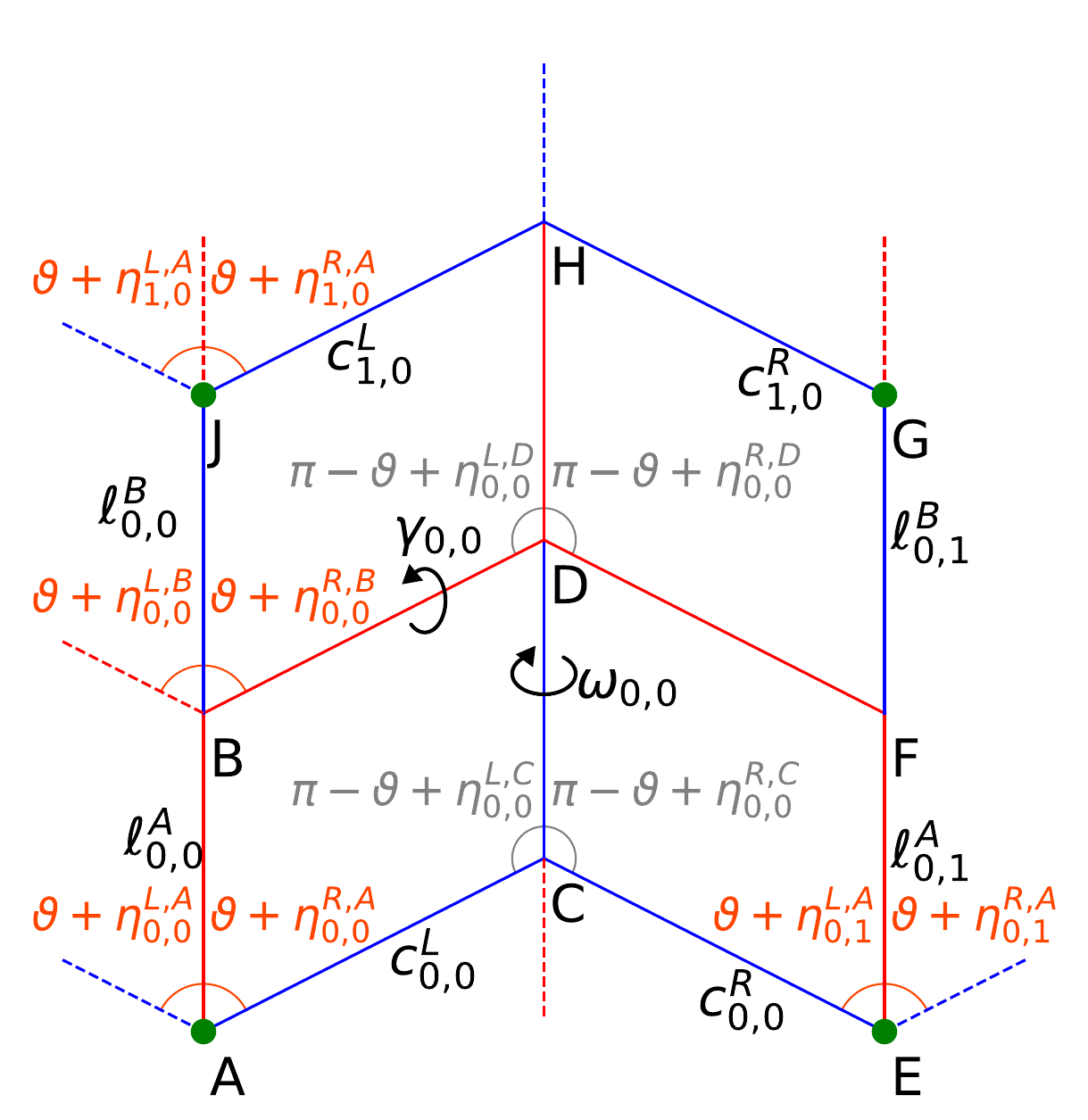}
		\caption{\label{fig:unit-cell-lengths-and-angles-perturbations}}
	\end{subfigure}
	\begin{subfigure}[b]{0.45\linewidth}
		\centering
		\includegraphics[width=\linewidth]
		{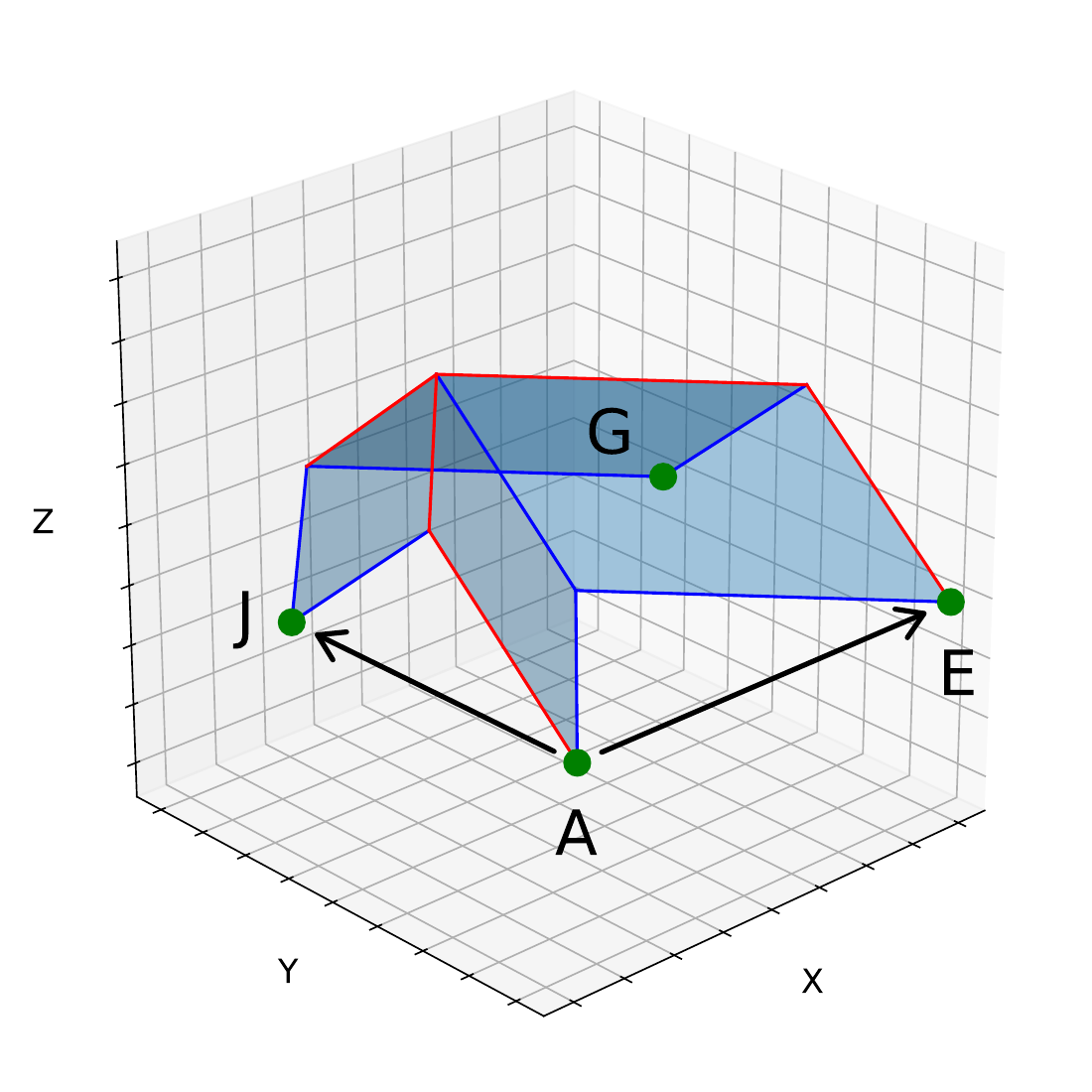}
		\caption{\label{fig:folded-unit-cell}}
	\end{subfigure}
	
	\caption{
		(a)
		A single unit cell used to analyze the angle perturbations and approximate the metric of the origami.
		The angle perturbations $\eta^{L,\sigma}_{i,j},\eta^{R,\sigma}_{i,j}$, for $\sigma=A,B,C,D$ are added to a Miura-Ori pattern characterized by a single angle $\vartheta$.
		$\omega_{0,0}$ is the folding angle around the crease line $CD$, also called the activation angle.
		(b)
		A folded unperturbed unit cell.	The marked green dots $A,E,J,G$ are the dots attached to the surface we try to approximate.
	}
\end{figure}

We consider our RFFQM as a grid of the basic unit cell, with a single cell consisting of $2\times 2$ panels, as shown in fig.~\ref{fig:unit-cell-lengths-and-angles-perturbations}.
For origami pattern with $(m,n)\in \{0,1...,2 N_y\}\times \{0,1...,2 N_x\}$ vertices we have in the unit cell grid $N_y$ rows and $N_x$ columns, where each unit cell is indexed by its $i\in \{0,1,...,N_y-1\}$ row and $j\in \{0,1,...,N_x-1\}$ column. 
Note that as a unit cell contains $2\times 2$ panels, each contains 4 vertices whose angle perturbation we denote by $\eta^{L,\sigma}_{i,j},\eta^{R,\sigma}_{i,j}$, for $\sigma=A,B,C,D$.
They are related to the previous notation $\beta^L_{m,n},\beta^R_{m,n}$ by:
\begin{align}
    \eta_{i,j}^{L,B} &      =\beta_{2i+1,2j}^{L} , & 
    \eta_{i,j}^{R,B} &      =\beta_{2i+1,2j}^{R}, &
    \eta_{i,j}^{L,D} &      =\beta_{2i+1,2j+1}^{L} ,  & 
    \eta_{i,j}^{R,D} &      =\beta_{2i+1,2j+1}^{R} 
    ; \nonumber
    \\
    \eta_{i,j}^{L,A} &      =\beta_{2i,2j}^{L} ,      &
    \eta_{i,j}^{R,A} &      =\beta_{2i,2j}^{R},  &
    \eta_{i,j}^{L,C} &      =\beta_{2i,2j+1}^{L} ,    & 
    \eta_{i,j}^{R,C} &      =\beta_{2i,2j+1}^{R}
    . \nonumber
\end{align}

\begin{figure}
	\centering
	\includegraphics[width=0.55\textwidth]
	{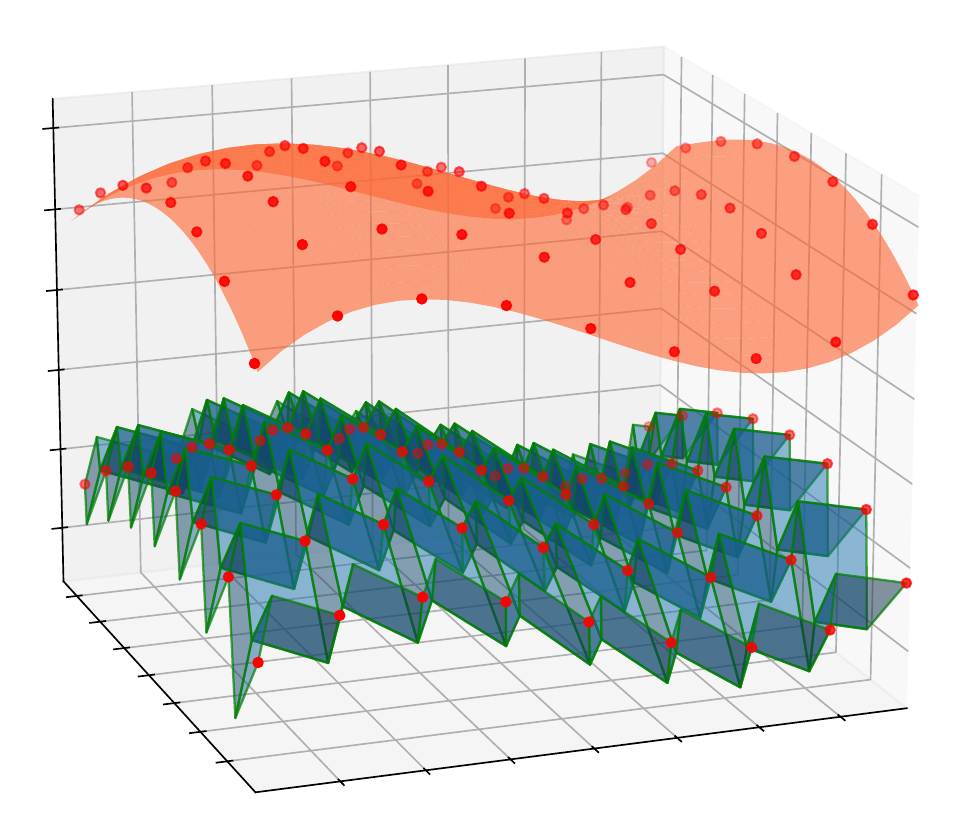}
	
	\caption{
		An illustration of surface approximation by RFFQM origami: A smooth surface (in orange) is assumed to be approximated by an RFFQM origami if all the points in the upper layer of the origami (marked in red dots) match those of the smooth surface. 
        We further assume that the typical wavelength of oscillations in the surface is larger than the size of a parallelograms.
	}
	\label{fig:rffqm-approximating-smooth-surface}
\end{figure}

For a specific activation angle, the pattern takes its shape in space and we look at it as approximating a smooth surface that passes through the outer vertices of the unit cells, denoted by $AEJG$ in fig.~\ref{fig:unit-cell-lengths-and-angles-perturbations}.
An example of such approximation can be seen in fig.~\ref{fig:rffqm-approximating-smooth-surface}.
We use curvilinear coordinates that follow the vectors $\vec{AE}$ and $\vec{AJ}$ as appear in fig.~\ref{fig:folded-unit-cell}, with integer values in coordinates $(j,i)$ that are mapped to point $A$ of the unit cell at row $i$ and column $j$ on the grid.

With these coordinates we can calculate the geometrical properties of the approximated surface by approximating all the derivatives with discrete differences.
The tangent space at point $(j,i)$ is spanned by $\mathbb{X}_{x},\mathbb{X}_{y}$ where in the discrete case they are simply:
\begin{align} \label{eq:tangent-vec-discrete-diff}
    \mathbb{X}_{x}&\approx\mathbb{X}(j+1,i)-\mathbb{X}(j,i)=\vec{AE}  ,
    \nonumber \\
    \mathbb{X}_{y}&\approx\mathbb{X}(j,i+1)-\mathbb{X}(j,i)=\vec{AJ}  .
\end{align}
The metric, which describes the intrinsic distance on the surface, is given by the inner product:
\begin{equation} \label{eq:metric-by-unit-cell-vectors}
	a_{\alpha\beta}=
	\begin{bmatrix}\vec{AE}\cdot\vec{AE} & \vec{AJ}\cdot\vec{AE}\\
		\vec{AJ}\cdot\vec{AE} & \vec{AJ}\cdot\vec{AJ}
	\end{bmatrix}
 .
\end{equation}
The second fundamental form in the discrete version is calculated using eq.~\eqref{eq:SFF-coordinates}:\footnotemark
\begin{align} \label{eq:SFF-in-RFFQM}
    b_{xx}&=\left(\vec{AE}\left(j+1,i\right)-\vec{AE}\left(j,i\right)\right)\cdot \vec{N},
    \nonumber \\
    b_{xy}=b_{yx}&=\left(\vec{AE}\left(j,i+1\right)-\vec{AE}\left(j,i\right)\right)\cdot \vec{N},
    \nonumber \\
    b_{yy}&=\left(\vec{AJ}\left(j,i+1\right)-\vec{AJ}\left(j,i\right)\right)\cdot \vec{N},
\end{align}
with $\vec{N}$ the normal vector:
\[
\vec{N}\left(j, i\right)=\frac{\vec{AE}\times\vec{AJ}}{\left\Vert \vec{AE}\times\vec{AJ}\right\Vert }.
\]

\footnotetext{In differential geometry, $\mathbb{X}_{\alpha}\cdot\vec{N}=0\implies\partial_{\beta}\left(\mathbb{X}_{\alpha}\cdot\vec{N}\right)=0=\mathbb{X}_{\alpha\beta}\cdot\vec{N}+\mathbb{X}_{\alpha}\cdot\vec{N}_{\beta}$ so we could also use $b_{\alpha\beta}=-\mathbb{X}_{\alpha}\cdot\vec{N}_{\beta}$ to calculate the second fundamental form. It is interesting to note that these two ways are not identical in the discrete approximation of the derivatives. We choose the option mentioned in the article out of convenience for the computations.}

The Gaussian curvature in this work can be calculated in two ways: 
the first one is by using the metric alone, using eq.~\eqref{eq:K-by-metric-gauss}, where again the derivatives are approximated by discrete difference:
\[
\partial_{x}a_{\alpha\beta}\approx a_{\alpha\beta}(j+1,i)-a_{\alpha\beta}(j,i),\quad
\partial_{y}a_{\alpha \beta}\approx a_{\alpha \beta}(j,i+1)-a_{\alpha \beta}(j,i),
\]
and derivatives of Christoffel symbols are obtained by the same rule.
The second way is by calculating the principal curvatures and taking their product. This way is more informative (though more difficult to compute) as the principal curvatures contain additional information about the shape.
For calculating the principal curvatures, we consider two approaches. 
First, we can do it directly from the second fundamental form. A second way is to approximate the principal curvatures by looking at the 2D cross section along the principal direction and calculate the radius of curvature of that curve. 
All approaches involve, in practice, ignoring some high-order effects; it turns out that the last one is the most accurate.

\subsection{Geometrical properties in the unperturbed Miura-Ori pattern} \label{sec:geo-properties-of-unperturbed-ori}
As a first example of our framework, we now study the simplest case: the unperturbed Miura-Ori pattern. In this case, the same unit cell of $2\times 2$ panels repeats itself along the entire pattern. Therefore, we only need to study a single unit cell.
We continue to use the notation as in fig.~\ref{fig:unit-cell-lengths-and-angles-perturbations}, where all the angle perturbations vanish $\eta \equiv 0$, and set the lengths: $\l^A_{0,0}=\l^B_{0,0}=L_0;c^L_{0,0}=c^R_{0,0}=C_0$.
To calculate the vectors $\vec{AE},\vec{AJ}$ we set coordinate system where the panel $ABDC$ remains on the $XY$ plane, with vector $\vec{AB}$ points in the $\hat{Y}$ direction.
The panel $CDFE$ to the right is rotated around the edge $CD$, which also points in $\hat{Y}$, by the activation angle $\omega=\omega_{0,0}$. We express the vectors in the flat configuration:
\[
    \vec{AC}^{\text{flat}}=C_{0}\begin{bmatrix}\sin\vartheta\\
		\cos\vartheta\\
		0
	\end{bmatrix};
	\quad
    \vec{CE}^{\text{flat}}=C_{0}\begin{bmatrix}\sin\vartheta\\
        -\cos\vartheta\\
        0
    \end{bmatrix}
,
\]
and the rotation matrix is a right-hand rotation around the vector $DC$:
\[
\vc{R}_{DC}(\omega)=
    \begin{bmatrix}\cos\omega & 0 & -\sin\omega\\
0 & 1 & 0\\
\sin\omega & 0 & \cos\omega
\end{bmatrix}.
\]
The vector $\vec{AE}$ is then:
\[
    \vec{AE}=\vec{AC}^{\text{flat}}+\vc{R}_{DC}\left(\omega\right)\vec{CE}^{\text{flat}}
    =
    C_{0}\begin{bmatrix}\left(1+\cos\omega\right)\sin\vartheta\\
	0\\
	\sin\omega\sin\vartheta
\end{bmatrix}
.
\]
Similarly, we calculate the vector $\vec{AJ}$:
\[
    \vec{AJ}=\vec{AB}^{\text{flat}}+\vc{R}_{DB}\left(-\gamma\right)\vec{BJ}^{\text{flat}}
    = 
    L_{0}\begin{bmatrix}\left(1-\cos\gamma\right)\sin\vartheta\cos\vartheta\\
	\sin^{2}\vartheta\cos\gamma+\cos^{2}\vartheta+1\\
	\sin\gamma\sin\vartheta
\end{bmatrix},
\]
with $\vc{R}_{DB}\left(-\gamma\right)$ a right-hand rotation around the vector $\vec{DB}$. The minus sign in the angle is to account for rotating the panel $BDHJ$ backwards instead of rotating $ABDC$ by $+\gamma$.
$\gamma$ is obtained by eq.~\eqref{eq:angles-around-vertex} after setting $\alpha^L=\alpha^R=\pi-\vartheta$, and assuming that $\vartheta<\pi/2$:
\begin{equation} \label{eq:gamma-by-omega-miura-ori}
    \gamma\left(\omega\right)=-\sgn\left(\omega\right)\arccos\left(\frac{\left(-1-\cos^{2}\left(\vartheta\right)\right)\cos\left(\omega\right)+\sin^{2}\left(\vartheta\right)}{-1-\cos^{2}\left(\vartheta\right)+\sin^{2}\left(\vartheta\right)\cos\left(\omega\right)}\right).
\end{equation}
From these calculations we obtain the metric entries:
\begin{align*}
    a_{xx}&=4C_{0}^{2}\cos^{2}\left(\frac{\omega}{2}\right)\sin^{2}(\vartheta)  ,
    \\
    a_{xy}&=0  ,
    \\
    a_{yy}&=L_{0}^{2}\frac{8\cos^{2}(\vartheta)}{-\cos(\omega)\sin^{2}(\vartheta)+\cos^{2}(\vartheta)+1}
    .
\end{align*}

Since the metric for each unit cell is equal, all the derivatives of the metric will vanish, and therefore the Gaussian curvature vanishes.
To calculate the second fundamental form, we need to look at two adjacent unit cells and calculate the difference in the vectors $\vec{AE},\vec{AJ}$. In this simple case, these differences also vanish since the vectors are parallel to each other along the entire pattern.
We conclude that the second fundamental form vanishes completely.
Indeed, the Miura-Ori remains flat for any activation angle.

\subsection{Geometrical properties in the perturbed case}
We can perform the same calculations also when the unit cell (fig.~\ref{fig:unit-cell-lengths-and-angles-perturbations}) is perturbed. The tangent vectors become:
{\small
\begin{equation*} % AE
    \vec{AE}=
\left(
\begin{array}{c}
 c_{0,0}^L \sin \left(\eta _{0,0}^{R,A}+\vartheta \right)+c_{0,0}^R \left(\cos \left(\omega _{0,0}\right) \cos \left(\eta _{0,0}^{L,C}+\eta _{0,0}^{R,A}\right) \sin \left(\vartheta -\eta _{0,0}^{R,C}\right)-\sin \left(\eta _{0,0}^{L,C}+\eta _{0,0}^{R,A}\right) \cos \left(\vartheta -\eta _{0,0}^{R,C}\right)\right) \\
 c_{0,0}^L \cos \left(\eta _{0,0}^{R,A}+\vartheta \right)-c_{0,0}^R \left(\cos \left(\omega _{0,0}\right) \sin \left(\eta _{0,0}^{L,C}+\eta _{0,0}^{R,A}\right) \sin \left(\vartheta -\eta _{0,0}^{R,C}\right)+\cos \left(\eta _{0,0}^{L,C}+\eta _{0,0}^{R,A}\right) \cos \left(\vartheta -\eta _{0,0}^{R,C}\right)\right) \\
 c_{0,0}^R \sin \left(\omega _{0,0}\right) \sin \left(\vartheta -\eta _{0,0}^{R,C}\right) \\
\end{array}
\right)
,
\end{equation*}
}
\begin{equation*} % AJ
    \vec{AJ}=
    \left(
\begin{array}{c}
 \ell _{0,0}^B \left(\sin \left(\eta _{0,0}^{L,B}+\vartheta \right) \cos \left(\eta _{0,0}^{R,B}+\vartheta \right)-\cos \left(\gamma _{0,0}\right) \cos \left(\eta _{0,0}^{L,B}+\vartheta \right) \sin \left(\eta _{0,0}^{R,B}+\vartheta \right)\right) \\
 \ell _{0,0}^A+\ell _{0,0}^B \left(\sin ^2\left(\frac{\gamma _{0,0}}{2}\right) \cos \left(\eta _{0,0}^{L,B}+\eta _{0,0}^{R,B}+2 \vartheta \right)+\cos ^2\left(\frac{\gamma _{0,0}}{2}\right) \cos \left(\eta _{0,0}^{L,B}-\eta _{0,0}^{R,B}\right)\right) \\
 \ell _{0,0}^B \sin \left(\gamma _{0,0}\right) \sin \left(\eta _{0,0}^{R,B}+\vartheta \right) \\
\end{array}
\right)
    ,
\end{equation*}
for $\gamma_{0,0}$ the horizontal folding angle, calculated from \eqref{eq:angles-around-vertex} assuming for simplicity that $\omega_{0,0}>0$:
\begin{equation*}
    \gamma_{0,0}=
    -\arccos\left(\frac{\cos \left(\omega _{0,0}\right) \left(-\cos \left(\vartheta -\eta _{0,0}^{L,D}\right) \cos \left(\vartheta -\eta _{0,0}^{R,D}\right)-1\right)+\sin \left(\vartheta -\eta _{0,0}^{L,D}\right) \sin \left(\vartheta -\eta _{0,0}^{R,D}\right)}{\cos \left(\omega _{0,0}\right) \sin \left(\vartheta -\eta _{0,0}^{L,D}\right) \sin \left(\vartheta -\eta _{0,0}^{R,D}\right)-\cos \left(\vartheta -\eta _{0,0}^{L,D}\right) \cos \left(\vartheta -\eta _{0,0}^{R,D}\right)-1}\right)
   .
\end{equation*}
The complete metric entries and the lengths of the upper-right edges appear in the supplementary Mathematica notebook.

Having non-uniform changes in the lengths and folding angles of different unit cells means that the metric entries also change along the pattern. Such non-uniform changes in the metric may give rise to a non-zero Gaussian curvature and open the possibilities to the designs of non-planar surfaces.

In order to calculate the Gaussian cuvature, we need to know, at least approximately, what are the lengths $c^L_{i,j},c^R_{i,j},\l^A_{i,j},\l^B_{i,j}$, the angle perturbations $\eta^{L,\sigma}_{i,j},\eta^{R,\sigma}_{i,j},\sigma=A,B,C,D$, and the folding angles $\omega_{i,j}$ for all the unit cells in the grid, as dictated by the marching algorithm.
In order to obtain closed formulas, we restrict ourselves to specific type of perturbations along the L-shaped boundary, which are slowly varying, as described below.
Once we obtain close formulas for the Gaussian and principal curvatures, we can inverse design surfaces by inverting these formulas.

\section{Slowly varying perturbations} \label{sec:slowly-varying-pert}

\subsection{Alternating perturbations}
\label{sec:angles}
Here we describe the perturbation that we impose on the L-shaped boundary of the crease pattern. 
Inspired by design patterns achieved in previous work \cite{dang2022inverse}, we choose to work with perturbations that sum up to $0$ for each vertex and alternate their values for 2 adjacent vertices on the same column.
As such, we call these patterns \emph{alternating angle patterns}.
The perturbations imposed on the boundary are written using the function $\delta:\{0,\frac{1}{2},1,...,N_x-\frac{1}{2},N_x\}\to \mathbb{R}$, by setting the $\eta$s from fig.~\ref{fig:unit-cell-lengths-and-angles-perturbations} according to: 
\begin{align} \label{eq:alternating-angles-by-pert-on-boundary}
& \text{left boundary:}  &
    \eta_{i,0}^{L,A} & =-\delta\left(0\right),  & 
    \eta_{i,0}^{R,A} & =\delta\left(0\right),   &
    \eta_{i,0}^{L,B} & =\delta\left(0\right) ,  &
    \eta_{i,0}^{R,B} & =-\delta\left(0\right);  
    \nonumber   \\
& \text{bottom boundary:} &
    \eta_{0,j}^{L,A} &=-\delta\left(j\right),   &
    \eta_{0,j}^{R,A} & =\delta\left(j\right),   & 
    \eta_{0,j}^{L,C} &=-\delta\left(j+\frac{1}{2}\right) , &
    \eta_{0,j}^{R,C} &=\delta\left(j+\frac{1}{2}\right) .
\end{align}

This form has the advantage that the marching algorithm yields a simple form for the angles along the entire pattern, since the angles on each column depend only on the perturbation on that column. The sector angles along the entire pattern are found to be:
\begin{equation} \label{eq:alternating-angles-by-pert}
\eta_{i,j}^{R,\sigma}=\begin{cases}
\delta\left(j\right) & \sigma=A\\
-\delta\left(j\right) & \sigma=B\\
\delta\left(j+\frac{1}{2}\right) & \sigma=C\\
-\delta\left(j+\frac{1}{2}\right) & \sigma=D
\end{cases},\quad\eta_{i,j}^{L,\sigma}=-\eta_{i,j}^{R,\sigma}.
\end{equation}

\begin{figure}
\centering
\sbox{\measurebox}{%
  \begin{minipage}[b]{0.58\textwidth}
  \subfloat[Alternating perturbation notation]{\label{fig:alternating-perturbation-notation}
  \includegraphics[width=\textwidth]{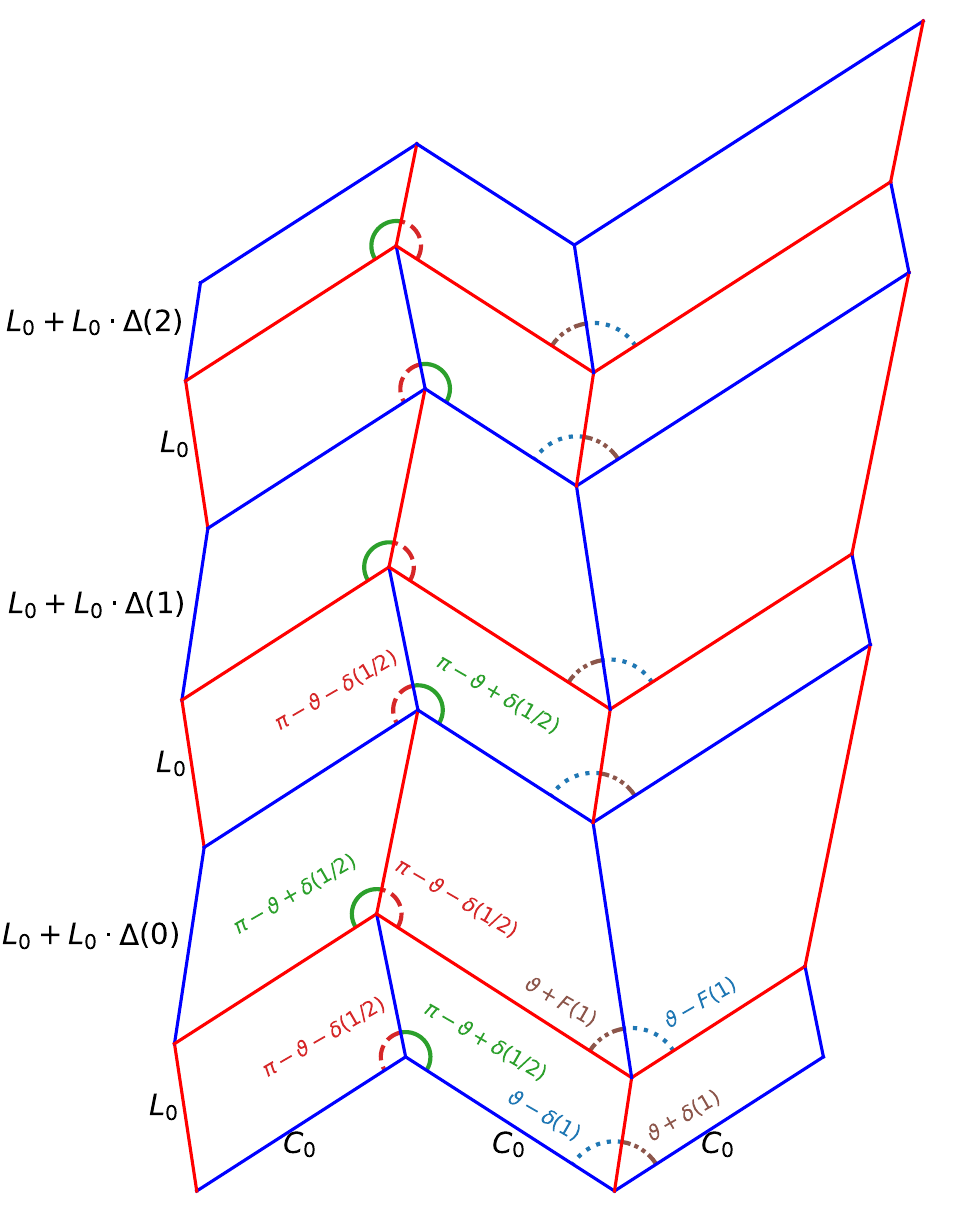}}
  \end{minipage}}
\usebox{\measurebox}
\begin{minipage}[b][\ht\measurebox][s]{0.4\textwidth}
\centering
\subfloat[Effect on a bottom vertex]{\label{fig:delta-pert-on-vertex}
\includegraphics[width=\textwidth]{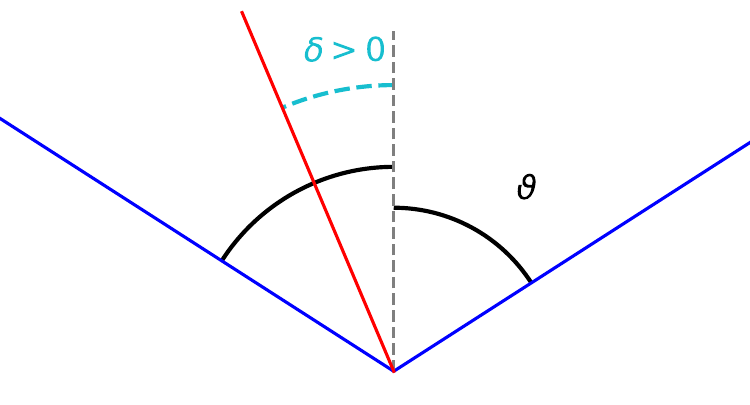}}

\vfill

\subfloat[Mars pattern]{\label{fig:MARS-example}
\includegraphics[width=\textwidth]{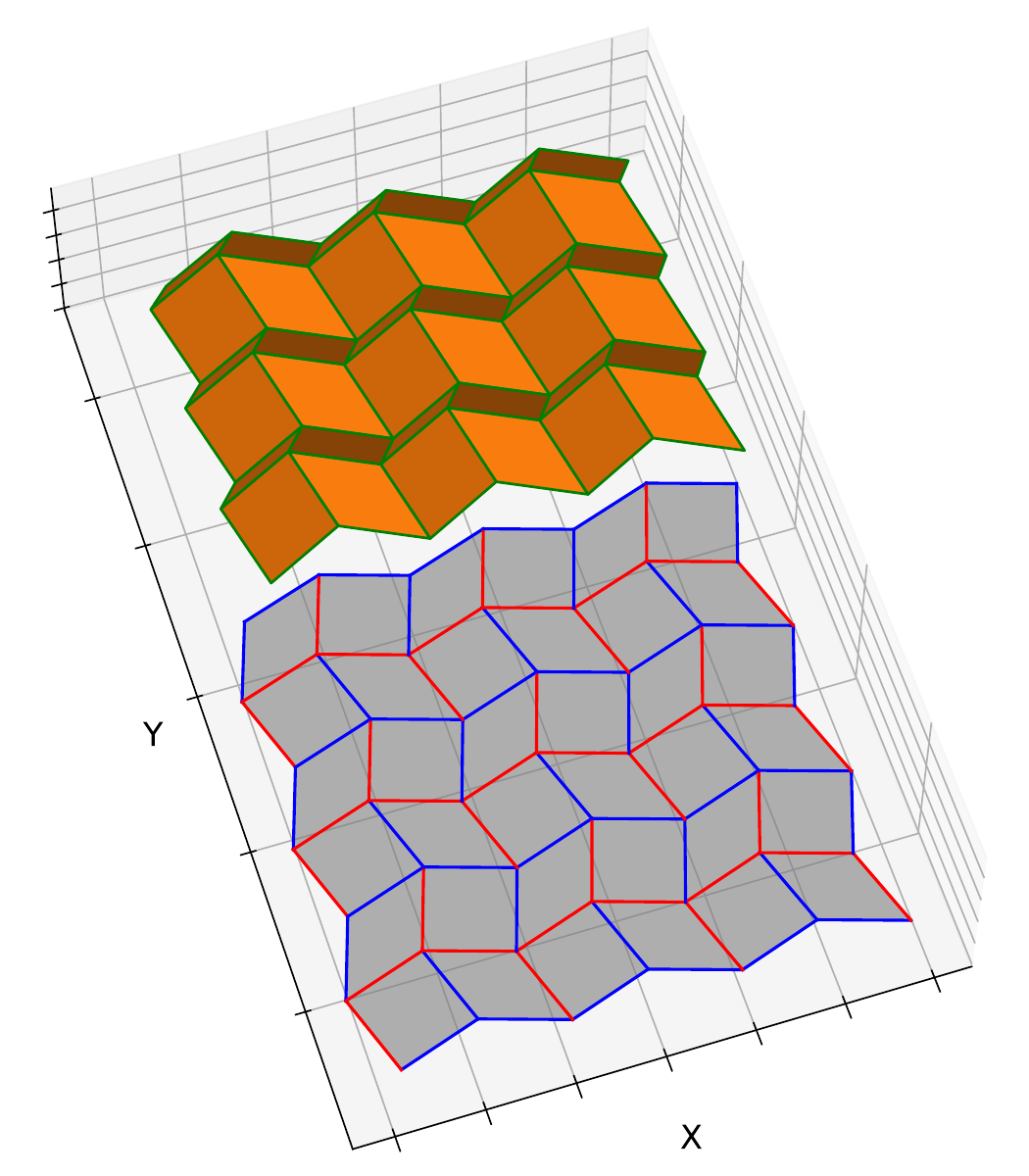}}
\end{minipage}
\caption{
    (a) The alternating perturbations considered in this paper: angle perturbations alternating in sign by a field $\delta$, and length modifications by a field $\Delta$.
    The angles with the same color have the same value. The alternating angles make the vertical crease lines break and form a zigzag.
    (b) The notation for $\delta$ perturbation on a vertex with even column on the bottom boundary. Positive $\delta$ makes the vertical edge inclined to the left and negative value makes the crease inclined to the right.
    (c) Mars pattern, which is a special case of the alternating perturbations, achieved by setting $\delta\equiv \pi/2-\vartheta, \Delta \equiv 0$.
}
\end{figure}

A short proof that these angles satisfy the compatibility condition appears in \ref{app:compatibility-of-alternating-angles}.
A positive $\delta$ on the bottom row vertices makes a tilt of the vertical edge to the left, and negative to the right, as illustrated in fig.~\ref{fig:delta-pert-on-vertex}. 
An example of a pattern according to this perturbation rule appears in fig.~\ref{fig:alternating-perturbation-notation}, where it can be seen that the vertical creases form a zigzag pattern along the columns.
We can look at alternating angle patterns as generalizations of the Mars design, introduced in \cite{barreto1997lines}.
To obtain the Mars pattern, we set the angle perturbation to be constant $\delta\equiv \frac{\pi}{2}-\vartheta$, resulting in alternating square and rhombus panels, as shown in fig.~\ref{fig:MARS-example}.

In addition to the angle perturbation, we add perturbations to the lengths at the left boundary on alternating vertical creases, so we set $\l^A_{i,0}=L_0$ for all $i$, and add perturbation prescribed by the function $\Delta:\{0,1,...,N_y-1\}\to \mathbb{R}$ by:
\begin{equation*}
    \l^B_{i,0}=L_0+L_0 \cdot \Delta\left( i \right)
    .
\end{equation*}
The horizontal crease lengths on the boundary are kept unperturbed, with $c^L_{0,j}=c^R_{0,j}=C_0$ for all $j$s.
The boundary perturbations we consider are shown in fig.~\ref{fig:alternating-perturbation-notation}.

In sec.~\ref{sec:pert-special-cases} below we discuss some special cases of these perturbations, like pure angle or pure length perturbations.
These special cases give some intuition about the effect of the perturbations, and they are also important for the analysis of the general case.

We chose to use the alternating angle perturbation since it is simple and yet rich enough to design nontrivial surfaces. It has several advantages that helped the inverse design process:
the marching algorithm for the angles is easy to solve, allowing us to predict exactly the sector angles along the pattern, and the marching algorithm for the lengths can be approximated by solving a set of simple equations that predict the crease lengths along the pattern.
In addition, our prediction for the angles and lengths do not diverge too quickly, meaning that small perturbations on the L-shaped boundary result in small perturbations of all the Miura-Ori panels, as the errors do not accumulate fast in marching algorithm. 
This is important for the validity of our (approximate) formulas of the geometric fields.

While for our perturbations the angles over all panels are easily obtained from the marching algorithm, for more general perturbations it is not the case, and it might be useful to simplify the marching algorithm. 
Thus, for completeness, we present in \ref{app:linearized-marching-algorithm} a linearized version of the marching algorithm for small angle perturbations.

\subsection{Origami geometry with smooth perturbation fields}

So far, the perturbations $\delta$ and $\Delta$ were allowed to change significantly between adjacent cells, resulting in large variations in the geometry of adjacent unit cells.
In this work, we want to avoid this and make sure that each neighborhood of unit cells contains cells that are similar to each other.
For this purpose, we assume that the perturbations are densely sampled from smooth counterparts fields.

For a pattern with $N_y\times N_x$ unit cells we use coordinate space $x,y\in U=\left[0,1\right]\times\left[0,1\right]$, and define the scaling parameters 
\begin{equation}\label{eq:chi_xi}
\chi=\frac{1}{N_x},\quad \xi=\frac{1}{N_y}
\end{equation}
that relate the discrete unit cell grid to the coordinates by the conversion:
\begin{equation*}
    x(j)=\chi j, \quad y(i)=\xi i.
\end{equation*}
The angle perturbation function $\delta$ is then sampled from a smooth function $\tilde{\delta}:[0,1]\to \mathbb{R}$ according to this conversion:
\begin{equation*}
    \delta(j)=s \tilde{\delta}\left(x\left(j\right)\right),
\end{equation*}
where $j\in\left\{0,\frac{1}{2},1,...,N_x-\frac{1}{2},N_x\right\}$, where we introduce the small parameter $s>0$ to quantify the smallness of the perturbation $\tilde \delta$.
The length perturbation on the left boundary $\Delta$ is treated as well as being sampled from a smooth perturbation function $\tilde{\Delta}:[0,1]\to\mathbb{R}$ according to
\begin{equation*}
    \Delta (i)=t \tilde{\Delta}\left(y\left(i\right)\right),
\end{equation*}
where $t>0$ is a small parameter.

In the inverse design process, we will first find the smooth perturbations $\tilde{\delta}$ and $\tilde{\Delta}$ and then we can choose the number of cells $N_x,N_y$ for which $\chi,\xi$ will be smaller than the oscillations of these functions.
Having more unit cells means that each unit cell should be smaller, so the folded origami pattern can capture better the oscillations of the targeted surface.
Thus, the lengths at the boundary are rescaled accordingly, with $C_0=\chi C_{tot}$ and $L_0 = \xi L_{tot}$.
To summarize, the boundary condition we apply given the scaling parameters $\chi, \xi$ are:
\begin{align}
    & \text{left boundary:}  &
    \l^A_{i,0} &= \xi L_{tot}, &
    \l^B_{i,0} & = \xi L_{tot}\left(1+t\tilde{\Delta}\left(y\left(i\right)\right)\right)
    ,
    \nonumber   \\
    & \text{bottom boundary:} &
    c_{0,j}^{L} &= \chi C_{tot},   &
    c_{0,j}^{R} &= \chi C_{tot}.
    \label{eq:lengths-boundary-condition}
\end{align}

\subsection{Lengths} \label{sec:slowly-varying-pert-lengths}
Once we choose the perturbation applied to the L-shaped boundary, all the crease lengths are dictated by the marching algorithm. This is useful for the direct problem, since we have all the data necessary to calculate the kinematics and shape of the folded origami. 
But for the inverse problem we need to be able to predict analytically the curvature on the entire pattern based on the boundary perturbations, hence we would prefer to have closed formulas for them (at least approximately) given the boundary data.
We obtain this by assuming that the lengths, dictated by the marching algorithm, change slowly along the pattern, and approximating the change between 2 consecutive lengths as derivatives, yielding a set of differential equations that we can solve.

For the discrete length fields $\l^A_{i,j},\l^B_{i,j},c^L_{i,j},c^R_{i,j}$ we present their smooth version counterparts, denoted by $\tilde{\l}_{A},\tilde{\l}_{B},\tilde{c}_L,\tilde{c}_R:U\to\mathbb{R}_{> 0}$. As the scaling parameters $\chi,\xi$ become smaller, the changes in the discrete length fields become smaller, and the approximation by smooth counterparts becomes better. In other words, we expect that in the limit $\chi,\xi\to 0$, we can avoid the iterative marching algorithm and approximate the lengths along the pattern as being sampled from their smooth counterpart:
\begin{align*}
    \forall i,j \quad \lim_{\chi,\xi\to0}\l_{i,j}^{A} & =\tilde{\l}_{A}\left(x\left(j\right),y\left(i\right)\right), \\
    \forall i,j \quad \lim_{\chi,\xi\to0}\l_{i,j}^{B} & =\tilde{\l}_{B}\left(x\left(j\right),y\left(i\right)\right), \\
    \forall i,j \quad \lim_{\chi,\xi\to0}c_{i,j}^{L} & 
    =\tilde{c}_{L}\left(x\left(j\right),y\left(i\right)\right), \\
    \forall i,j \quad \lim_{\chi,\xi\to0}c_{i,j}^{R} & = 
    \tilde{c}_{R}\left(x\left(j\right),y\left(i\right)\right).
\end{align*}

We focus again on a single unit cell (fig.~\ref{fig:unit-cell-lengths-and-angles-perturbations}) and express the lengths at the right and upper edges $\l^A_{0,1},\l^B_{0,1},c^L_{1,0},c^L_{1,0}$ as functions of the left and bottom edges $\l^A_{0,0},\l^B_{0,0},c^L_{0,0},c^R_{0,0}$.
These are calculated by Euclidean geometry and appear in the supplementary Mathematica notebook.

Beginning with the $\tilde{\l}_A(x,y)$ field, the discrete difference $\l^A_{i,j+1}-\l^A_{i,j}$ is used to approximate its derivative:
\begin{equation*}
    \partial_{x}\tilde{\l}_A\left(x,y\right)\approx
    \frac{\tilde{\l}_A\left(x+\chi,y\right)-\tilde{\l}_A\left(x,y\right)}{\chi}
    =\frac{\l^A_{i,j+1}-\l^A_{i,j}}{\chi}
    .
\end{equation*}
Substituting the angles according to eq.~\eqref{eq:alternating-angles-by-pert} and taking the limit $\chi\to 0$ we get the equation:
\begin{equation} \label{eq:l_A-derivative}
    \partial_{x}\tilde{\l}_A\left(x,y\right) = 
    \frac{s \tilde{\delta }'(x) \sin \left(2 s \tilde{\delta }(x)\right) \tilde{\ell }_A(x,y)}{\cos \left(2 s \tilde{\delta }(x)\right)-\cos (2 \vartheta )}
    .
\end{equation}
For an equation without taking the $\chi\to 0$ limit we refer the reader to the supplementary Mathematica notebook.
The solution for this equation with boundary condition $\tilde{\l}_A(0,y)=\xi L_{tot}$ and for general $\tilde{\delta}(x)$ is given by:
\begin{equation}
    \tilde{\l}_A\left(x,y\right) =
    \xi L_{tot}
    \frac{  \sqrt{\cos \left(2 s \tilde{\delta }(0)\right)-\cos (2 \vartheta )}}{\sqrt{\cos \left(2 s \tilde{\delta }(x)\right)-\cos (2 \vartheta )}}
    ,
\end{equation}
Once again, this becomes our approximation for the vertical lengths of the origami by sampling $\l^A_{i,j}\approx \tilde{\l}_A(\chi j,\xi i)$.

We repeat the process for the other vertical lengths $\tilde{\l}_B$, and approximate the derivative:
\begin{equation*}
    \partial_{x}\tilde{\l}_B\left(x,y\right)\approx
    \frac{\tilde{\l}_B\left(x+\chi,y\right)-\tilde{\l}_B\left(x,y\right)}{\chi}
    =\frac{\l^B_{i,j+1}-\l^B_{i,j}}{\chi}
    .
\end{equation*}
In the limit $\chi\to 0$ we find the derivative of the same form as for $\l_A$, from eq.~\eqref{eq:l_A-derivative}:
\begin{equation*}
    \partial_{x}\tilde{\l}_B\left(x,y\right) =
    \frac{s \tilde{\delta }'(x) \sin \left(2 s \tilde{\delta }(x)\right) }{\cos \left(2 s \tilde{\delta }(x)\right)-\cos (2 \vartheta )} \tilde{\ell }_B(x,y)
    .
\end{equation*}
The solution to this equation, using the boundary condition $\tilde{\l}_B(0,y)=\xi L_{tot}\left(1+t\tilde{\Delta}\left(y\right)\right)$ is:
\begin{equation}
    \tilde{\l}_B(x,y) =
    \xi L_{tot} \left( 1+t \tilde{\Delta}\left(y\right) \right)
    \frac{\sqrt{\cos \left(2 s \tilde{\delta }(0)\right)-\cos (2 \vartheta )} }{\sqrt{\cos \left(2 s \tilde{\delta }(x)\right)-\cos (2 \vartheta )}}
    .
\end{equation}

We now study the two horizontal length fields $c^L_{i,j}$ and $c^R_{i,j}$, and their continuous counterparts $\tilde{c}_L$ and $\tilde{c}_R$.
We transform the discrete difference $c^L_{i+1,j}-c^L_{i,j}$ to approximated derivative using the scaling variable $\xi$ in the $y$ direction:
\begin{equation*}
    \partial_{y}\tilde{c}_L\left(x,y\right)\approx
    \frac{\tilde{c}_L\left(x,y+\xi \right)-\tilde{c}_L\left(x,y\right)}{\xi}
    =\frac{c^L_{i+1,j}-c^L_{i,j}}{\xi}
    .
\end{equation*}
We substitute the fields we already found for $\tilde{\l}_A(x,y)$ and $\tilde{\l}_B(x,y)$, and approximate it up to linear order in $\chi$:
\begin{equation*}
    \partial_{y} \tilde{c}_L \left(x,y\right)\approx 
    \chi L_{tot} \frac{s \tilde{\delta}'(x)\sqrt{\cos\left(2s\tilde{\delta}(0)\right)-\cos(2\vartheta)}}{2\sqrt{\cos\left(2s\tilde{\delta}(x)\right)-\cos(2\vartheta)}}\left(\frac{1+t\tilde{\Delta}(y)}{\sin\left(\vartheta-s\tilde{\delta}(x)\right)}-\frac{1}{\sin\left(\vartheta+s\tilde{\delta}(x)\right)}\right)
    +O(\chi^{2})
    .
\end{equation*}
With boundary condition $\tilde{c}_L(0,y)=\chi C_{tot}$, its solution to a leading-order is obtained by integrating along the $y$ coordinate:
\begin{equation}
    \tilde c_L(x,y)\approx 
    \chi C_{tot}+
    \chi L_{tot} \frac{s \tilde{\delta}'(x)\sqrt{\cos\left(2s\tilde{\delta}(0)\right)-\cos(2\vartheta)}}{2\sqrt{\cos\left(2s\tilde{\delta}(x)\right)-\cos(2\vartheta)}}\left(\frac{y+t\int_{0}^{y}\tilde{\Delta}\left(r\right)\,dr}{\sin\left(\vartheta-s\tilde{\delta}(x)\right)}-\frac{y}{\sin\left(\vartheta+s\tilde{\delta}(x)\right)}\right)
    .
\end{equation}
    
The last field to find is $\tilde{c}_R(x,y)$ for which we use the same procedure as $\tilde{c}_L(x,y)$. The approximated equation is similar:
\begin{equation*}
    \partial_{y} \tilde{c}_R(x,y)\approx 
\chi L_{tot} \frac{s \tilde{\delta}'(x)\sqrt{\cos\left(2s\tilde{\delta}(0)\right)-\cos(2\vartheta)}}{2\sqrt{\cos\left(2s\tilde{\delta}(x)\right)-\cos(2\vartheta)}}\left(\frac{1+t\tilde{\Delta}(y)}{\sin\left(\vartheta+s\tilde{\delta}(x)\right)}-\frac{1}{\sin\left(\vartheta-s\tilde{\delta}(x)\right)}\right)
    +O(\chi^{2})
    ,
\end{equation*}
and we use the same boundary condition $\tilde{c}_R(0,y)=\chi C_{tot}$, to get the solution to a leading-order:
\begin{equation}
    \tilde{c}_R(x,y)\approx 
    \chi C_{tot}+
\chi L_{tot}\frac{s\tilde{\delta}'(x)\sqrt{\cos\left(2s\tilde{\delta}(0)\right)-\cos(2\vartheta)}}{2\sqrt{\cos\left(2s\tilde{\delta}(x)\right)-\cos(2\vartheta)}}\left(\frac{y+t\int_{0}^{y}\tilde{\Delta}\left(r\right)\,dr}{\sin\left(\vartheta+s\tilde{\delta}(x)\right)}-\frac{y}{\sin\left(\vartheta-s\tilde{\delta}(x)\right)}\right)
% +O(\chi^{2})
    .
\end{equation}

\subsection{Folding angles} \label{sec:slowly-varying-pert-angles}
For fully characterizing the kinematics of any unit cell we also need to estimate the folding angles $\omega_{i,j},\gamma_{i,j}$ (as in fig.~\ref{fig:unit-cell-lengths-and-angles-perturbations}). 
From the analysis of folding angle around a vertex (eq.~\eqref{eq:angles-around-vertex}) we conclude that for any RFFQM pattern, $\omega$ is constant along columns, with $\omega_{i,j}=\omega_{0,j}$ and $\gamma$ is constant along rows, with $\gamma_{i,j}=\gamma_{i,0}$.
This holds since $\gamma_{i,j}$ (calculated from \eqref{eq:angles-around-vertex}) depends only on the folding angle $\omega_{i,j}$ and the sector angles around point $D$, with $\alpha^L_{i,j}=\pi-\vartheta+\eta_{i,j}^{L,D},\alpha^R_{i,j}=\pi-\vartheta+\eta_{i,j}^{R,D}$ (see fig.~\ref{fig:unit-cell-lengths-and-angles-perturbations}), and in the alternating angle perturbations $\eta_{i+1,j}^{L,D}=\eta_{i,j}^{L,D},\eta_{i+1,j}^{R,D}=\eta_{i,j}^{R,D}$, so $\gamma_{i+1,j}=\gamma_{i,j}$.
The constant value of $\gamma$ is calculated according to the 4-vertex relation \eqref{eq:angles-around-vertex} for the left-bottom unit cell:
\begin{equation}
    \tilde{\gamma}(x,y)\equiv
-\arccos\left(-\frac{\left(\cos \left(\omega _{0,0}\right)-1\right) \cos \left(2 s \tilde{\delta }(0)\right)+2 \cos \left(\omega _{0,0}\right)+\cos (2 \vartheta ) \left(\cos \left(\omega _{0,0}\right)+1\right)}{\left(\cos \left(\omega _{0,0}\right)-1\right) \cos \left(2 s \tilde{\delta }(0)\right)-\cos (2 \vartheta ) \left(\cos \left(\omega _{0,0}\right)+1\right)-2}\right)
+O(\chi)
    ,
\end{equation}

To find the values of $\omega_{i,j}$ everywhere, we again approximate the discrete difference as the derivative of a smooth field $\tilde{\omega}:[0,1]\to [-\pi,\pi]$.
The equation for finite $\chi$ is given in the supplementary Mathematica notebook. 
In the limit $\chi\to 0$ we obtain:
\begin{equation*}
    \partial_x \tilde{\omega}(x) = 
    \lim_{\chi\to 0} \frac{\omega_{i,j+1}-\omega_{i,j}}{\chi} =
    s \tilde{\delta }'(x) \tan \left(s \tilde{\delta }\left(x\right)\right) \sin \left(\tilde{\omega } \left(x\right)\right)
    ,
\end{equation*}
whose solution, for the boundary condition $\tilde{\omega}(0)=\omega_{0,0}$ is
\begin{equation}
    \tilde{\omega}(x)=
    2\text{arccot}\left(\cot\left(\frac{\omega_{0,0}}{2}\right)\frac{\cos\left(s\tilde{\delta}(x)\right)}{\cos\left(s\tilde{\delta}(0)\right)}\right)
.
\end{equation}

Equipped with all this data on the entire pattern, we can approximate the metric at each point and then calculate the Gaussian curvature and principal curvatures along the pattern. We provide our results in the next section, including some examples of inverse design of surfaces.

\section{Forward design: geometric fields of smoothly perturbed Miura-Ori} \label{sec:forward-design-geometric-fields}
Now that we obtained expressions for all the kinematic fields, we use our calculations for the lengths fields and evaluate the metric entries for a general unit cell at $y=\xi i,x=\chi j$.
In our rescaled coordinates with $\chi,\xi$ parameters, the metric entries should also be rescaled according to:
\begin{equation*}
    a=\matrixII{\frac{a_{xx}}{\chi^2}}{\frac{a_{xy}}{\chi \xi}}{\frac{a_{xy}}{\chi \xi}}{\frac{a_{yy}}{\xi^2}}
    .
\end{equation*}
We could at this point calculate the Gaussian curvature by plugging the metric and its derivatives in Gauss formula from eq.~\eqref{eq:K-by-metric-gauss}.
Similarly, we could obtain the second fundamental form and the principal curvatures.
The expressions obtained are, however, too complicated for our inverse design goal.
Instead, we Taylor expand the lengths and the metric entries in the perturbations factors $s,t$.
In order to understand the leading order in $s$ and $t$ for the various geometric fields, we first analyze the special cases with $s=0$ or $t=0$, that is, pure length or pure angle perturbations.

\subsection{Special cases} \label{sec:pert-special-cases}
We begin by studying the effect of each of the angle and length modifications on the resulting folded origami surfaces separately.
This gives important insights on the leading-order behavior of the Gaussian curvature in the general case.
In particular, we show that for small perturbation parameters $s,t$, the lowest order contribution to the curvature will be of order 2, proportional to $st$, while the terms $s^2,t^2$ will not contribute in this order.
In this discussion we assume that $s$ and $t$ are of the same order of magnitude, while $\chi,\xi$ are both much smaller.

\paragraph{$\delta \equiv \const;\Delta \equiv0$}
If we set $\delta \equiv \const$ and $\Delta \equiv0$ we get patterns that remain flat for any activation angle, as can be seen in the special example of Mars design from fig.~\ref{fig:MARS-example}.

\begin{figure}
    \centering
    \subfloat[\label{fig:pert-M-only}$\delta \equiv 0$, $\Delta \nequiv 0$]{
        \centering
        \includegraphics[width=0.59\linewidth]{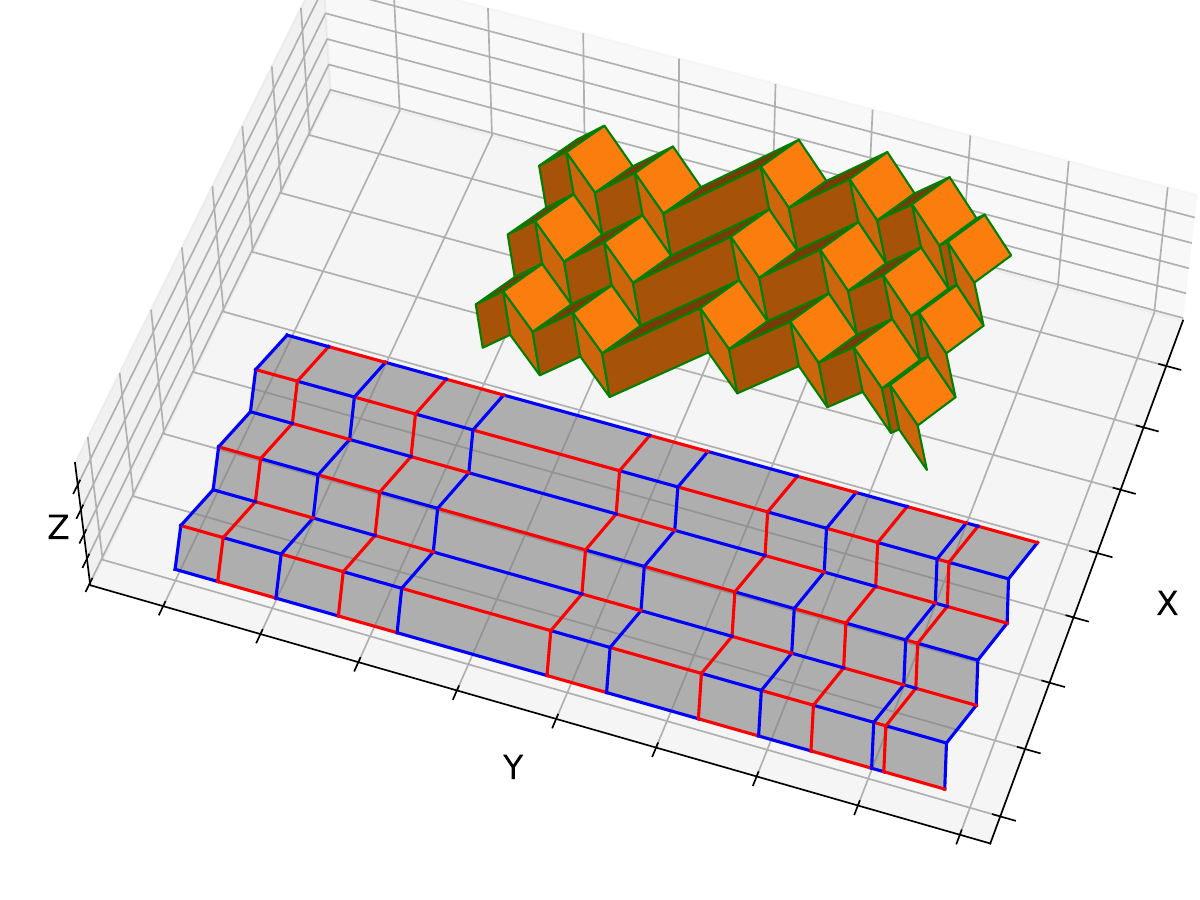}
    }
    \subfloat[\label{fig:pert-delta-only}$\delta \nequiv 0$, $\Delta\equiv 0$]{
        \centering
        \includegraphics[width=0.39\linewidth]{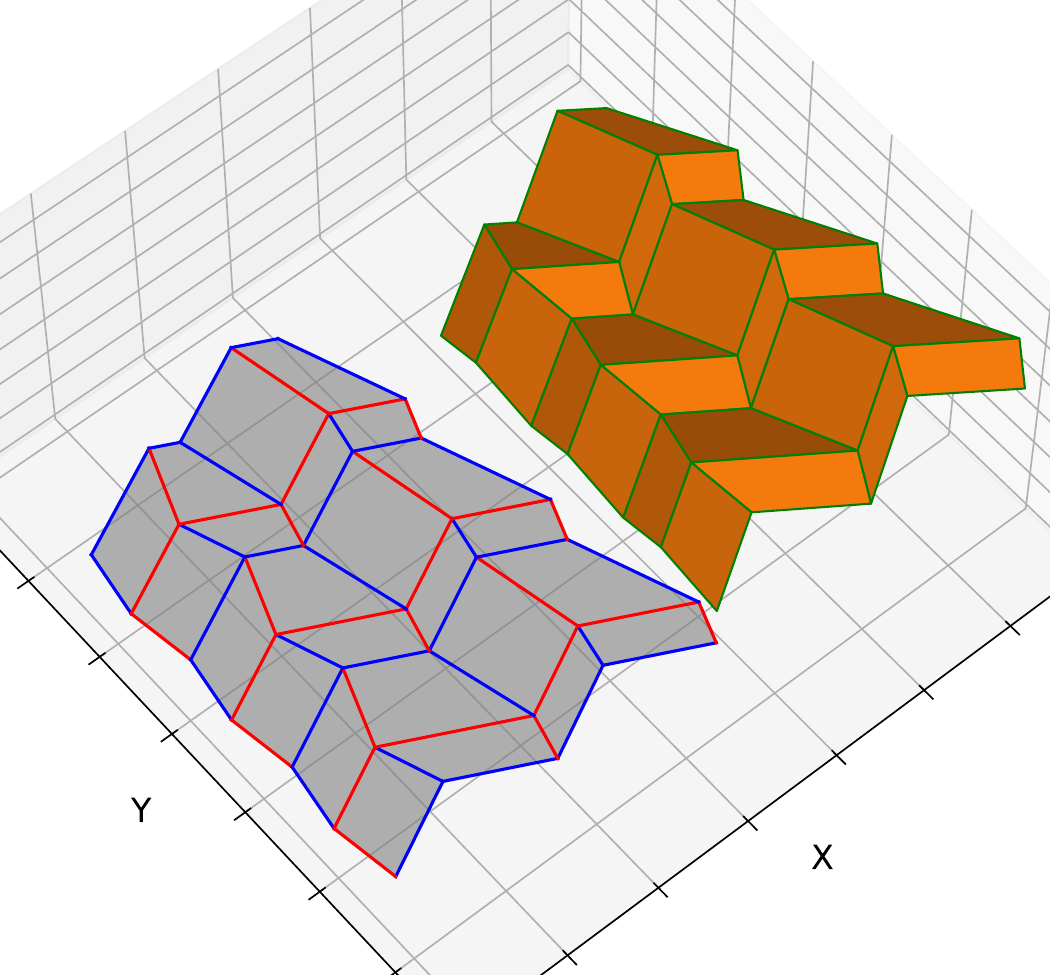}
    }
	
    \caption{
        Examples of surfaces achieved by the perturbations fields $\delta,\Delta$ applied separately.
        (a) is an example that has zero Gaussian curvature, showing that $\delta$ is necessary in order to get non-zero Gaussian curvature.
        In (a), the change in the vertical lengths induced by $\Delta (i)$ creates a zigzag pattern that can be used to approximate a smooth curve in the $YZ$ plane.
        Example (b) may have non-zero Gaussian curvature but it is of high order $O(s^4)$.
    }
\end{figure}

\paragraph{$\delta \equiv 0$}
If we allow $\Delta $ to vary, while $\delta \equiv 0$, then we obtain patterns that are curved along the $y$ direction but have translational symmetry along the $x$ direction, as seen in fig.~\ref{fig:pert-M-only}.
This kind of boundary condition can achieve curved lines when looking at the $YZ$ cross section of the folded pattern, with vertical lengths that form a zigzag shape. Other design procedures used this zigzag mechanism for the inverse design with RFFQM \cite{dang2022deployment, song2017design}.
Such surfaces have vanishing Gaussian curvature since their second fundamental form satisfies $b_{xx}\equiv b_{xy}\equiv 0$, and hence its determinant vanishes.
This means that the Gaussian curvature cannot have terms proportional to $t,t^2,t^3,...$ because they will contribute to the curvature even when $s=0$, which is equivalent to setting $\delta \equiv 0$.

\paragraph{$\Delta \equiv 0$}
We focus on 2 columns from a pattern in which $\Delta \equiv 0$ and $\delta $ is allowed to vary, as in fig.~\ref{fig:pert-delta-only}. 
The analysis here is more complicated since we do not have translational symmetry along the $y$ axis, as the horizontal lengths $c^L_{i,j},c^R_{i,j}$ change along the columns.
We can observe though that since $\Delta \equiv 0$, the vertical lengths of the left boundary are equal,
and because the horizontal folding angles $\gamma_{i,j}$ are constant, the tangent vector $\mathbb{X}_y$, marked by $\vec{AJ}$ in fig.~\ref{fig:folded-unit-cell}, also remains constant.
As a result, $b_{yy}$ vanishes along the entire first column.
For other columns, the vertical lengths may change, but for each column $j=j_0$ they always alternate between 2 values: $\l^A_{i,j_0}=\l^A_{0,j_0};\l^B_{i,j_0}=\l^B_{0,j_0}$. This can be observed in fig.~\ref{fig:pert-delta-only}, where all the blue crease lines at the same column have the same length, and similarly for the red crease lines, while the lengths still change between different columns. As a result, the vertical tangent vector that depends on $\l^A_{i,j_0}$ and $\l^B_{i,j_0}$ is constant along other columns as well. We conclude from that, that $b_{yy}$ vanishes for all the columns, so on the entire pattern, $b_{yy}\equiv 0$.

The term $b_{xy}$ in these patterns does not necessarily vanish, yet we can approximate the term $b_{xy}$ and see that it is at least of order $s^2$.
This is obtained by following the kinematics of a general unit cell from (as explained in \ref{app:kinematic-calculations}), using the data for the lengths and angles, and calculating the difference between 2 horizontal tangent vectors $\mathbb{X}_{xy}\approx\frac{\vec{AE}\left(x,y+\xi\right)-\vec{AE}\left(x,y\right)}{\xi}$, while setting $t=0$.
We find that $\mathbb{X}_{xy}\propto s^2$ so $b_{xy}$ is of order $s^2$.
Since $b_{yy}=0$, the Gaussian curvature is proportional to $\det(b_{\alpha \beta})\propto b_{xy}^2$, hence we expect the Gaussian curvature in these patterns to be at least of order The detailed calculations appear in the supplementary Mathematica notebook.

The argument above shows that if $t=0$, then the contribution to the Gaussian curvature would be of $O(s^4)$, and therefore the terms proportional to $s,s^2,s^3$ cannot contribute to the curvature.

\subsection{Approximating Gaussian curvature from the metric}
We approximate all the unit cell variables $\tilde{\l}_A,\tilde{\l}_B,\tilde{c}_L,\tilde{c}_R,\tilde{\omega},\tilde{\gamma}$ from sec.~\ref{sec:slowly-varying-pert-lengths} up to order $O(s^3,t^3)$.
The complete expressions appear in \ref{app:smooth-fields-metric-data}.
We present here a simplified version, in which terms proportional of $s^2$ and $t^2$, which do not contribute to the leading-order curvature are not written in full detail, but as functions $T_i$  that may depend on the position $x,y$, on the perturbations $\tilde{\delta},\tilde{\Delta}$ and on the constants $L_{tot},C_{tot},\omega_{0,0},\vartheta$ but have no dependence over $s,t,\xi,\chi$.
With this notation, the unit cells data up to $O\left(s^{3},t^{3},\chi^{2},\xi^{2}\right)$ is given by:
\begin{align}
    \tilde{\l}_A(x,y)
    &\approx\xi L_{tot}+\xi s^{2} T_{1};
    \nonumber   \\
    \tilde{\l}_B(x,y) &\approx
    \xi L_{tot}\left(1+t\tilde{\Delta}\left(y\right)\right)+\xi s^{2}T_{2};
    \nonumber   \\
    \tilde{c}_L(x,y) &\approx
    \chi C_{tot}+\frac{1}{2}\chi stL_{tot}\csc(\vartheta)\tilde{\delta}'(x)\int_{0}^{y}\tilde{\Delta}(r)\,dr+s^{2}\chi T_{3};
    \nonumber   \\
    \tilde{c}_R(x,y) & \approx
    \chi C_{tot}+\frac{1}{2}\chi stL_{tot}\csc(\vartheta)\tilde{\delta}'(x)\int_{0}^{y}\tilde{\Delta}(r)\,dr+s^{2}\chi T_{4};
    \nonumber   \\
    \tilde{\omega}(x,y)&\approx \omega_{0,0}+s^{2}T_{5}.
\end{align}
We use these to calculate the metric, approximating to order $O\left(s^{3},t^{3},\chi,\xi\right)$:
\begin{align}
    a_{xx}& \approx
    4C_{tot}^{2}\sin^{2}(\vartheta)\cos^{2}\left(\frac{\omega_{0,0}}{2}\right)+4stC_{tot}L_{tot}\sin(\vartheta)\cos^{2}\left(\frac{\omega_{0,0}}{2}\right)\tilde{\delta}'(x)\left(\int_{0}^{y}\tilde{\Delta}(r)\,dr\right)+s^{2}T_{6}
    ;
    \nonumber   \\
    a_{yy} & \approx
    L_{tot}^{2}\frac{8\cos^{2}\left(\vartheta\right)\left(1+t\tilde{\Delta}(y)\right)}{2-\sin^{2}\left(\vartheta\right)\cos\left(\omega_{0,0}\right)-\sin^{2}\left(\vartheta\right)}+s^{2}T_{7}+t^{2}T_{8}
    ;
    \nonumber   \\
    a_{xy}&\approx
    2stC_{tot}L_{tot}\sin(\vartheta)\tilde{\delta}(x)\tilde{\Delta}(y)+s^{2}T_{9}
    .
\end{align}
The metric derivatives and its inverse appear in the supplementary Mathematica notebook. Using Gauss formula from eq.~\eqref{eq:K-by-metric-gauss},
the approximated Gaussian curvature we find is: 
\begin{equation} \label{eq:K-by-pert}
    K(x,y)\approx
    \frac{st\tan(\vartheta)\sec(\vartheta)\tan^{2}\left(\frac{\omega_{0,0}}{2}\right)\tilde{\delta}'(x)\tilde{\Delta}'(y)\left(2\csc^{2}(\vartheta)-\cos\left(\omega_{0,0}\right)-1\right)}{16C_{tot}L_{tot}}
   +O\left(s^3,t^3,\chi,\xi\right)
   .
\end{equation}

\subsection{Approximating the second fundamental form} \label{sec:SFF-linear-approx}
\begin{figure}
	\centering
	\includegraphics[width = 0.6\linewidth]
	{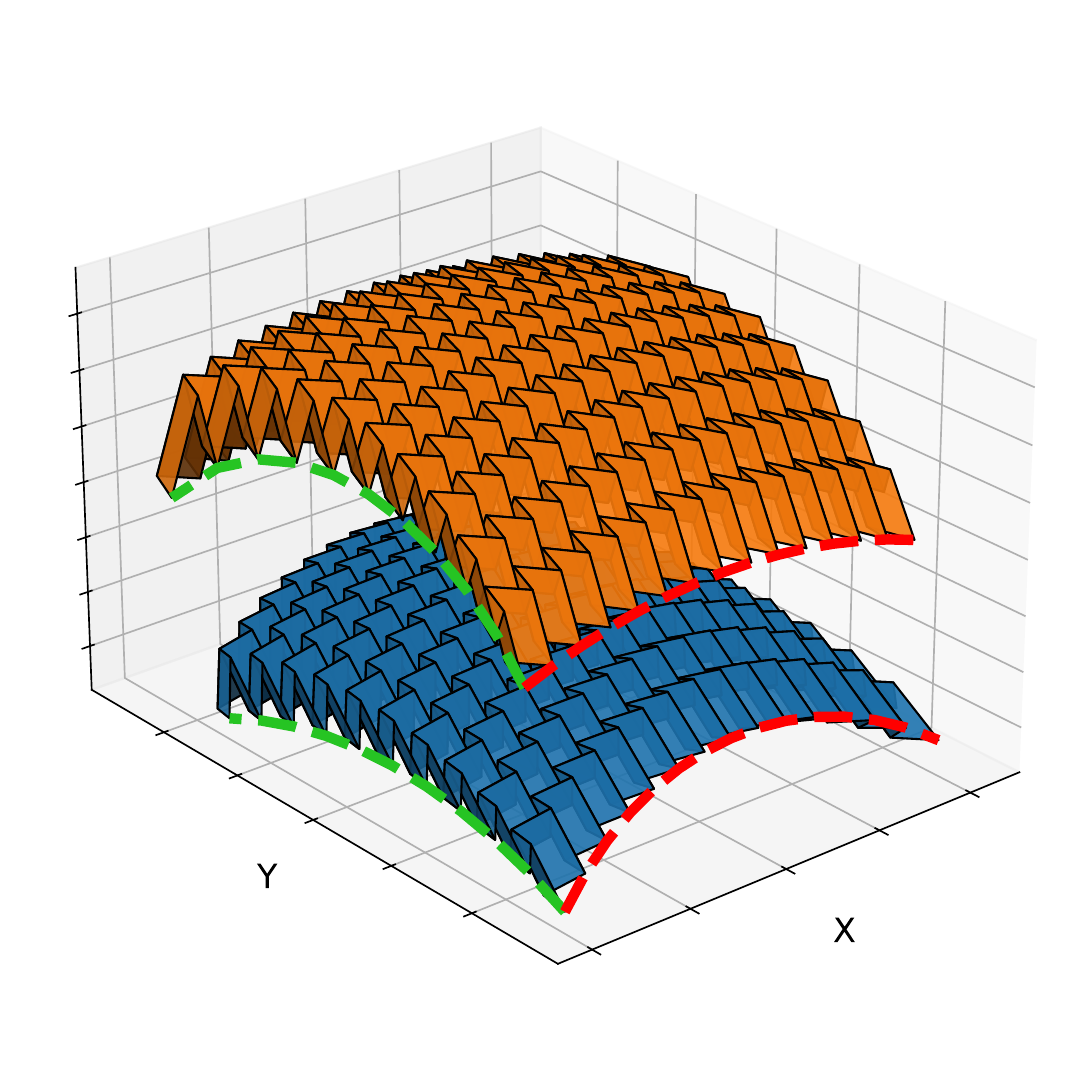}
	
	\caption{
	Designs of 2 surfaces with the same constant Gaussian curvature but with different principal curvatures.
	The red and green dashed line indicate the principal curvatures $\kappa_x$ and $\kappa_y$, respectively.
	}
	\label{fig:cap-different-curvatures}
\end{figure}

Having the expression for the Gaussian curvature opens the door for designing surfaces by dictating their curvature map.
But, the Gaussian curvature alone is only enough to determine the product of the perturbations $\tilde{\delta}'(x) \cdot \tilde{\Delta}'(y)$, as seen in eq.~\eqref{eq:K-by-pert}.
Keeping the product constant while choosing different values for the perturbations indeed produce different surfaces that have the same Gaussian curvature. The surfaces differ though by their principal curvatures. Fig.~\ref{fig:cap-different-curvatures} shows origami designs of 2 surfaces with the same constant Gaussian curvature. 
To account for this difference, and allow more control over the designing process, we need to approximate the second fundamental form as well.

\begin{figure}
	\centering
	\includegraphics[width = 0.5\linewidth]
	{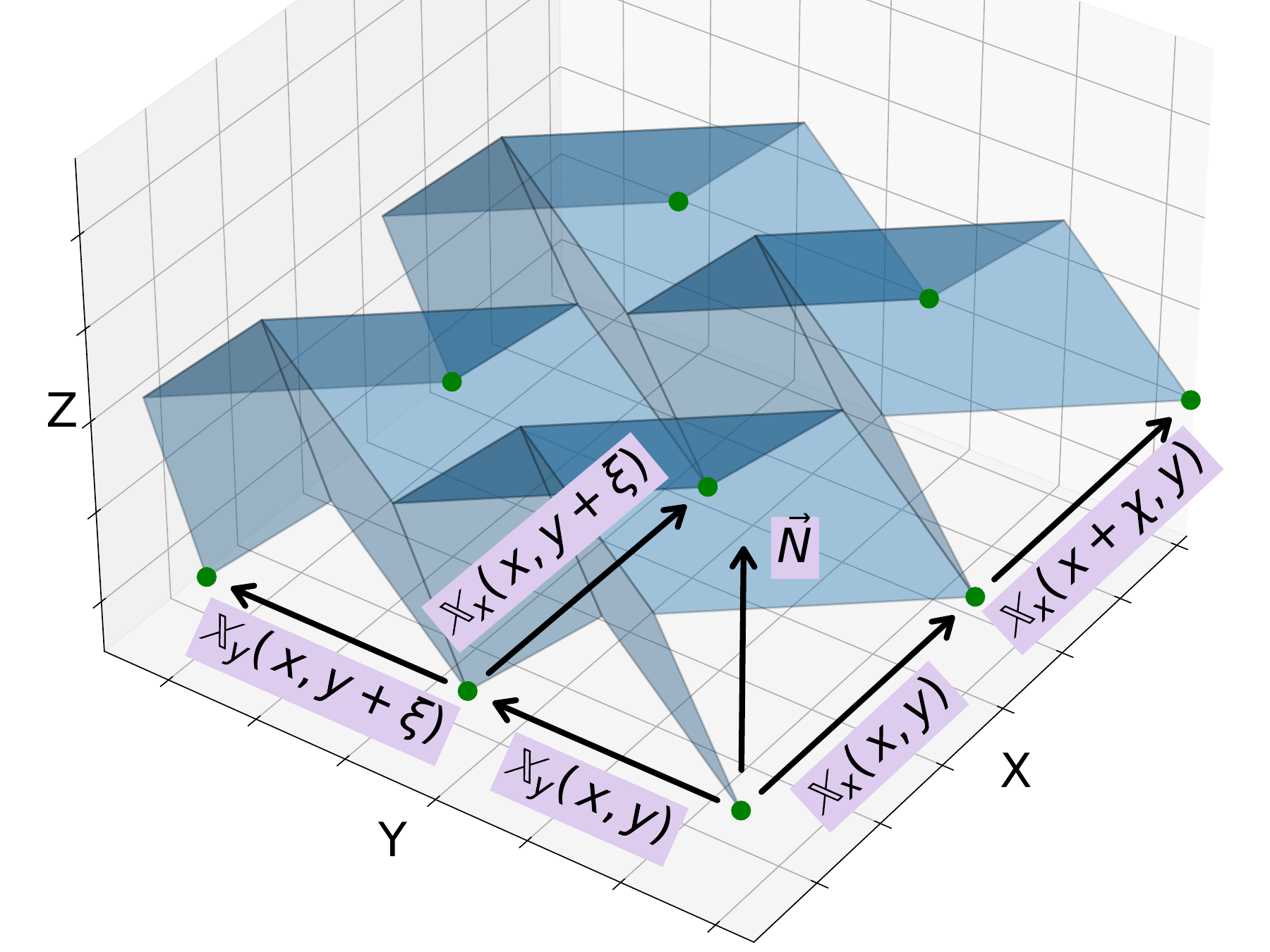}
	
	\caption{
        The vectors used to approximate the second fundamental form. The tangent vectors for the different unit cells are $\mathbb{X}_x, \mathbb{X}_y$ and the normal vector for the cell at $x,y$ is marked by $\vec N$.
    }
	\label{fig:sff-vectors}
\end{figure}

The second fundamental form includes derivatives of the tangent vectors, and for that we need to look at a unit cell and its neighboring cells. We approximate the entries by the discrete difference according to eq.~\eqref{eq:SFF-in-RFFQM}, for the tangent vectors as appear in fig.~\ref{fig:sff-vectors}.
Since we need to approximate the 2nd derivatives of the configuration $\mathbb{X}$ we approximate the tangent vectors $\mathbb{X}_x,\mathbb{X}_y$ up to linear order in $\chi,\xi$:
Then the second derivatives are taken up to zeroth order in these scaling parameters:
\begin{align*}
    \mathbb{X}_{xx}&\approx\frac{\mathbb{X}_{x}\left(x+\chi,y\right)-\mathbb{X}_{x}\left(x,y\right)}{\chi}+O\left(\chi,\xi\right),\\\mathbb{X}_{yy}&\approx\frac{\mathbb{X}_{y}\left(x+\chi,y\right)-\mathbb{X}_{y}\left(x,y\right)}{\xi}+O\left(\chi,\xi\right),\\\mathbb{X}_{xy}&\approx\frac{\mathbb{X}_{x}\left(x,y+\xi\right)-\mathbb{X}_{x}\left(x,y\right)}{\xi}+O\left(\chi,\xi\right).
\end{align*}

We use our approximation for the lengths, pattern angles and folding angles, as provided in sec.~\ref{sec:slowly-varying-pert-lengths}, to write the second fundamental form entries for a general unit cell at coordinate $x,y$.
We continue to treat the perturbations parameters $s,t$ as small variables and approximate the expressions up to $O(s^2,t^2)$.

% SFF vectors notation

The resulting approximated second fundamental form is:
{
\begin{align}
    b_{xx} & \approx 
    -sC_{tot}\sin\left(\frac{\omega_{0,0}}{2}\right)\tilde{\delta}'(x)\sqrt{2}\sqrt{2-\sin^{2}\left(\vartheta\right)\cos\left(\omega_{0,0}\right)-\sin^{2}\left(\vartheta\right)}
   ;   \nonumber    \\
    b_{yy} & \approx 
	-\frac{2tL_{tot}\sin(\vartheta)\sin\left(\frac{\omega_{0,0}}{2}\right)\tilde{\Delta}'(y)}{\sqrt{2}\sqrt{2-\sin^{2}\left(\vartheta\right)\cos\left(\omega_{0,0}\right)-\sin^{2}\left(\vartheta\right)}}
   ;   \nonumber    \\
   b_{xy} & \approx 
	0
   .
\end{align}
}
By eq.~\eqref{eq:shape-op-by-forms} and using the metric entries appearing in \ref{app:smooth-fields-metric-data}, we find the approximate shape operator to be diagonal, with $C_{xy}=0$ and the diagonal terms are:
{
\begin{align}
    C_{xx} & \approx \kappa_{x} \approx
    \frac{s\csc(\vartheta)\tan\left(\frac{\omega_{0,0}}{2}\right)\sec\left(\frac{\omega_{0,0}}{2}\right)\tilde{\delta}'(x)\sqrt{2}\sqrt{2\csc^{2}\left(\vartheta\right)-\cos\left(\omega_{0,0}\right)-1}}{4C_{tot}}
	\label{eq:kx-approx}
   ;\\
   C_{yy} & \approx \kappa_{y} \approx
	\frac{t\tan^{2}\left(\vartheta\right)\sin\left(\frac{\omega_{0,0}}{2}\right)\tilde{\Delta}'(y)\sqrt{2}\sqrt{2\csc^{2}\left(\vartheta\right)-\cos\left(\omega_{0,0}\right)-1}}{8L_{tot}}
	\label{eq:ky-approx}
   .
\end{align}
}
As a diagonal matrix, the diagonal entries $C_{xx},C_{yy}$ of $C$ are its eigenvalues, which give the principal curvatures $\kappa_{x},\kappa_{y}$ along the $x,y$ directions, respectively (that is, the $x,y$ directions are, to a leading order, the principle directions).
Their product agrees with our previous result of the approximated Gaussian curvature from eq.~\eqref{eq:K-by-pert}.

\subsection{Weakly nonlinear approximation of the principal curvatures} \label{sec:principal-curvatures-approx}

Our results for the principal curvatures show that, in leading-order, the perturbation field $\delta(j)$ creates curvature $\kappa_x$ in the $x$ direction and the field $\Delta(i)$ for the lengths creates curvature $\kappa_y$ in the $y$ direction.
Thus, in the leading order, the Gaussian curvature of a surface obtained by the profiles $\delta$ and $\Delta$ is a product of $\kappa_x$ of the surface $\delta\ne 0,\Delta=0$, and $\kappa_y$ of the surface $\delta= 0,\Delta \ne 0$.
In this section, we analyze in more detail these two simpler cases, and obtain expressions to $\kappa_x$ and $\kappa_y$ that go beyond first order.
As it turns out, the product of these expressions yields better results for inverse design problems than eq.~\eqref{eq:K-by-pert}.

\begin{figure}
	\centering
	\begin{subfigure}[b]{0.45\linewidth}
		\centering
		\includegraphics[width=\linewidth]
		{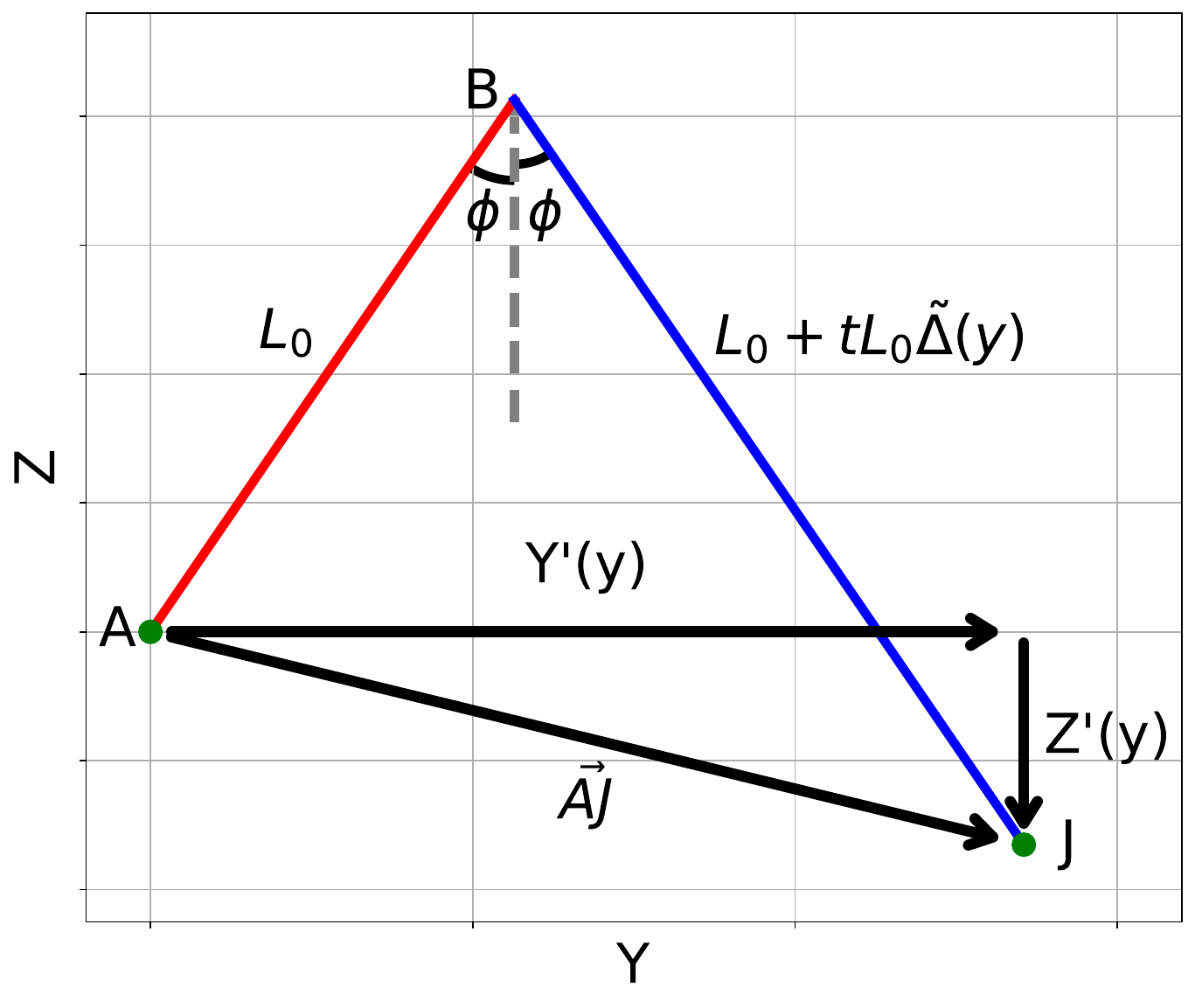}
		\caption{\label{fig:yz-zigzag-cross-section}}
	\end{subfigure}
	\begin{subfigure}[b]{0.45\linewidth}
		\centering
		\includegraphics[width=\linewidth]{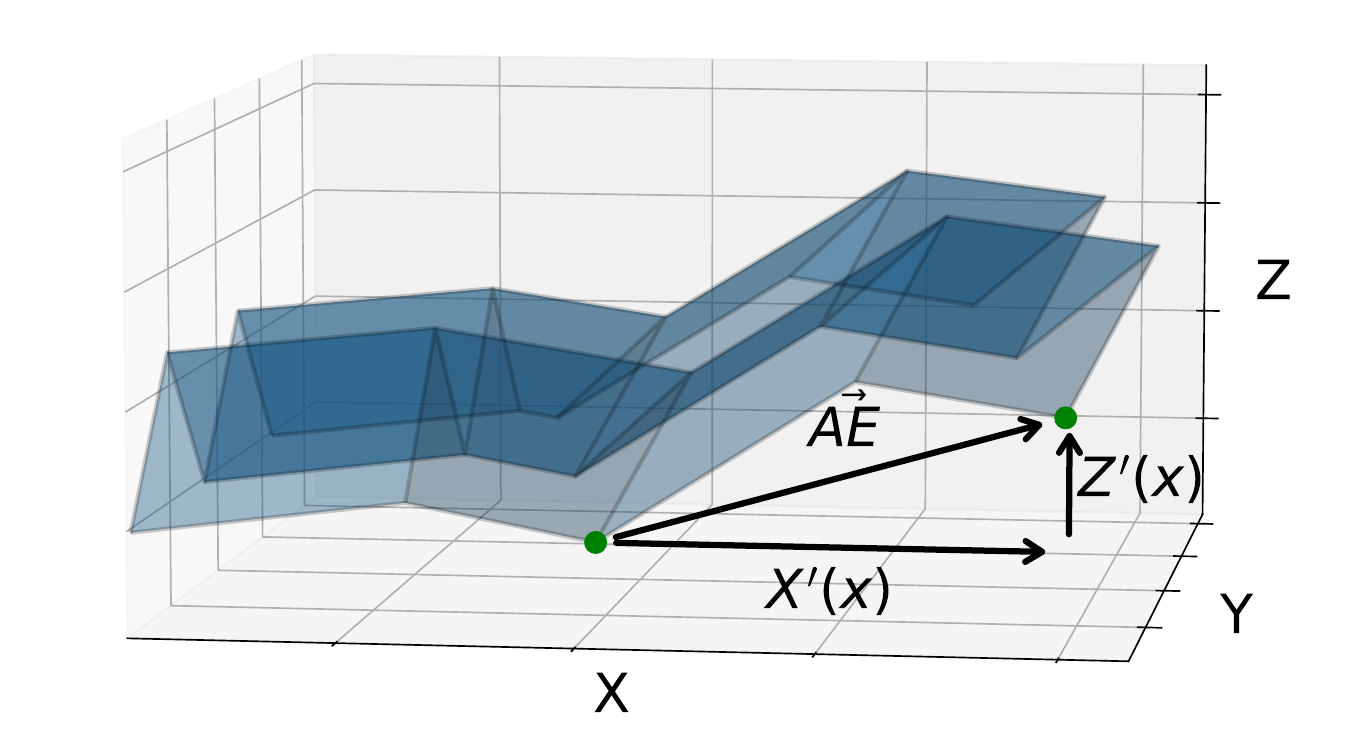}
		\caption{\label{fig:xz-cross-section}}
	\end{subfigure}
	\caption{
		(a) a cross section in the $YZ$ plane of a single unit cell with $\tilde{\Delta}(y)>0$. The zigzag pattern is characterized by the single angle $\phi$ that is related to the folding angle $\omega$.
		The displacement vector $\vec{AJ}$ is split to its components $Y'(y),Z'(y)$.
		(b) a grid of $2\times 2$ unit cells, the first has no perturbation, and the second is perturbed with constant $\delta$. The unperturbed cell is used to fix the orientation in space, and the perturbed cell is used to analyze the displacement $\vec{AE}$ as a function of the perturbation.
		We note that $\vec{AE}$ has also small component in $Y$ direction that is not visible in the plot.
	}
\end{figure}

\paragraph{Principal curvature in y direction}
In the case $\delta= 0,\Delta \ne 0$, we can completely characterize the resulting surfaces and calculate precisely the curvature along the $y$ direction.
An example of surface of this kind is seen in fig.~\ref{fig:pert-M-only} where we can see that it has translational symmetry along the $x$ direction but the perturbations can create zigzag patterns in the $YZ$ plane that follow a curve in that plane.
To analyze the principal curvature $\kappa_y$ in this case, it suffices to look at the cross section of the pattern in the $YZ$ plane, as in fig.~\ref{fig:yz-zigzag-cross-section}.

For a single unit cell, the displacement $\vec{AJ}$ is assumed to describe the derivative of a smooth curve in the $YZ$ plane.
We look at such curve in parametrized form
\begin{equation*}
    y\mapsto \begin{bmatrix}Y(y)\\Z(y)\end{bmatrix}
,
\end{equation*}
with the same parameter $y$ of our origami coordinates.
The vector $\vec{AJ}$ is the displacement of the curve after a discrete step, and we express it by its $YZ$ components:
\begin{equation*}
    \vec{AJ}=\begin{bmatrix}Y\left(y+\xi\right)\\
Z\left(y+\xi\right)
\end{bmatrix}-\begin{bmatrix}Y\left(y\right)\\
Z\left(y\right)
\end{bmatrix}=\xi L_{tot}\begin{bmatrix}\left(2+t\tilde{\Delta}\left(y\right)\right)\sin\phi\\
-t\tilde{\Delta}\left(y\right)\cos\phi
\end{bmatrix},
\end{equation*}
for $\phi$ the zigzag angle as appeared in fig.~\ref{fig:yz-zigzag-cross-section}, which is related to the folding angle by: 
\begin{equation} \label{eq:phi-by-gamma}
    \cos(\phi)=-\cos (\gamma ) \sin ^2(\vartheta )-\cos ^2(\vartheta )   
    .
\end{equation}
Dividing by $\xi$ will give the approximation for the derivative of the curve:
\begin{equation*}
    \frac{\vec{AJ}}{\xi} = 
    \begin{bmatrix}Y'\left(y\right)\\
Z'\left(y\right)
\end{bmatrix}=L_{tot}\begin{bmatrix}\left(2+t\tilde{\Delta}\left(y\right)\right)\sin\phi\\
-t\tilde{\Delta}\left(y\right)\cos\phi
\end{bmatrix}
.
\end{equation*}
The curvature $\kappa_y$ is the inverse of the radius of curvature, which for parametrized curve is given by 
\begin{equation*}
R_{y}=\left|\frac{\left(\left(Y'\left(y\right)\right)^{2}+\left(Z'\left(y\right)\right)^{2}\right)^{3/2}}{Y'(y)Z''(y)-Z'(y)Y''(y)}\right|,
\end{equation*}
from which we obtain 
\begin{equation}
    \kappa_y=
    \frac{t\sin(2\phi)\tilde{\Delta}'(y)}{L_{tot}\left(t^{2}\tilde{\Delta}(y)^{2}+4\sin^{2}(\phi)\left(1+t\tilde{\Delta}(y)\right)\right)^{3/2}}
    .
\end{equation}
The sign of $\kappa_y$ is determined to match our previous sign convention of the approximated principal curvature from eq.~\eqref{eq:ky-approx}.
Substituting $\phi$ from \eqref{eq:phi-by-gamma} and $\gamma$ from \eqref{eq:gamma-by-omega-miura-ori} yields the curvature
\begin{equation} \label{eq:improved-ky-approx}
\kappa_y=
    \frac{
    t\sin^{2}\left(\vartheta\right)\cos\left(\vartheta\right)\sin\left(\frac{\omega_{0,0}}{2}\right)\tilde{\Delta}'\left(y\right)\sqrt{2}\sqrt{2\csc^{2}\left(\vartheta\right)-\cos\left(\omega_{0,0}\right)-1}
    }{8L_{tot}\left(\cos^{2}\left(\vartheta\right)+t\cos^{2}\left(\vartheta\right)\tilde{\Delta}\left(y\right)+\frac{1}{8}t^{2}\tilde{\Delta}\left(y\right)^{2}\sin\left(\vartheta\right)\left(2\csc^{2}\left(\vartheta\right)-\cos\left(\omega_{0,0}\right)-1\right)\right)^{3/2}},
\end{equation}
and approximating $t$ up to linear term indeed reproduces our previous result for the approximated principal curvature from eq.~\eqref{eq:ky-approx}.

\paragraph{Principal curvature in x direction}
The calculation of $\kappa_x$, even in the case $\Delta \equiv0$, is more complicated, and, unlike $\kappa_y$, is only approximate, since an origami pattern with $\Delta \equiv0$ has transnational symmetry only to a leading order.
To analyze this case we focus on 2 unit cells adjacent to each other in the $x$ direction, as shown in fig.~\ref{fig:xz-cross-section}. 
The first unit cell is kept unperturbed, with $\delta(0)=\delta(\frac{1}{2})=0$, while the other is perturbed by some constant $\delta(1)=\delta(\frac{3}{2})=\delta(2)=\delta_0$. 
The unperturbed cell is used to define the axes in space, where it is aligned so that $\vec{AE}(0,0)\parallel \hat{X}$ and $\vec{AJ}(0,0)\parallel \hat{Y}$. Since $\Delta \equiv 0$, the normal points to $\hat{Z}$.
Next, we calculate the tangent vector $\vec{AE}(1,0)$ that appears in the figure, and find its $X$ component and $Z$ component. Here though, $\vec{AE}(1,0)$ has also $Y$ component that we neglect in order to simplify the analysis, which is valid because it is of higher order in $s$.
This emphasizes, though, that the effect of the angle perturbation $\delta$ is more complex and not fully characterized here.

We treat the vector $\vec{AE}(1,0)$ as sitting on a curve in $XZ$ plane, so its components are the derivatives of a parametrized curve
\begin{equation*}
    x\mapsto \begin{bmatrix}X(x)\\Z(x)\end{bmatrix}
.
\end{equation*}
Replacing the constant $\delta_0$ with a general $\tilde{\delta}(x)$, the components are:
\begin{align*}
    X'(x) & \approx
    2 C_{tot} \cos \left(\frac{\omega_{0,0}}{2}\right) \sin (\vartheta )
    + O(s^2)
    , \\
    Z'(x) & \approx
    -s C_{tot}\sin\left(\frac{\omega_{0,0}}{2}\right)\sin\left(\vartheta\right)\tilde{\delta}(x) \sqrt{2} \sqrt{2\csc^{2}\left(\vartheta\right)-\cos\left(\omega_{0,0}\right)-1}+O(s^{2})
    .
\end{align*}
Using again the formula for radius of curvature of a curve in plane, we find $\kappa_x$ to be:
\begin{equation} \label{eq:improved-kx-approx}
    \kappa_x \approx
\frac{s\csc\left(\vartheta\right)\tan\left(\frac{\omega_{0,0}}{2}\right)\sec\left(\frac{\omega_{0,0}}{2}\right)\tilde{\delta}'\left(x\right)\sqrt{2}\sqrt{2\csc^{2}\left(\vartheta\right)-\cos\left(\omega_{0,0}\right)-1}}{4C_{tot}\left(1+\frac{1}{2}s^{2}\tan^{2}\left(\frac{\omega_{0,0}}{2}\right)\tilde{\delta}\left(x\right)^{2}\left(2\csc^{2}\left(\vartheta\right)-\cos\left(\omega_{0,0}\right)-1\right)\right)^{3/2}}
    .
\end{equation}
Taking only linear term in $s$ in eq.~\ref{eq:improved-kx-approx} agrees with our previous approximation of the principal curvature by the second fundamental form from eq.~\eqref{eq:kx-approx}.

In the inverse design process we will use the formulas presented here for the principal curvatures (eq.~\eqref{eq:improved-ky-approx} and eq.~\eqref{eq:improved-kx-approx}) to design surfaces, as they are fairly simple yet yield much better results compared to the linearized ones. 
If the inverse design is done by a target Gaussian curvature field, we will use the product of these principal curvature as approximation to the Gaussian curvature.

\section{Inverse design algorithm and results} \label{sec:inverse-design-results}
We now describe the inverse design algorithm, followed by a detailed example of inverse designing a spherical cap with constant Gaussian curvature. 
We compare two approaches to design the surface: the linearized approximation to the curvatures from sec.~\ref{sec:SFF-linear-approx} and the weakly nonlinear approximation to the principal curvatures from sec.~\ref{sec:principal-curvatures-approx}, showing that indeed the latter is more accurate.

\paragraph{Inverse design algorithm}
\begin{enumerate}
    \item \textbf{Create Geometric data:} To target a specific geometry we first set the functional form of $\kappa_x$ and $\kappa_y$, varying along $x$ and $y$ correspondingly. We do it either by prescription, or by deriving directly from a specific target surface.
%    
%    Our method enables to inverse design surfaces whose principal curvatures $\kappa_x$ and $\kappa_y$ each depend on a single variable ($x$ and $y$, respectively), and whose principle directions also diagonalize the metric.

    \item \textbf{Setting and Solving the Equation:} The principal curvatures  are substituted in \eqref{eq:improved-ky-approx} and \eqref{eq:improved-kx-approx} to obtain two uncoupled differential equations for the perturbation fields $t\tilde{\Delta}$ and $s\tilde{\delta}$. We solve these equations for desired values of the parameters $C_{tot},L_{tot}$ which set the total size of the flat pattern, the unperturbed angle $\vartheta$,  and the activation angle $\omega_{0,0}$ at which the desired geometry will be achieved.

%    \item \textbf{Obtaining the smooth boundary perturbations:} Given this information, we set the length parameters $C_{tot},L_{tot}$ that are roughly the total size of the flat pattern
%    the unperturbed angle $\vartheta$ and the activation angle $\omega_{0,0}$ to use.
%    We substitute the parameters in the expressions for the principal curvatures \eqref{eq:improved-ky-approx} and \eqref{eq:improved-kx-approx} to obtain two uncoupled differential equations for the perturbation fields $t\tilde{\Delta}$ and $s\tilde{\delta}$.
    \item \textbf{Discrete Sampling Of The Boundary Perturbations:} Upon solving these equations on the segment $\left[0,1\right]$ for initial conditions $t\tilde{\Delta}(0),s\tilde{\delta}(0)$ to our choice, we set the scaling parameters $\chi,\xi$ which  determine the discretization, i.e. the number of  unit cells in the pattern. The boundary angles are then set by eq.~\eqref{eq:alternating-angles-by-pert-on-boundary} and the lengths by eq.~\eqref{eq:lengths-boundary-condition}.
 
%    \item \textbf{Discrete sampling of the boundary perturbations:} Given the smooth perturbations, we choose the scaling parameters $\chi,\xi$ that should be smaller than the oscillations in the perturbation fields $\tilde{\delta}$ and $\tilde{\Delta}$, respectively.
%    They will determine the total number of unit cells in the pattern (according to eq.~\eqref{eq:chi_xi}) and the smoothness of the folded origami.
%    The boundary angles are set by eq.~\eqref{eq:alternating-angles-by-pert-on-boundary} and the lengths by eq.~\eqref{eq:lengths-boundary-condition}.
    
    \item \textbf{Recovering The Origami Crease Pattern} Given the lengths and angles on the boundaries, we use the (exact) marching algorithm to obtain the full crease pattern.
    The angle perturbations are given by eq.~\eqref{eq:alternating-angles-by-pert}, and the lengths (to a leading order) are given in sec.~\ref{sec:slowly-varying-pert-lengths}.
    
    \item \textbf{Folding The Pattern:} Given the full pattern, we fold the origami by the predetermined activation angle $\omega_{0,0}$, yielding an approximation to the target surface prescribed by its principal curvatures.
\end{enumerate}

\paragraph{Failure Scenarios}
This procedure may fail in the following cases:
\begin{enumerate}
    \item If the solutions for the perturbations diverge in the segment $\left[0,1\right]$ or if the solutions are not in the valid range for the perturbations:
    \begin{align}
    \forall x\in\left[0,1\right]&\quad-\vartheta<s\tilde{\delta}\left(x\right)<\vartheta,
    \nonumber   \\
    \forall y\in\left[0,1\right]&\quad t\tilde{\Delta}\left(y\right)>-1.
    \end{align}
    
    \item If the marching algorithm yields nonpositive crease lengths.
\end{enumerate}

\begin{figure}
    \centering
    
    \subfloat[\label{fig:inverse-design-example}]{
		\centering
		\includegraphics[width=0.40\linewidth]
        {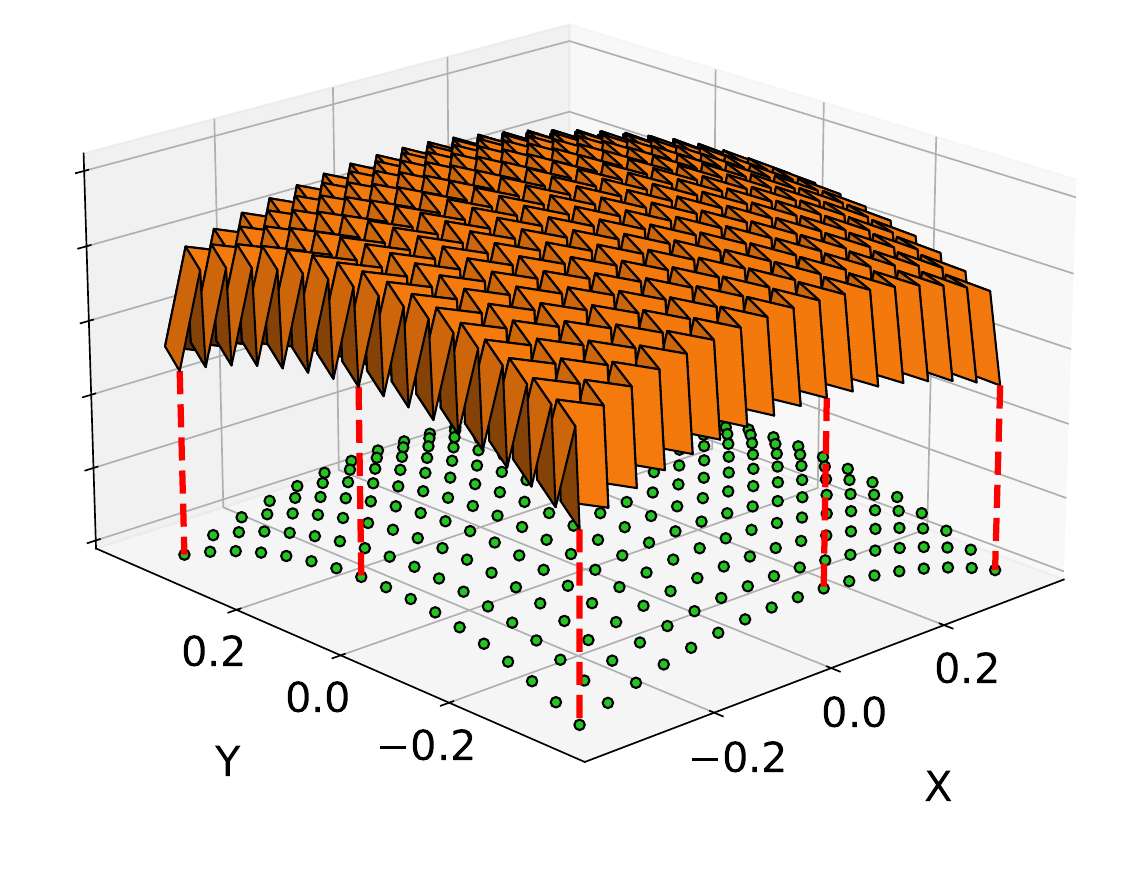}
	}
    \subfloat[\label{fig:inverse-design-actual-K}]{
        \centering
		\includegraphics[height=0.25\linewidth]
        {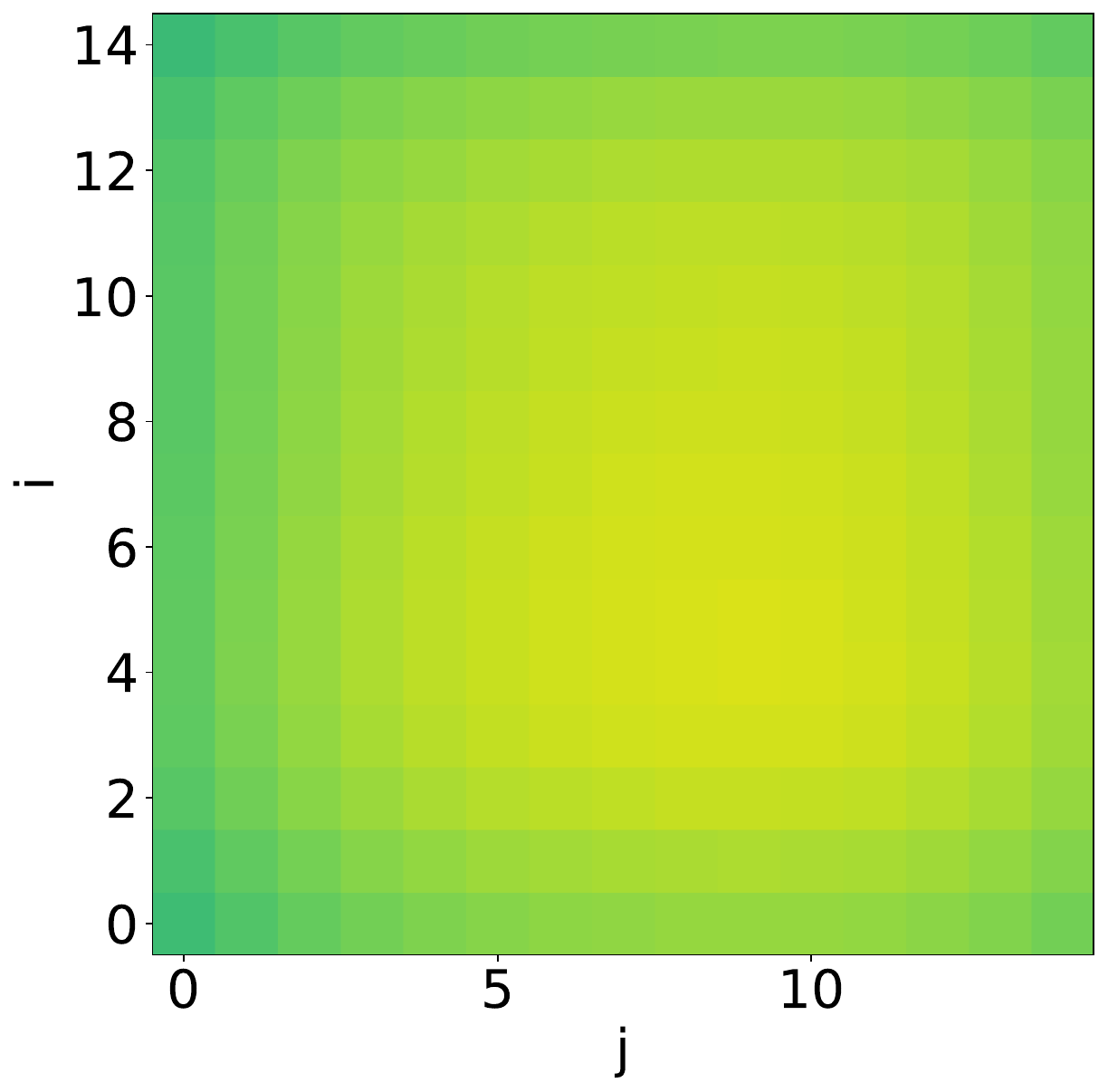}
	}
    \subfloat[\label{fig:inverse-design-comparison-linearized}]{
      \includegraphics[height=0.25\linewidth]
        {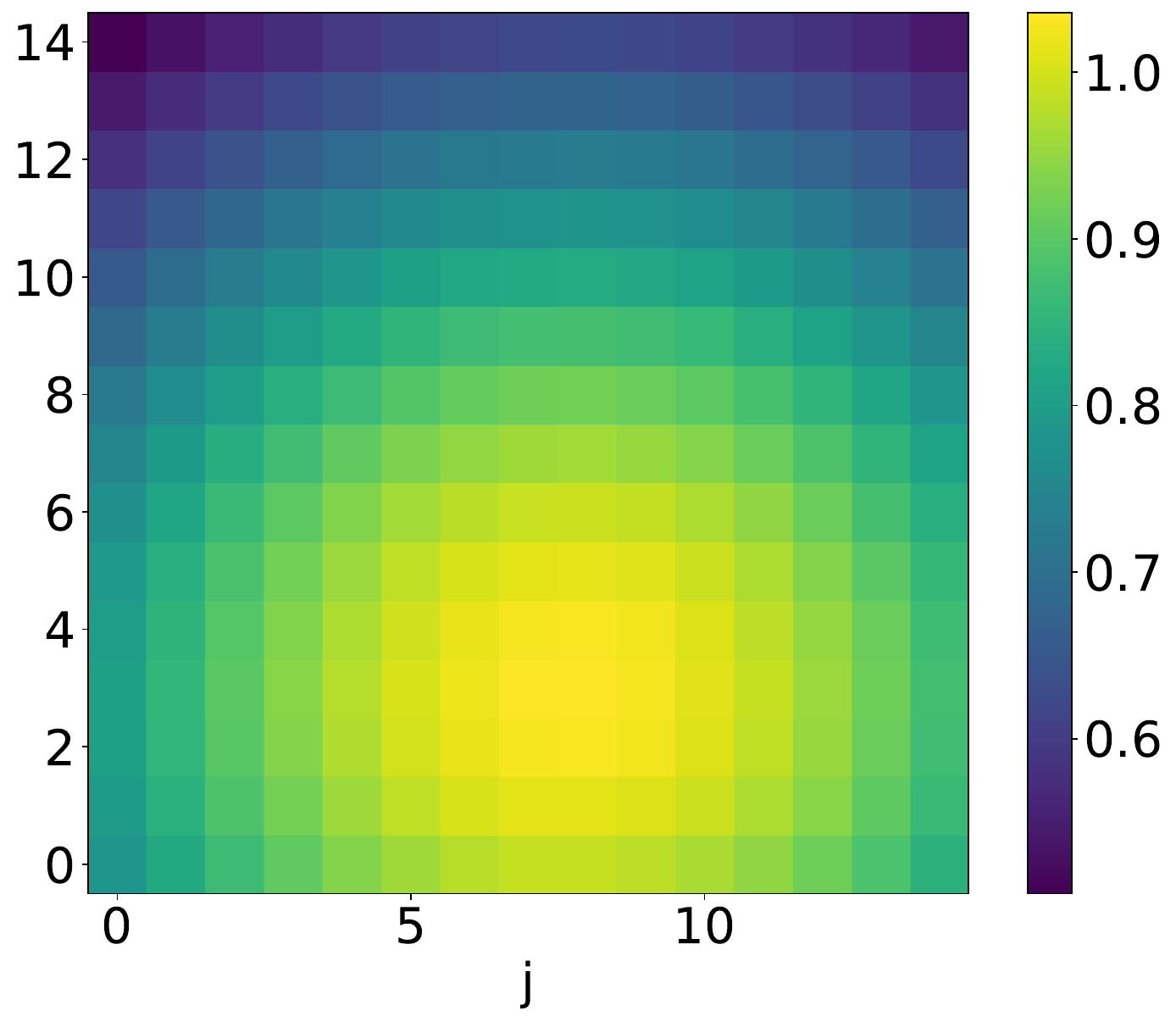}
	}
	
	\caption{
        (a) An origami pattern achieved by the inverse design procedure.
        Our improved approximation for the curvatures, discussed in sec.~\ref{sec:principal-curvatures-approx}, was used to design a surface with targeted constant curvatures $\kappa_x\equiv \kappa_y \equiv 1$.
        The green dots are the dots on the unit cells that should sit on the target surface, shifted downward for visibility purposes.
        (b) The actual Gaussian curvature of the resulting surface, shown as heat map where the curvature is computed numerically for each unit cell with $i,j$ coordinates on the grid.
        Due to the numerical evaluation of the metric derivatives, there are $(N_y-1)\times (N_x-1)=15\times 15$ pixels in the heat map.
        We can see that the resulting surface has curvature of $\approx 1$ at the middle of the pattern, and the curvature drops to $\approx 0.9$ at the edges.
        (c) The curvature map of another surface, achieved by inverse designing the same target surface but using the linear approximation to the curvatures from sec.~\ref{sec:SFF-linear-approx} to find the perturbation fields. Here we see that the curvature is $\approx 1$ only at small region and it drops to $\approx 0.6$ at the edges.
        This shows that our weakly nonlinear approximation of the principal curvatures indeed improves our ability to design surfaces when the perturbations are relatively large.
    }
\end{figure}

\paragraph{Example: Nonlinear Design of a spherical cap}
In this example we approximate a surface with constant Gaussian curvature. In equations \eqref{eq:improved-ky-approx}--\eqref{eq:improved-kx-approx}  we substitute $\kappa_x(x)\equiv \kappa_y(y)\equiv 1$ and set system size by $C_{tot}=1,L_{tot}=2$. The unperturbed parallelograms angle is chosen to be $\vartheta=0.45 \pi$, and the activation angle $\omega_{0,0}=0.76 \pi$.\footnote{Generally, values close to $\pi/2$ are good for the inverse design of surfaces with significant curvature due to the divergence of $\tan(\vartheta)\sec(\vartheta)$ term in eq.~\eqref{eq:K-by-pert}. }
Upon solving the perturbations fields $s \tilde \delta(x)$ and $t \tilde{\Delta}(y)$ with boundary conditions $s \tilde{\delta}(0)=\SI{-0.15}{\radian}$ and $t \tilde{\Delta}(0)=-0.1$, and setting the scaling parameters $\chi$ and $\xi$ to $\chi=\xi=1/16$, meaning there are $16\times 16$ unit cells in the origami pattern, we construct the creases pattern, and fold it to the predetermined activation angle.
%The initial conditions for the solutions that we choose to be: $s \tilde{\delta}(0)=\SI{-0.15}{\radian}$ and $t \tilde{\Delta}(0)=-0.1$.
%The initial conditions were chosen by simple trial and error in order to find perturbation fields whose magnitudes remain small.
%With the boundary data, we build the pattern and fold it to the predetermined activation angle.
The resulting origami appears in fig.~\ref{fig:inverse-design-example}.

To test our predictions, the fundamental forms of the origami are computed numerically for each unit cell as discussed in sec.~\ref{sec:differential-geometry-of-miura-ori} and the Gaussian curvature is then evaluated for each cell using eq.~\eqref{eq:gaussian-by-fundamental-forms}.
Fig.~\ref{fig:inverse-design-actual-K} shows the Gaussian curvature of the resulting surface. We can see that the actual origami has roughly the desired constant curvature, with $\approx 1$ at the middle of the pattern and $\approx 0.9$ at the edges of the pattern.

\begin{figure}
    \centering
    \includegraphics[width=0.85\linewidth]{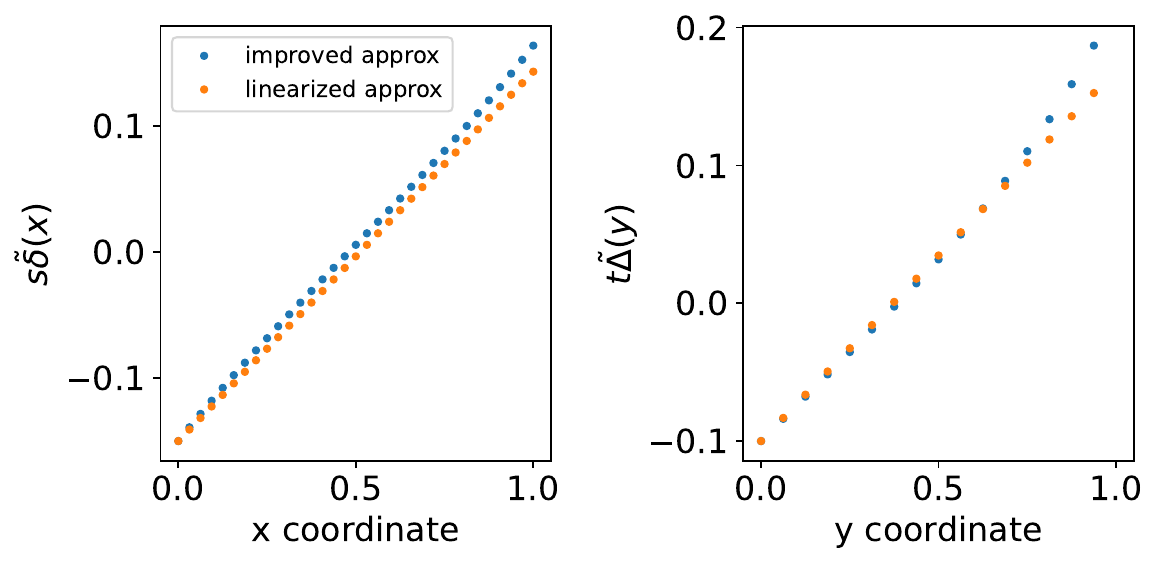}
    \caption{
    The perturbation fields $s \tilde{\delta}$ and $t \tilde{\Delta}$ used for the inverse design of a spherical cap.
    The angle perturbation $\tilde{\delta}$ is sampled for every horizontal vertex on $x\in \{0, \frac{1}{2}\chi, \chi, ... , 1-\frac{1}{2} \chi , 1\}$ and the length field $\tilde{\Delta}$ is sampled for every unit cell with $y\in \{ 0, \xi, ..., 1-\xi\}$.
    Blue dots are the fields found by solving the weakly nonlinear approximation for the principal curvatures from sec.~\ref{sec:principal-curvatures-approx} and the orange dots are the fields achieved by solving the linearized approximation from sec.~\ref{sec:SFF-linear-approx}, for the same initial conditions and the same origami pattern's parameters.
    It is interesting to see that while the perturbations in both methods are quite close to each other, the patterns created by them are quite different as shown in the Gaussian curvature heatmaps fig.~\ref{fig:inverse-design-actual-K}--\ref{fig:inverse-design-comparison-linearized}.
    }
    \label{fig:inverse-design-example-pert-comparison}
\end{figure}

\paragraph{Example: Linear Design of a spherical cap}
We could also design the same spherical cap but use the linearized expressions for the principal curvatures, written in eqs.~\eqref{eq:kx-approx}--\eqref{eq:ky-approx}, to find the perturbation fields.
It is clear that for constant principal curvatures, both fields should grow linearly where the slope is determined by the origami pattern's parameters. For the same parameters described above, and the same initial conditions, we find that the perturbations should be $s \tilde{\delta}(x)=0.293 x + s \tilde{\delta}(0)$ and $t \tilde{\Delta}(y)=0.269 y + t \tilde{\Delta}(0)$.
In fig.~\ref{fig:inverse-design-example-pert-comparison} we show these linear perturbations compared to the more accurate perturbations used to create the surface in fig.~\ref{fig:inverse-design-example}.
We see there the slight deviation of the numerical solutions to the improved approximation from the linear fields obtained here.

A second pattern is built using these simplified perturbation fields and after folding the pattern to the same target activation angle, the Gaussian curvature is calculated for each unit cell.
The heat map in fig.~\ref{fig:inverse-design-comparison-linearized} shows the Gaussian curvature of this pattern, where we see that despite the target Gaussian curvature of constant with value $1$, the actual surface has this curvature only at small region in the middle, and the value drops to $\approx 0.6$ at the edges, much worse compared to the weakly nonlinear approximation method of the principal curvatures from sec.~\ref{sec:principal-curvatures-approx}.

\paragraph{Further examples}

\begin{figure}
    \centering
    \begin{tabular}{@{} c @{}  c @{}  c @{}}
    
    \begin{subfigure}{0.32\linewidth}
        \centering
        \includegraphics[width=\linewidth]{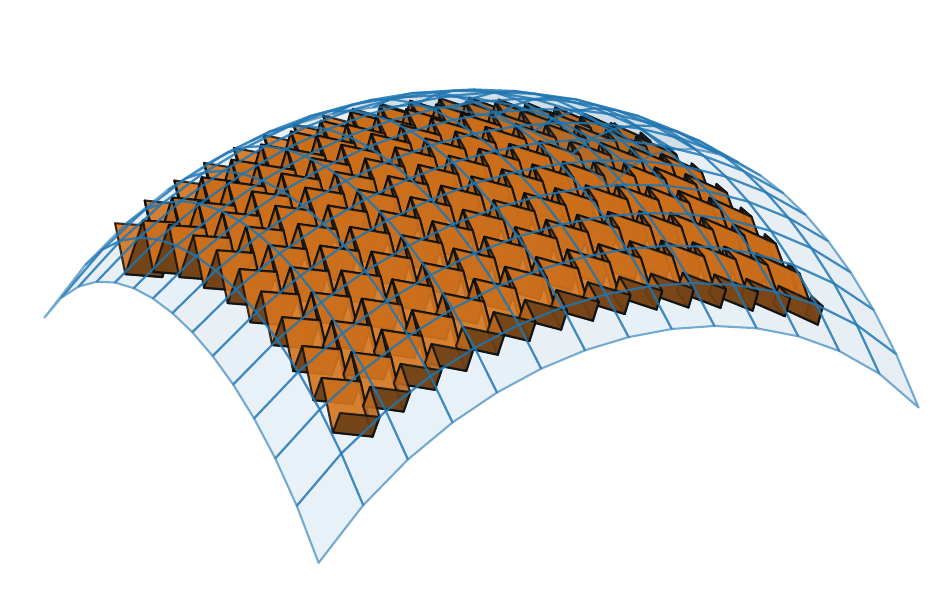}
        \caption{Elliptic geometry: $\kappa_x=\kappa_y>0$}
    \end{subfigure}
    &
    \begin{subfigure}{0.32\linewidth}
        \centering
        \includegraphics[width=\linewidth]{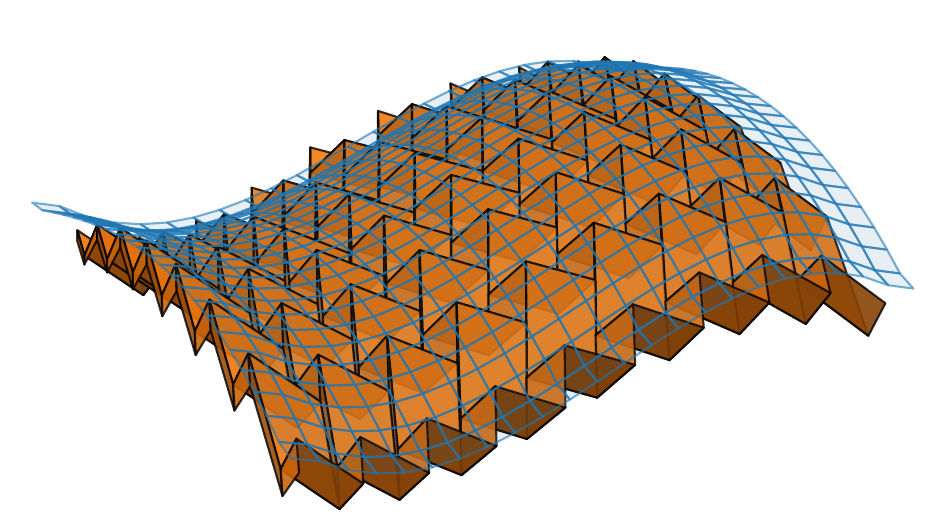}
        \caption{Spatially varying geometry
    }
    \end{subfigure}
    &
    \begin{subfigure}{0.32\linewidth}
        \centering
        \includegraphics[width=\linewidth]{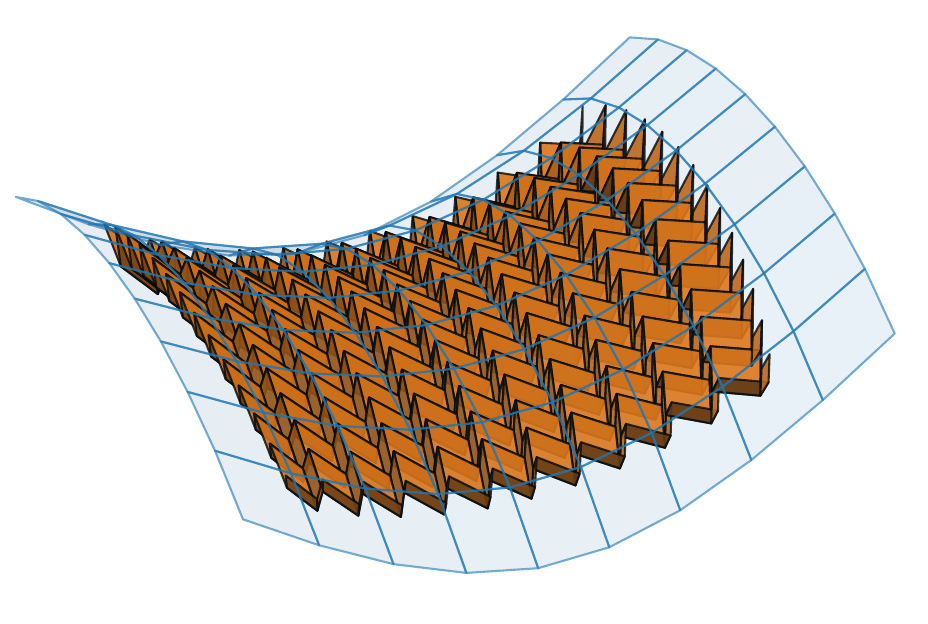}
        \caption{Hyperbolic geometry: $\kappa_y=-\kappa_x>0$}
    \end{subfigure}
    \\
    \begin{subfigure}{0.315\linewidth}
        \centering
        \includegraphics[width=\linewidth]{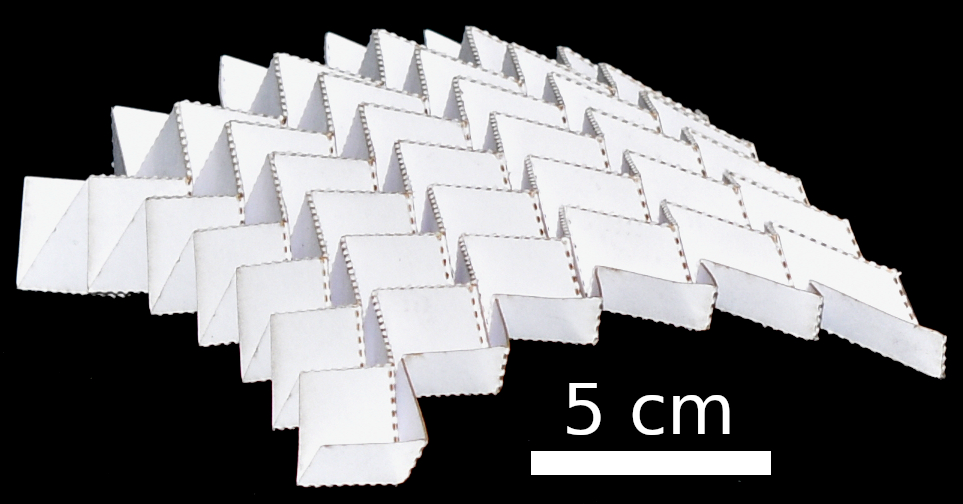}
        \caption{}
    \end{subfigure}
    &
    \begin{subfigure}{0.315\linewidth}
        \centering
        \includegraphics[width=\linewidth]{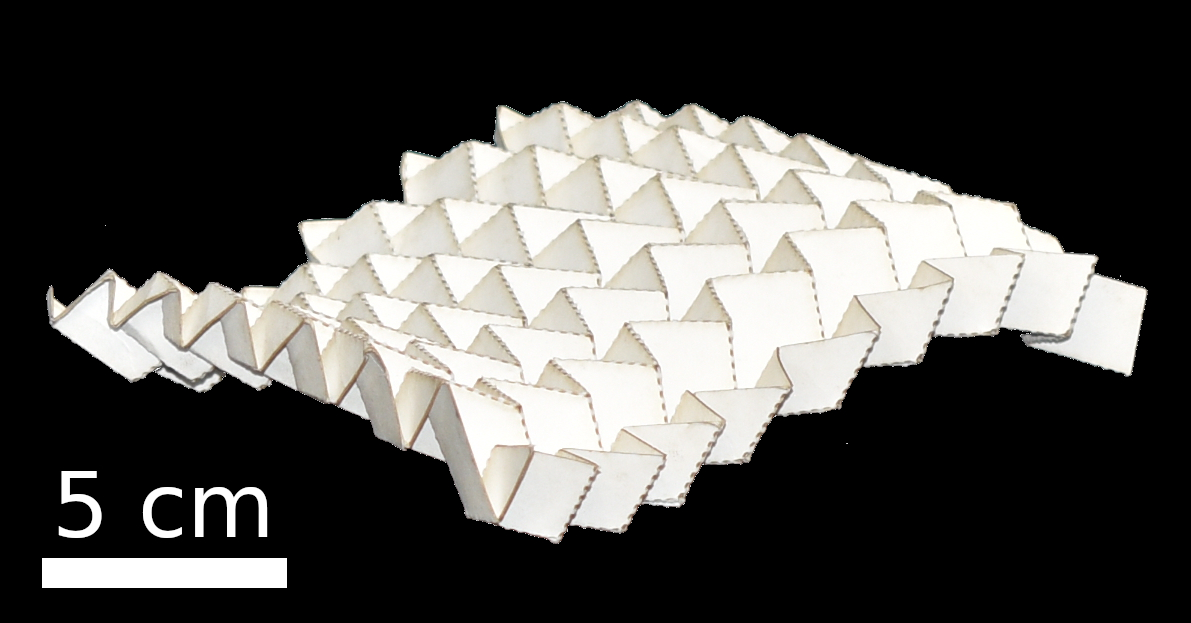}        
        \caption{}
    \end{subfigure}
    &
    \begin{subfigure}{0.315\linewidth}
        \centering
        \includegraphics[width=\linewidth]{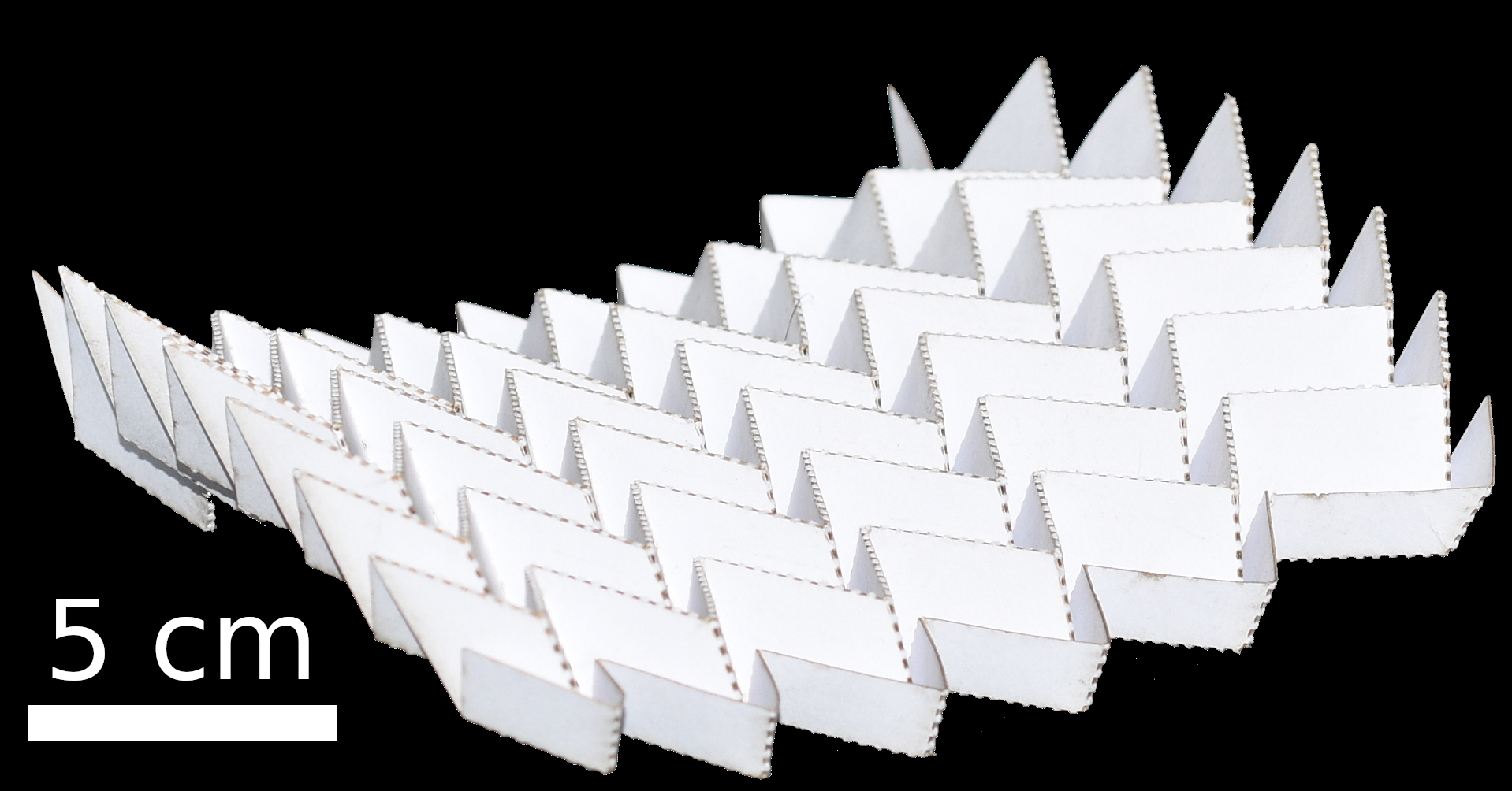}
        \caption{}
    \end{subfigure}
    \\
    \begin{subfigure}{0.27\linewidth}
        \centering
        \includegraphics[width=\linewidth]{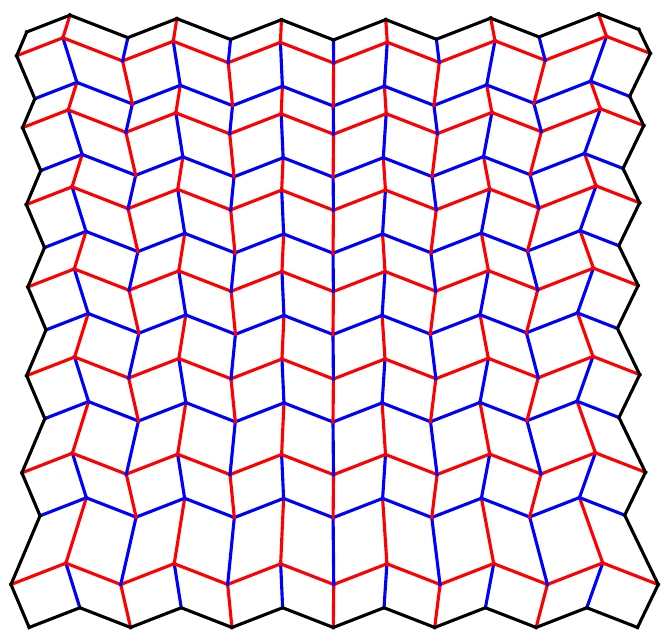}
        \caption{}
    \end{subfigure}
    &
    \begin{subfigure}{0.33\linewidth}
        \centering
        \includegraphics[width=\linewidth]{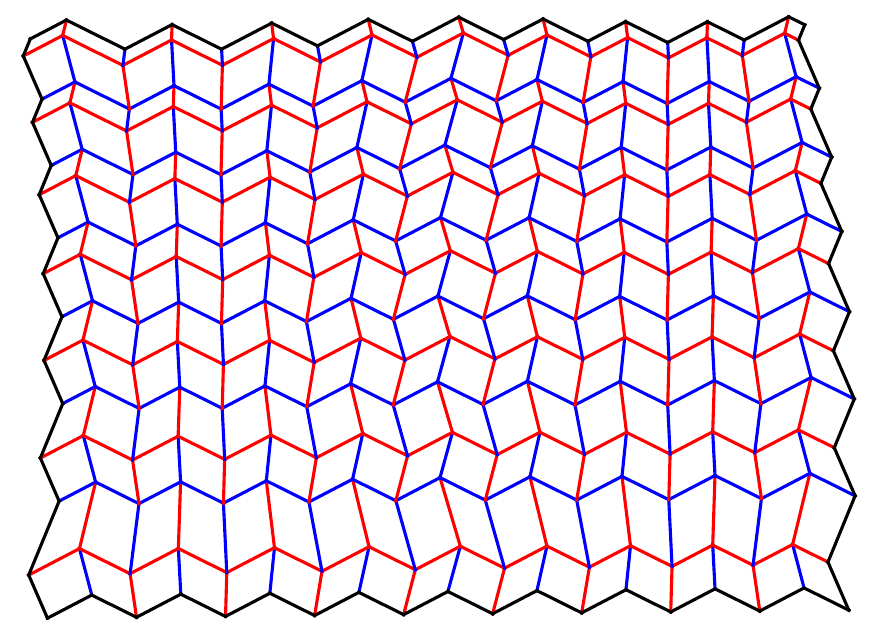}
        \caption{}
    \end{subfigure}
    &
    \begin{subfigure}{0.27\linewidth}
        \centering
        \includegraphics[width=\linewidth]{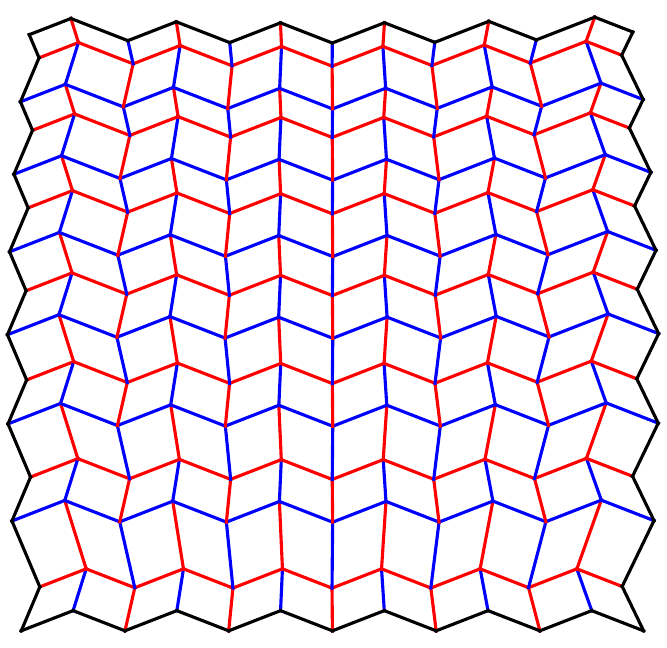}
        \caption{}
    \end{subfigure}
    
    \end{tabular}

    \begin{tabular}{@{}c@{}@{}c@{}}
        \subfloat[Vase: $\kappa_x=\const>0, \kappa_y\propto \cos(\pi y)$]{
        \centering
        \includegraphics[width=0.49\linewidth]{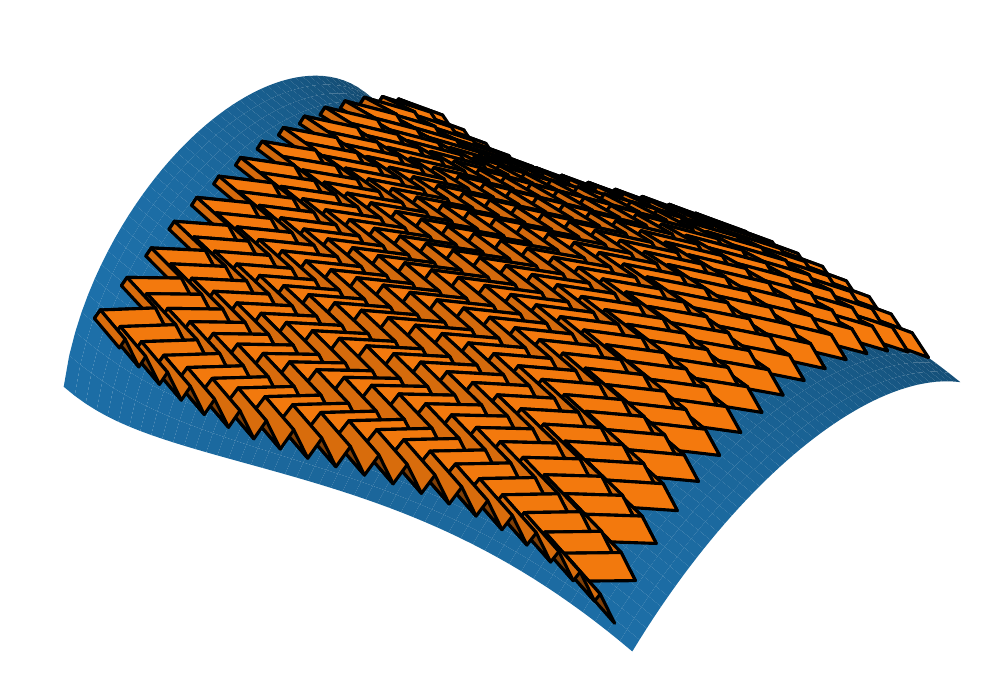}}
        &
        \subfloat[Sinusoidal curvatures: $\kappa_x \propto \sin(2\pi x), \kappa_y\propto \sin(2\pi y)$]{
        \centering
        \includegraphics[width=0.49\linewidth]{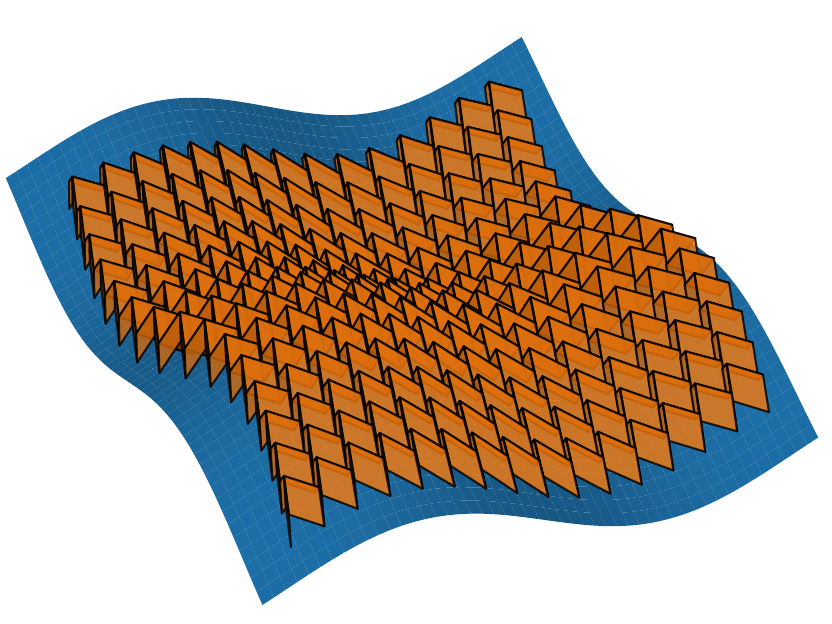}
        }
    \end{tabular}
	
    \caption{
        Examples of inverse designs of some simple surfaces with principal curvature of the form $\kappa_x(x)$ and $\kappa_y(y)$ where the perturbation fields $\tilde \delta$ and $\tilde \Delta$ were obtained using our weakly nonlinear approximation to the principal curvatures.
        (a) is a surface with constant positive Gaussian curvature, with principal curvatures that are $\kappa_x=\kappa_y=\const>0$.
        (b) is a surface with both negative and positive Gaussian curvature, obtained by targeting a surface with $\kappa_x\propto \tanh\left(-3\left(x - 0.5\right)\right), \kappa_y=\const>0$.
        (c) is a surface with constant and negative Gaussian curvature with principal curvatures $\kappa_x={-}\kappa_y=\const<0$.
        (d)--(f) are physical realizations of the surfaces in (a)--(c) with a sheet of paper. Examples (d) and (f) have less unit cells than their computational counterparts (a) and (c), respectively. (g)--(i) are the flat crease patterns used for the folded pattern (d)--(f), respectively.
        (j) is a part of a vase, with curvatures $\kappa_x=\const>0, \kappa_y\propto \cos(\pi y)$ and (k) is a surface with $\kappa_x \propto \sin(2\pi x), \kappa_y\propto \sin(2\pi y)$.
    }
    \label{fig:more-design-examples}
\end{figure}

This procedure for the inverse design was used to build the origami patterns that appear in fig.~\ref{fig:more-design-examples}.
The exact parameters $C_{tot},L_{tot},\vartheta,\omega_{0,0},t \tilde{\Delta}(0), s \tilde{\delta}(0)$ were chosen by short trial and error to create valid patterns with relative small perturbations.

\begin{figure}
    \centering

    \subfloat[\label{fig:inverse-design-periodic}Surface with periodic principal curvatures]{
        \centering
        \includegraphics[width = 0.8\linewidth]
    	{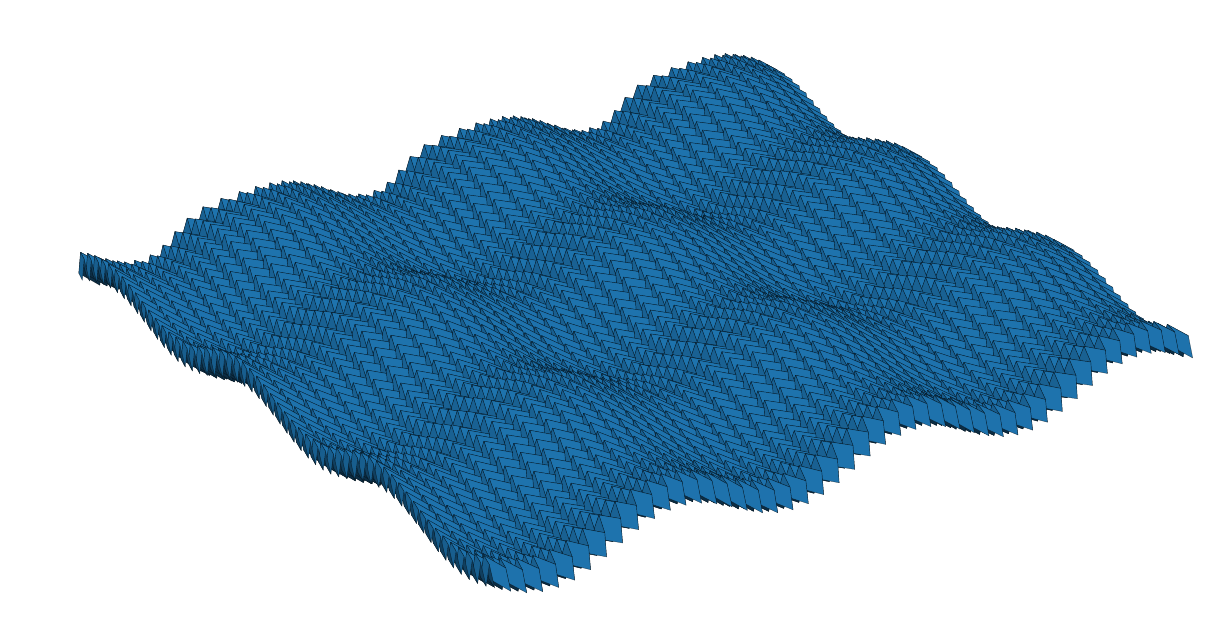}
    }
    
    \subfloat[Gaussian curvature comparison
            \label{fig:inverse-design-periodic-K-comparison}]{
		\centering
		\includegraphics[width=0.40\linewidth]{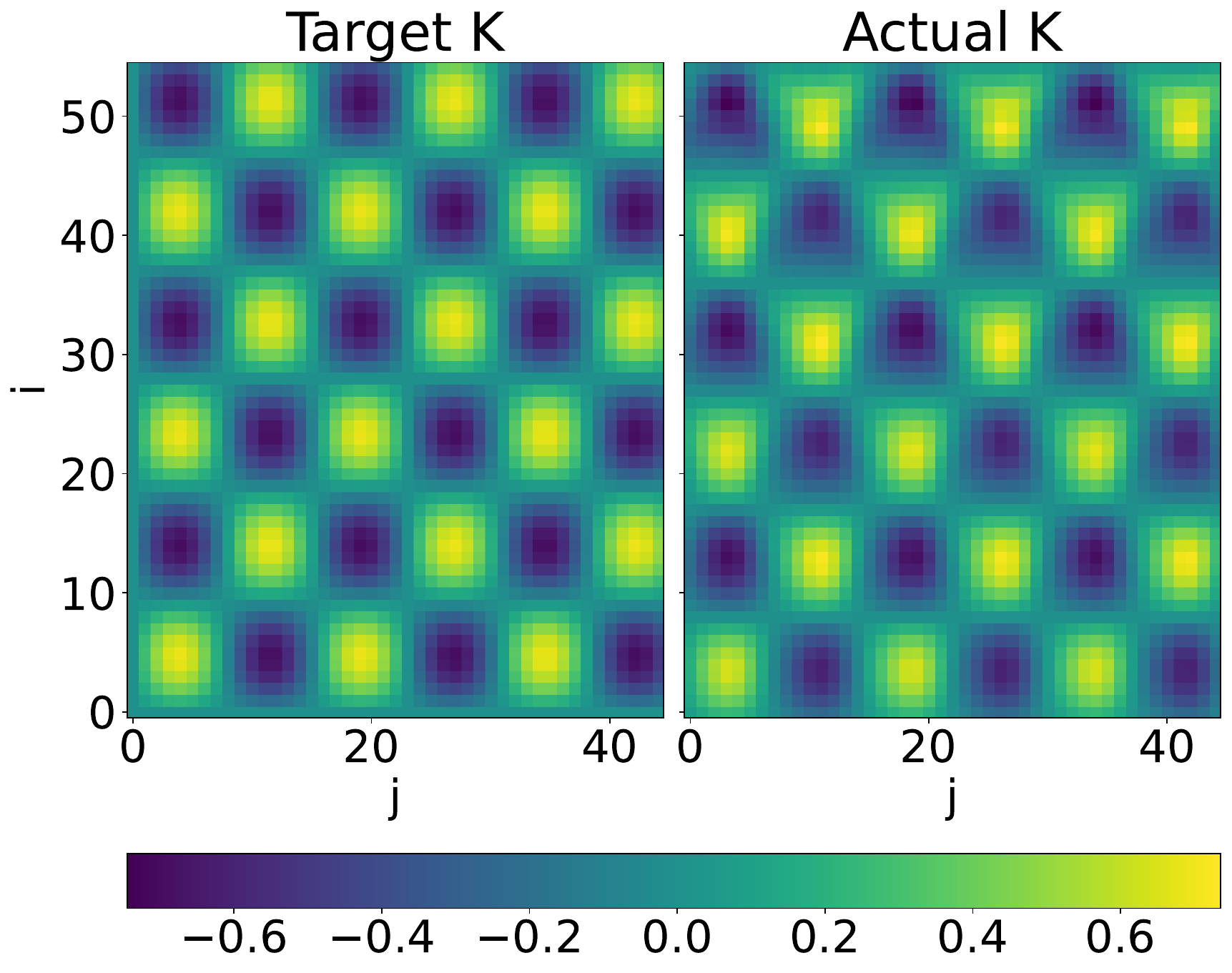}
	}
    \subfloat[Mean curvature comparison
            \label{fig:inverse-design-periodic-H-comparison}]{
		\centering
		\includegraphics[width=0.40\linewidth]{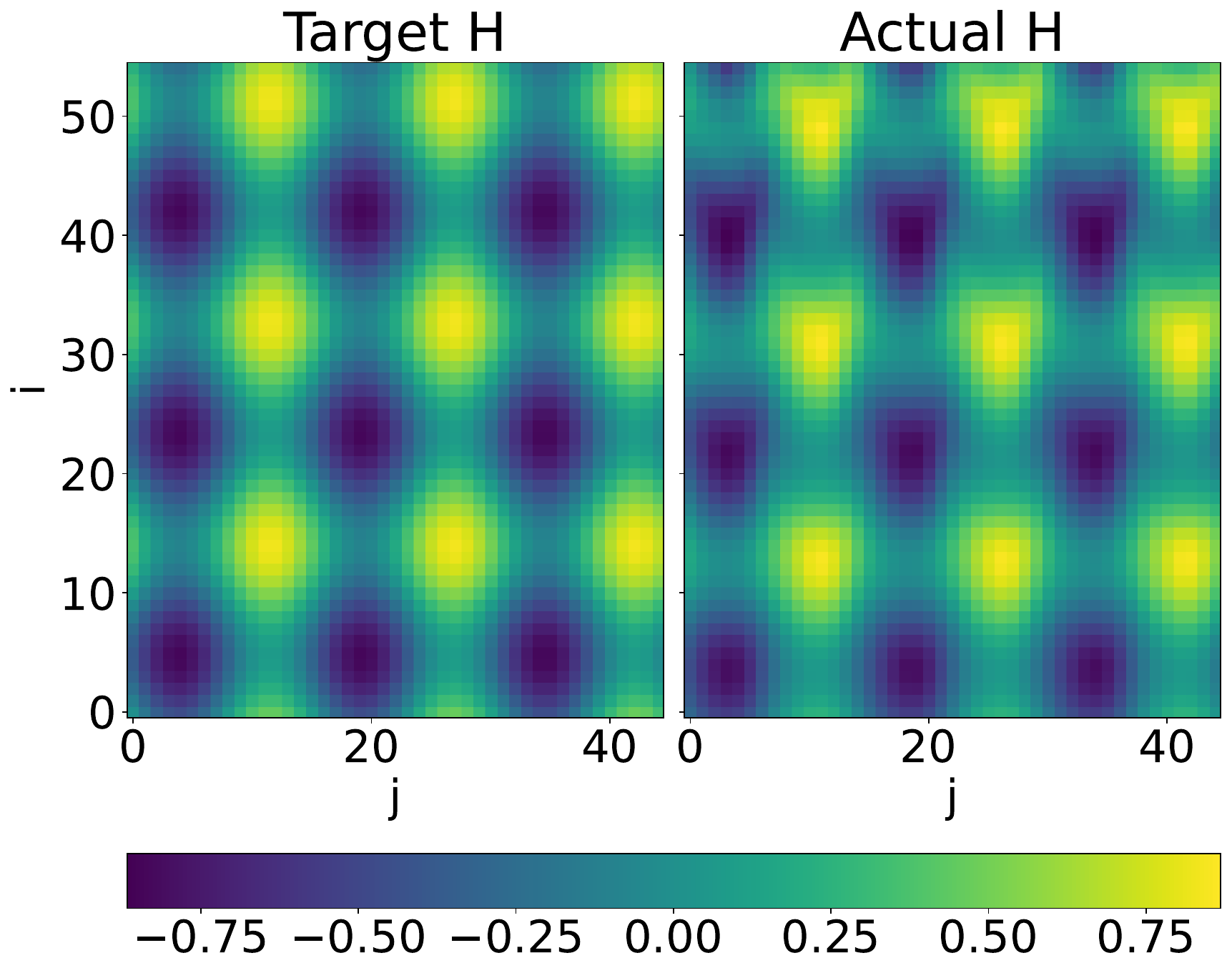}
	}
	
	\caption{
        Inverse designing a surface with periodic curvatures of the form: $\kappa_x\propto \sin \left(3\cdot 2 \pi x \right), \kappa_y\propto \sin \left(3\cdot 2 \pi y \right)$.
        (a) The resulting surface that shows the periodic mountains and valleys. We note that the origami pattern itself is not periodic.
        (b) Comparison between the target Gaussian curvature and the actual curvature obtained by numerically evaluating the Gaussian curvature for each cell in the resulting surface.
        (c) Comparison between the mean curvature $H=\frac{1}{2}(\kappa_x+\kappa_y)$ of the target surface and the resulting surface, calculated for each unit cell.
        We can see that both the Gaussian and the mean curvatures agree nicely with the target curvatures when $i$ is small, but that the curvature maps become more distorted at the upper part of the pattern, with $i\gtrsim 35$.
        If the target curvatures are smaller, then the perturbations needed are also smaller and the curvatures agree more nicely.
    }
\end{figure}

Fig.~\ref{fig:inverse-design-periodic} shows the design of eggs-tray like surface, with periodic principal curvatures. A comparison between the actual Gaussian curvature and the target is shown in fig.~\ref{fig:inverse-design-periodic-K-comparison}, and comparison between the actual and target mean curvatures is shown in fig.~\ref{fig:inverse-design-periodic-H-comparison}.

\section{Discussion}\label{sec:discussion}
In this work we developed a continuum geometric framework for the inverse design of perturbed Miura-Ori. 
Our framework considers slowly varying perturbations that are used to obtain closed formulas for the lengths and angles along the pattern that otherwise had to be obtained by iteratively solving the marching algorithm.
These perturbations are then used to approximate the classical geometric fields of the folded origami pattern such as its metric and curvatures.
Once these expressions for the geometric fields are found, we could easily address the inverse design problem by inverting these expressions and find the required perturbations given the target geometric fields of the surface we try to design.
We employed the alternating perturbation subclass of RFFQM patterns and provided several patterns that solve a list of representative inverse problems, including surfaces with constant and undulating Gaussian curvature, showing the effectiveness of our framework.\footnote{A comment about uniqueness: Even within the subclass of alternating perturbations, the relation between the pattern and the emerging geometry is not unique. For example, both classical MO pattern the Mars pattern (Fig.~\ref{fig:MARS-example}) give rise to flat configurations. 
However, for this subclass the space of possible patterns (for a given geometry and number of pannels) is quite limited: In our algorithm the pattern is obtained uniquely after fixing the initial data $\vartheta, \omega_{0,0}, \tilde{\Delta}(0)$ and $\tilde{\delta}(0)$, which are parameters that can be optimized about (see examples in the previous section).}

\paragraph{Approximations in the coarse-grained approach}
The coarse-grained approach we followed in this paper involves, in principle, two types of approximations: given perturbation fields, deriving PDEs for the angles $\eta_{i,j}$, the vertical lengths $\ell_{i,j}$ and the horizontal lengths $c_{i,j}$ of the panels (sec.~\ref{sec:slowly-varying-pert-lengths}) and the folding angles $\omega_{i,j}, \gamma_{i,j}$ (sec.~\ref{sec:slowly-varying-pert-angles}), and we solve it to a leading order. 
We then use these solutions to obtain the contents of the first and second fundamental forms, and principle curvatures, again to leading orders (sec.~\ref{sec:forward-design-geometric-fields}).
The geometrical expressions we obtain are thus either accurate only up to quadratic order of the perturbation fields (in the linear approximation), or, in the weakly nonlinear approximation, assume that the effect of the perturbations is independent of each other.
An immediate question that arises is about the accuracy of these approximations. While we did not provide an explicit bound on the deviation from the target geometries, this work calls for a rigorous study of estimating the level of approximation.

\paragraph{More general slowly varying GMOs}
A natural generalization of this work is to consider perturbations along the L-shaped boundary that are more general than the ones used in this work (alternating angle perturbations on the bottom boundary, length perturbations on the left boundary), in which case one could still obtain (more complicated) equations for the origami angles and lengths discussed above.
While this has the potential to obtain an even richer variety of surfaces, we note two special features of the perturbations considered in this work, which will have to be dealt more carefully in a more general design: 
\begin{itemize}
    \item The fields $\eta_{i,j}$ and are immediately solved exactly, and the equation that we solve for $\ell_{i,j}$ and $\omega_{i,j}$ is exact; the omission of higher order terms occurs only for the fields $c_{i,j}$ and $\gamma_{i,j}$.
    Thus, a more general perturbation might require more approximations (possibly restricting the validity of the obtained solutions), or alternatively solving much more complicated equations in order to obtain exact solutions to some of the fields.
    \item The geometry of solutions of some special cases (sec.~\ref{sec:pert-special-cases} is either exact or can be easily understood in the leading order. 
    This enables us to easily identify the leading order in which the Gauss and principle curvatures should appear.
    In more general perturbations the identification of the leading order terms might be more convoluted, which might lead to more complicated equations (which will eventually need to be inverted for the inverse design procedure). 
\end{itemize}

\paragraph{Mechanics of GMOs}
Another aspect of the origami that we did not discuss here is the elastic energy of the panels. While we assumed throughout this work that the panels are rigid and the folded configurations have no elastic energy, it is interesting to study the mechanics of these generalized Miura-Ori under some external constraints, such as forces or displacement.
Previous works have studied the continuum mechanical description of the classical Miura-Ori \cite{nassar2018fitting, xu2023derivation}, and the fundamental forms that we studied in the perturbed case can help develop the theory of elasticity of these origami surfaces, in the spirit of \cite{efrati2009elastic}.
In other words, in this work we discuss an isometric limit --- stress- and torque-free configurations. By developing a theory of coarse-grained mechanics for these GMOs, allowing stretching and bending of faces, and penalizing for the bending of the folds, one could obtain an even richer variety of inverse design possibilities for a given set of MO perturbations. The results of the current work can serve as the basis for a future mechanical theory, where the set of isometries characterized here will serve as the set relative to which deformations are measures.

\paragraph{Limitations of GMO design}
While it is known that generic origami can recover any surface in 3D \cite{demaine2017origamizer}, the limitations on the geometries induced by generalized Miura-Ori are far from being understood. Our analytical method, in contrast to previous numerical approaches, reveals limitations in targeting desired geometries. For example, the condition on the perturbation field $-\vartheta < \delta(i) < \vartheta$ forms a solvability condition for the expression relating geometric properties to the imposed perturbation. Therefore, we anticipate that a nonlinear extension of our framework, valid for large perturbations and leading to analytical expressions relating geometry to perturbations, will uncover solvability conditions for the specific inverse design problem under study.

\paragraph{Similar coarse-grained approach in other systems}
In conclusion, we emphasize again that our methodology is not limited to the alternating perturbations studied in this work, and that our approach could be adapted to studying other families of perturbations within the same continuous field approach, including those studied in \cite{zhou2015design, wang2016folding, song2017design, dang2022deployment}. 
A similar methodology can be further extended to other types of discrete geometric mechanism, e.g., curvygami, kirigami, and other shape morphing mechanisms and materials (for example, to the kirigami structure with slowly-varying unit cells recently studied in \cite{peng2024programming}).
Doing so with origami and other inherently discrete systems will lead to effective continuous description of these systems, with the benefit of porting methods and tools from continuum theories of matter and differential geometry to the field of origami, kirigami and more.

\section*{Methods}
This manuscript is accompanied with a Wolfram Mathematica notebook recovering  central calculations in this work:
\url{https://github.com/AlonSardas/InverseDesignOriNotebook}.
 
\section*{Acknowledgment}
This work was supported by Israel Science Foundation Grants No.\ 1441/19, 1296/19, and 2304/24, U.S.-Israel Binational Science Foundation Grant No.\ 2022076 and the Vigevani Foundation.

\clearpage
\appendix

\section{Kinematic calculations} \label{app:kinematic-calculations}
For the first and second fundamental forms, we need to calculate the tangent vectors of the unit cells.
These are calculated by following the kinematics of the pattern, similar to what we showed in sec.~\ref{sec:geo-properties-of-unperturbed-ori}, with the notation from fig.~\ref{fig:unit-cell-lengths-and-angles-perturbations}.
The tangent vectors of the left-bottom unit cell are evaluated by
\[
    \vec{AE}_{0, 0}
    =\vec{AC}^{\text{flat}}+\vc{R}_{DC}\left(\omega_{0,0}\right)\vec{CE}^{\text{flat}}
    ,
\]
\[
    \vec{AJ}_{0, 0}
    =\vec{AB}^{\text{flat}}+\vc{R}_{DB}\left(-\gamma_{0,0}\right)\vec{BJ}^{\text{flat}},
\]
where $\vc{R}_{DC}(\omega)$ is the right-hand rotation matrix around the axis $\vec{DC}^{\text{flat}}$ by the angle $\omega$.
To calculate the vector $\vec{AE}_{1, 0}=\vec{JG}$, we rotate the panel $DHFG$ around $\vec{DH}$ and then fold again the 2nd row panels around the crease $\vec{DB}$.
We obtain:
\begin{equation*}
    \vec{AE}_{1,0}=\vc{R}_{DB}\left(-\gamma_{0,0}\right)\left(\vec{JH}^{\text{flat}}+\vc{R}_{DH}\left(\omega_{0,0}\right)\vec{HG}^{\text{flat}}\right).
\end{equation*}
We will need to consider the adjacent unit cells to calculate the tangent vectors $\vec{AE}_{0,1}$ and $\vec{AJ}_{1,0}$. Their values, according to the kinematics are:
\begin{equation*}
    \vec{AJ}_{1,0}=\vc{R}_{DB}\left(-\gamma_{0,0}\right) \vc{R}_{HJ}\left(-\gamma_{HJ}\right)\left(\vec{AB}_{1,0}^{\text{flat}}+\vc{R}_{{DB}_{1,0}}\left(-\gamma_{1,0}\right)\vec{BJ}_{1,0}^{\text{flat}}\right),
\end{equation*}
\begin{equation*}
    \vec{AE}_{1,0}=
    \vc{R}_{DC}\left(\omega_{0,0}\right) \vc{R}_{FE}\left(\omega_{EF}\right) \left(\vec{AC}_{0,1}^{\text{flat}}+\vc{R}_{DC_{0,1}}\left(\omega_{0,1}\right)\vec{CE}_{0,1}^{\text{flat}}\right)
    ,
\end{equation*}
with $\gamma_{HJ},\omega_{EF}$ the folding angles around the crease lines $\vec{HJ},\vec{EF}$.

\section{Compatibility of alternating angles} \label{app:compatibility-of-alternating-angles}
Here we verify that the alternating angles described by the perturbations in eq.~\eqref{eq:alternating-angles-by-pert} satisfy the compatibility conditions for RFFQM.
For that we rewrite the compatibility condition using the perturbed angles notation $\beta^L_{m,n},\beta^R_{m,n}$.

\begin{figure}
    \centering
    \includegraphics[width=0.4\linewidth]
    {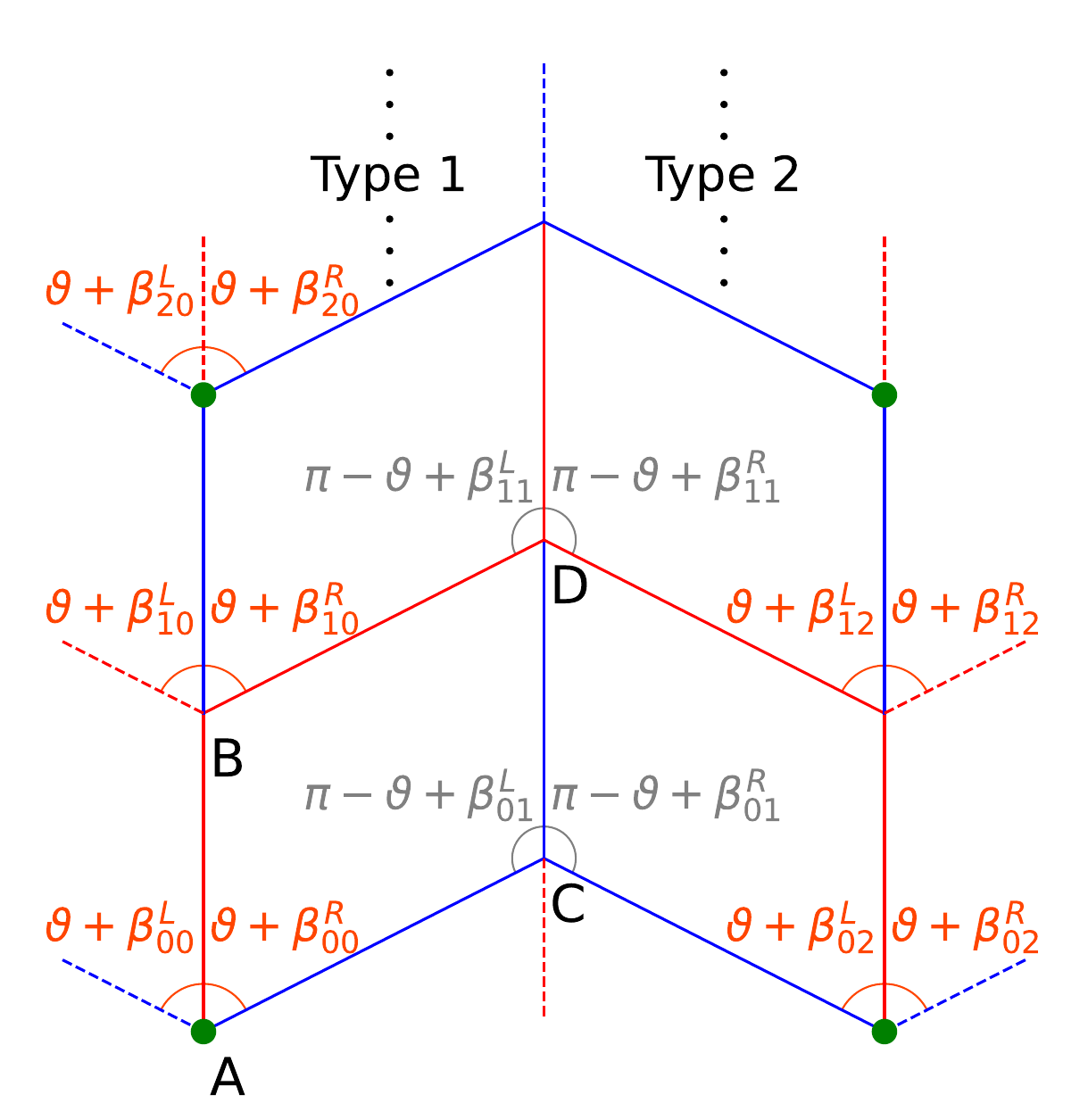}
 
 \caption{
     A part of the origami crease pattern that indicates the angle perturbations $\beta^L_{m,n},\beta^R_{m,n}$ for each vertex of the crease.
     There are 2 types of panels that are characterized by their base Miura-Ori angles. These are the same along the columns and alternate between columns.
    }
\label{fig:crease-pattern-angles}
\end{figure}

After setting the sector angles of the pattern by perturbations to Miura-Ori, given in eq.~\eqref{eq:crease-angles-by-pert}, we can distinguish 2 types of parallelograms in the pattern: the first is with angles
\begin{equation*}
    \begin{bmatrix}
    \alpha^L_{A} & \alpha^R_{A}\\
    \alpha^L_{B} & \alpha^R_{B}\\
    \alpha^L_{C} & \alpha^R_{C}\\
    \alpha^L_{D} & \alpha^R_{D}
    \end{bmatrix}=\begin{bmatrix}
    \vartheta+\beta^L_{A} & \vartheta+\beta^R_{A}\\
    \vartheta+\beta^L_{B} & \vartheta+\beta^R_{B}\\
    \pi-\vartheta+\beta^L_{C} & \pi-\vartheta+\beta^R_{C}\\
    \pi-\vartheta+\beta^L_{D} & \pi-\vartheta+\beta^R_{D}
    \end{bmatrix},
\end{equation*}
and the other is similar, with the change of $\vartheta\to \pi -\vartheta$.
These 2 types are the same along the columns and the type is alternated in each column, as seen in fig.~\ref{fig:crease-pattern-angles}.
For both types, the sum of angles inside the quadrangle from eq.~\eqref{eq:angles-in-quadrangles} becomes:
\begin{equation} \label{eq:angles-in-quadrangle-pert-only}
    \beta^R_{D}=\beta^R_{A}-\beta^L_{B}+\beta^L_{C}.
\end{equation}

This condition is satisfied for the pattern eq.~\eqref{eq:alternating-angles-by-pert}, as eq.~\eqref{eq:alternating-angles-by-pert} implies that for any $m,n$ we have $\beta^R_{m,n}=-\beta^L_{m,n}=\beta^L_{m+1,n}=-\beta^R_{m+1,n}$, and thus for each quadrangle with vertices $A,B,C,D$ as in fig.~\ref{fig:panel-compatibility}, $\beta_D^R=\beta_C^L$ and $\beta_{A}^{R}=\beta_{B}^{L}$. 

Next, we use the fold angle multiplier mechanism, following the work done by \cite{feng2020designs}, to express the smooth folding compatibility condition from eq.~\eqref{eq:alpha-D-L-by-mu-abc}.
This equation, when assuming non-degeneracy $\alpha_D^L + \alpha_D^R\neq \pi$, is equivalent to demanding:
\begin{equation}
    \frac{\cos\left(\frac{1}{2}\left(\alpha_{D}^{L}-\alpha_{D}^{R}\right)\right)}{\cos\left(\frac{1}{2}\left(\alpha_{D}^{L}+\alpha_{D}^{R}\right)\right)}\frac{\cos\left(\frac{1}{2}\left(\alpha_{C}^{L}+\alpha_{C}^{R}\right)\right)}{\cos\left(\frac{1}{2}\left(\alpha_{C}^{L}-\alpha_{C}^{R}\right)\right)}\frac{\cos\left(\frac{1}{2}\left(\alpha_{B}^{L}+\alpha_{B}^{R}\right)\right)}{\cos\left(\frac{1}{2}\left(\alpha_{B}^{L}-\alpha_{B}^{R}\right)\right)}\frac{\cos\left(\frac{1}{2}\left(\alpha_{A}^{L}-\alpha_{A}^{R}\right)\right)}{\cos\left(\frac{1}{2}\left(\alpha_{A}^{L}+\alpha_{A}^{R}\right)\right)}=1
    .
\end{equation}
Substituting the perturbed angles, for the first type of panels we get:
\begin{equation*}
    \frac{\cos\left(\frac{1}{2}\left(\beta_{D}^{L}-\beta_{D}^{R}\right)\right)}{\cos\left(\vartheta-\frac{1}{2}\left(\beta_{D}^{L}+\beta_{D}^{R}\right)\right)}\frac{\cos\left(\vartheta-\frac{1}{2}\left(\beta_{C}^{L}+\beta_{C}^{R}\right)\right)}{\cos\left(\frac{1}{2}\left(\beta_{C}^{L}-\beta_{C}^{R}\right)\right)}\frac{\cos\left(\vartheta+\frac{1}{2}\left(\beta_{B}^{L}+\beta_{B}^{R}\right)\right)}{\cos\left(\frac{1}{2}\left(\beta_{B}^{L}-\beta_{B}^{R}\right)\right)}\frac{\cos\left(\frac{1}{2}\left(\beta_{A}^{L}-\beta_{A}^{R}\right)\right)}{\cos\left(\vartheta+\frac{1}{2}\left(\beta_{A}^{L}+\beta_{A}^{R}\right)\right)}=1
    ,
\end{equation*}
and for the second type:
\begin{equation*}
    \frac{\cos\left(\frac{1}{2}\left(\beta_{D}^{L}-\beta_{D}^{R}\right)\right)}{\cos\left(\vartheta+\frac{1}{2}\left(\beta_{D}^{L}+\beta_{D}^{R}\right)\right)}\frac{\cos\left(\vartheta+\frac{1}{2}\left(\beta_{C}^{L}+\beta_{C}^{R}\right)\right)}{\cos\left(\frac{1}{2}\left(\beta_{C}^{L}-\beta_{C}^{R}\right)\right)}\frac{\cos\left(\vartheta-\frac{1}{2}\left(\beta_{B}^{L}+\beta_{B}^{R}\right)\right)}{\cos\left(\frac{1}{2}\left(\beta_{B}^{L}-\beta_{B}^{R}\right)\right)}\frac{\cos\left(\frac{1}{2}\left(\beta_{A}^{L}-\beta_{A}^{R}\right)\right)}{\cos\left(\vartheta-\frac{1}{2}\left(\beta_{A}^{L}+\beta_{A}^{R}\right)\right)}=1.
\end{equation*}
In the alternating angles, $\beta_{m,n}^R=-\beta_{m,n}^L$, and these conditions for both types become:
\begin{equation*}
    \frac{\cos\left(\beta_{D}^{L}\right)}{\cos\left(\beta_{C}^{L}\right)}\frac{\cos\left(\beta_{A}^{L}\right)}{\cos\left(\beta_{B}^{L}\right)}=1.
\end{equation*}
This condition is satisfied for the alternating pattern \eqref{eq:alternating-angles-by-pert}, since the pattern \eqref{eq:alternating-angles-by-pert} directly applies that in each quadrilateral, $\beta_{D}^{L}=-\beta_{C}^{L}$ and $\beta_{A}^{L}=-\beta_{B}^{L}$.

We conclude that the alternating angles in the form \eqref{eq:alternating-angles-by-pert} are compatible along the entire crease, meaning that if we impose those angles on the boundary alone (appears in \eqref{eq:alternating-angles-by-pert-on-boundary}) then the marching algorithm will yield the alternating angles for all the vertices in the pattern.

\section{Linearized compatibility condition} \label{app:linearized-marching-algorithm}
We would like to study the compatibility conditions of a panel under the assumption that the angles are close to Miura-Ori pattern, with small perturbations $\beta^L_{m,n},\beta^R_{m,n}\ll \vartheta$. These perturbations are added to the base Miura-Ori angles for each vertex along the panel, as shown in fig.~\ref{fig:crease-pattern-angles}.
Linearizing the compatibility conditions might help to discover other types of angle perturbations to Miura-Ori under long-wavelength continuous formulation.

The first condition for the angles is for the angle sum in a quadrangle from eq.~\eqref{eq:angles-in-quadrangle-pert-only} that is already linear in the perturbations.

For the second condition, we focus on the angles for the first type of panels from fig.~\ref{fig:crease-pattern-angles} and calculate $\mu_{ABC}$ from eq.~\eqref{eq:mu-abc}:
\begin{equation*}
    \mu_{ABC}=\frac{\cos\left(\frac{1}{2}\left(\beta_{A}^{R}-\beta_{A}^{L}\right)\right)}{\cos\left(\vartheta+\frac{1}{2}\left(\beta_{A}^{R}+\beta_{A}^{L}\right)\right)}\frac{\cos\left(\vartheta+\frac{1}{2}\left(\beta_{B}^{R}+\beta_{B}^{L}\right)\right)}{\cos\left(\frac{1}{2}\left(\beta_{B}^{R}-\beta_{B}^{L}\right)\right)}\frac{\cos\left(\vartheta-\frac{1}{2}\left(\beta_{C}^{R}+\beta_{C}^{L}\right)\right)}{\cos\left(\frac{1}{2}\left(\beta_{C}^{R}-\beta_{C}^{L}\right)\right)}.
\end{equation*}
We extract the angle $\beta^L_D$ using eq.~\eqref{eq:alpha-D-L-by-mu-abc},
and after Taylor expansion to linear order of all the perturbed angles we find the relation:
\begin{equation*}
    \beta^L_{D}\approx\beta^L_{A}-\beta^R_{B}+\beta^R_{C}.
\end{equation*}
Since this formula does not depend on $\vartheta$, it applies also to the second quadrangle type.

Repeating these calculations along the marching algorithm, we obtain the following recurrence relation:
\begin{align} \label{eq:recurrence-relation}
	\beta^R_{m+1,n+1} & =\beta^R_{m,n}-\beta^L_{m+1,n}+\beta^R_{m,n+1}  ,\nonumber \\
	\beta^L_{m+1,n+1} & =\beta^R_{m,n+1}-\beta^R_{m+1,n}+\beta^L_{m,n}  .
\end{align}
Such recurrence relation may be used as a simple alternative to the marching algorithm.
If the L-shaped boundary angles of the crease are given in form of perturbations to a Miura-Ori, this relation is enough to determine all the angles of the crease pattern, as long as the perturbations remain small enough.

\section{Approximated smooth fields metric data} \label{app:smooth-fields-metric-data}
We use the metric entries with the smooth fields approximation for the angles and lengths as appear in sec.~\ref{sec:slowly-varying-pert-lengths} and sec.~\ref{sec:slowly-varying-pert-angles}.
With these we find the metric entries for a general unit cell at $x=\chi j,y=\xi i$. 

The unit cell data approximated up to $O\left(s^3,t^3\right)$:
\begin{align*}
    \tilde{\l}_A(x,y) & \approx 
    \xi L_{tot}\left(1+\frac{1}{2}s^{2}\csc^{2}(\vartheta)\left(\tilde{\delta}(x)^{2}-\tilde{\delta}(0)^{2}\right)\right)
    ;\\
    \tilde{\l}_B(x,y) & \approx
    \xi L_{tot}\left(1+t\tilde{\Delta}(y)+\frac{1}{2}s^{2}\csc^{2}(\vartheta)\left(\tilde{\delta}(x)^{2}-\tilde{\delta}(0)^{2}\right)\right)
    ;\\
    \tilde{c}_L(x,y) & \approx
    \chi C_{tot}+\frac{1}{2}\chi L_{tot}s\csc(\vartheta)\tilde{\delta}'(x)\left(t\int_{0}^{y}\tilde{\Delta}(r)\,dr+2sy\cot(\vartheta)\tilde{\delta}(x)\right)
    ;\\
    \tilde{c}_R(x,y) & \approx
    \chi C_{tot}+\frac{1}{2}\chi L_{tot}s\csc(\vartheta)\tilde{\delta}'(x)\left(t\int_{0}^{y}\tilde{\Delta}(r)\,dr-2sy\cot(\vartheta)\tilde{\delta}(x)\right)
    ;\\
    \tilde{\omega}(x,y) & \approx
    \omega_{0,0}+\frac{1}{2}s^{2}\sin\left(\omega_{0,0}\right)\left(\tilde{\delta}(x)^{2}-\tilde{\delta}(0)^{2}\right)
    .
\end{align*}

The approximated metric entries, their derivatives and the metric inverse appear in the supplementary Mathematica notebook.

\clearpage 

\renewcommand \v \oldv

%\bibliographystyle{elsarticle-harv} 
%\bibliography{references}

\end{document}